\documentclass[11pt]{article}
\usepackage{enumerate}
\usepackage{graphicx}
\RequirePackage{natbib}
\usepackage{amssymb}
\usepackage{algorithm}
\usepackage{amsmath}
\usepackage{setspace}
\usepackage{soul}
\usepackage{color}
\usepackage[left=2.54cm,top=2.54cm,right=2.54cm,bottom=2.54cm]{geometry}
\usepackage{enumerate}
\usepackage[utf8]{inputenc}
\usepackage{amsthm}
\newtheorem{theorem}{Theorem}

\newtheorem{lemma}{Lemma}
\theoremstyle{definition}
\newtheorem{definition}{Definition}
\usepackage{smile}
\usepackage{cases}
\usepackage{bbm}
\usepackage[colorlinks,
linkcolor=red,
anchorcolor=blue,
citecolor=blue
]{hyperref}

\usepackage{xr}
\def\T{\mathrm{\scriptstyle T}} %%%transpose operator
\usepackage{fullpage}

\usepackage{tikz}

\begin{document}
	\title{Exponential Family Graphical Models: Correlated Replicates and Unmeasured Confounders, with Applications to fMRI Data}
	\author{Yanxin Jin, Yang Ning, and Kean Ming Tan}
    \maketitle
 \bibliographystyle{plain} 

\begin{abstract}
Graphical models have been used extensively for modeling brain connectivity networks.
However, unmeasured confounders and correlations among measurements are often overlooked during model fitting, which may lead to spurious scientific discoveries. 
Motivated by functional magnetic resonance imaging (fMRI) studies, we propose a novel method for constructing brain connectivity networks with correlated replicates and latent effects.  
In a typical fMRI study, each participant is scanned and fMRI measurements are collected across a period of time.
In many cases, subjects may have different states of mind that cannot be measured during the brain scan: for instance, some subjects may be awake during the first half of the brain scan, and may fall asleep during the second half of the brain scan.   
To model the correlation among replicates and latent effects induced by the different states of mind, 
we assume that the correlated replicates within each independent subject follow a one-lag  vector autoregressive model, and that the latent effects induced by the unmeasured confounders are piecewise constant.
The proposed method results in a convex optimization problem which we solve using a block coordinate descent algorithm.
Theoretical guarantees are established for parameter estimation.
We demonstrate via extensive numerical studies that our method is able to estimate latent variable graphical models with correlated replicates more accurately than existing methods.

\end{abstract}


%%%%%%%%%%%%%%%%%%%%%%%%%%%%%%%%%%%%
%%%%%%%%%%%%%%%%%%%%%%%%%%%%%%%%%%%%
% Introduction
%%%%%%%%%%%%%%%%%%%%%%%%%%%%%%%%%%%%
%%%%%%%%%%%%%%%%%%%%%%%%%%%%%%%%%%%%
\section{Introduction}
\label{intro}
Undirected graphical models have been used extensively in various scientific domains to represent conditional dependence relationships between pairs of variables.  
In a graph, each node represents a random variable, and an edge connecting a pair of nodes indicates that the pair of variables is conditionally dependent, given all of the other variables. 
For instance, in a brain connectivity network, each node represents a brain region, and an edge between two nodes indicate that the two brain regions are conditionally dependent.
Many methods were proposed for estimating graphical models under various model assumptions.  
In particular, Gaussian graphical models have been studied extensively
\citep{meinshausen2006high,yuan2007model,friedman2008sparse,rothman2008sparse,cai2011constrained,sun2013sparse,tanetal2015,renetal2016,lin2016estimation}. 
To relax the Gaussianity assumption, exponential graphical models in which the node-conditional distribution for each variable belongs to an exponential family distribution were proposed \citep{ravikumar2010high,yang2015graphical,chen2014selection,yang2018semiparametric}.
More recently, several authors considered nonparametric graphical models without imposing any distributional assumption on the random variables \citep{voormanetal2014,janofsky2015exponential,sun2015learning,tan2019layer}.  
The literature on graphical models is vast: we refer the reader to \citet{drton2017structure} for a comprehensive list of references. 

In this paper, we focus on estimating brain connectivity networks using fMRI data. 
There are two major challenges presented by fMRI data: correlated replicates for each independent subject and the presence of unmeasured confounders.  
Firstly, each independent subject is scanned over a period of time, and therefore yields a series of correlated brain scans.
Moreover, while the fMRI brain scans are taken over time, the subjects may have different states of mind or head motion, which can be interpreted as unmeasured confounders.  
 For instance, certain subjects may be awake during the first half of the brain scan, and may fall asleep during the second half of the brain scan. Different brain regions may be active or inactive, depending on whether the subject is awake or asleep. 
Thus, it is of utmost importance to model the correlation across replicates and the latent effects induced by the unmeasured confounders to obtain an accurate conditional independence graph.  

Most existing methods for estimating conditional independence graph assume that all relevant variables are observed. 
However, this assumption is often violated in many scientific studies in which certain variables are not measured either due to cost constraints, ethical issues, or that they are simply unmeasurable. 
For instance, in the context of fMRI studies, some variables such as the state of mind during the fMRI scan is unmeasurable.     
Not taking into account the unmeasured confounders during model fitting will yield a graph with spurious edges between pairs of variables.
In the context of Gaussian graphical models, \citet{chandrasekaran2010latent} showed that marginalizing over the unmeasured confounders will yield a dense conditional independence graph of the observed variables even when the true underlying graph for the observed variables is sparse.    
To address this issue, various methods were proposed for modeling latent variable graphical models under various assumptions on the unmeasured confounders  \citep{chandrasekaran2010latent,tan2016replicates,fan2017high,wu2017graphical}.

However, the aforementioned work mainly focused on estimating a conditional independence graph based on independent realizations of a common random vector.   
In many scientific settings, data can be collected over time from multiple independent subjects.
For instance, in the context of fMRI studies, brain scans are taken every 1.5 seconds, yielding highly correlated replicates. 
Some authors assumes that the graph evolves across time, i.e., time-varying graphical models, but these work do not model the correlation across replicates  \citep{kolar2010estimating, hanneke2010discrete, sarkar2006dynamic, guo2007recovering, zhou2010time}.
To take into account the correlated replicates, several authors have modeled the correlation by assuming that the replicates follow a  vector autoregressive (VAR) process, and that the resulting graphical model is invariant over time \citep{qiu2016joint, hall2016inference, basu2015regularized}.

\begin{figure}[!t]
\centering
   \subfigure[]{\includegraphics[scale=0.35]{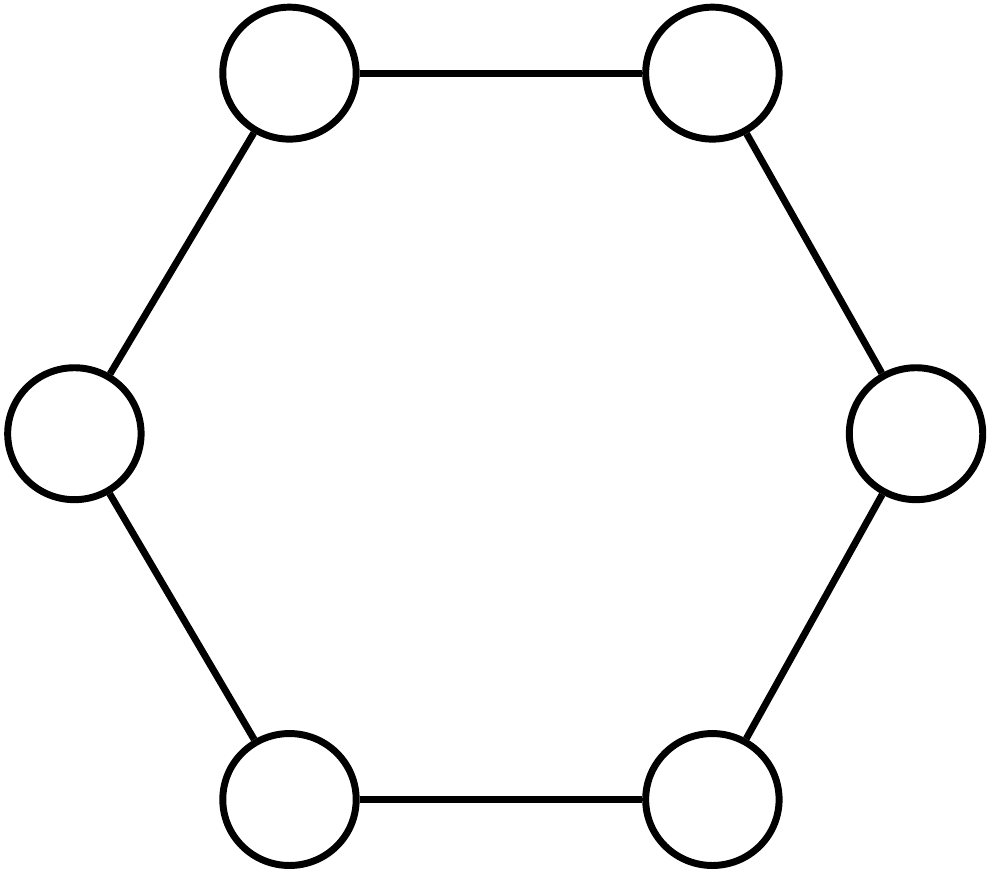}}\qquad
   \subfigure[]{\includegraphics[scale=0.35]{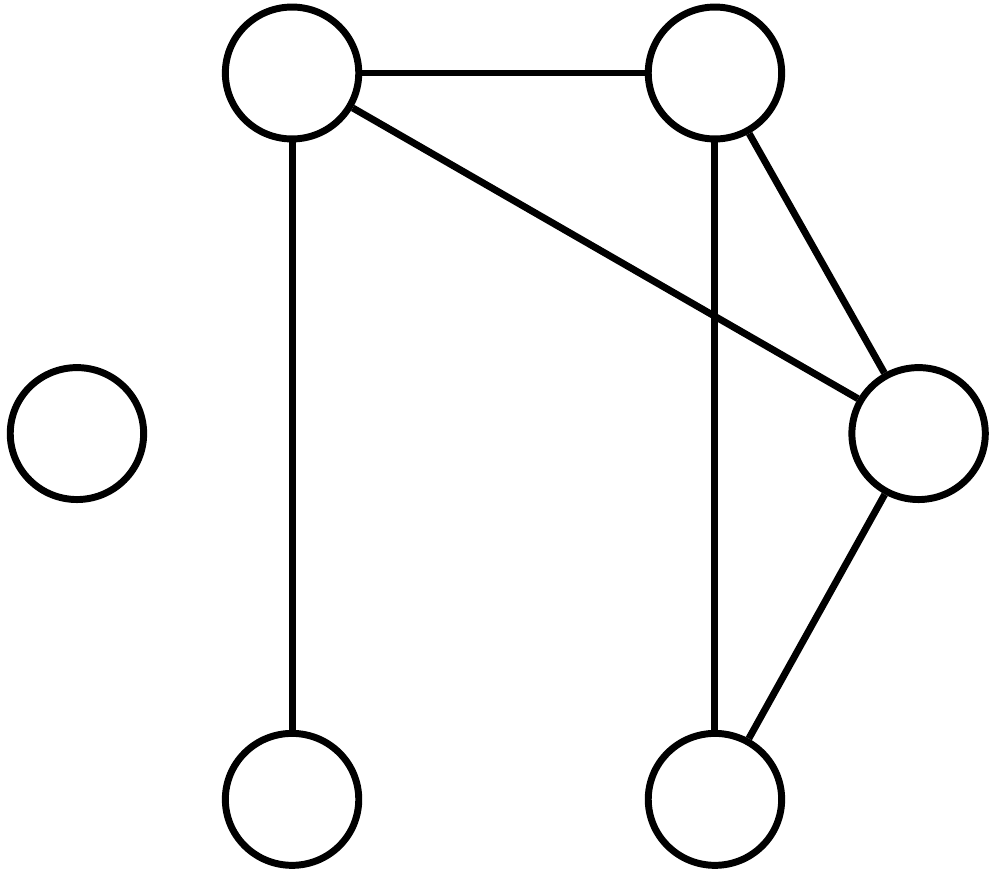}}\qquad
   \subfigure[]{\includegraphics[scale=0.35]{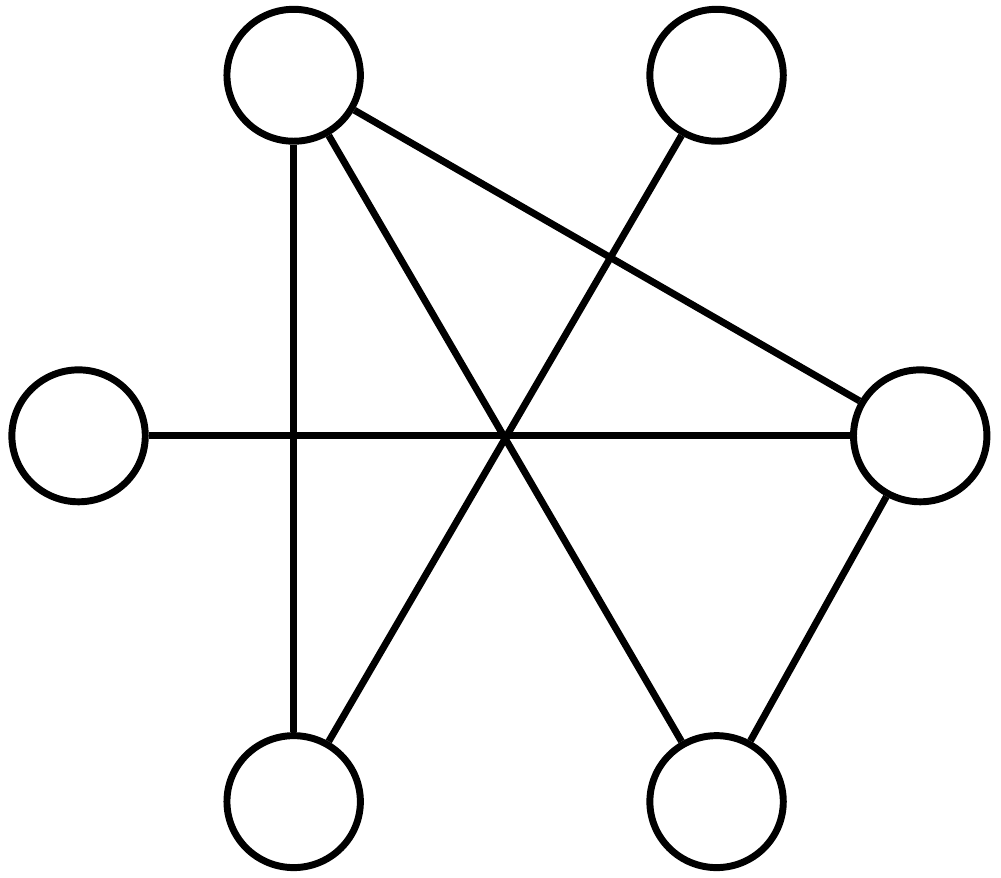}}\qquad
   \subfigure[]{\includegraphics[scale=0.35]{into_simu_full.pdf}}
\caption{A toy example on a Gaussian graphical model with unmeasured confounders and correlated replicates. 
Panels (a), (b), (c), and (d) correspond to the true underlying graph, estimated graphs by  \cite{friedman2008sparse},  \cite{chandrasekaran2010latent}, and our proposed method, respectively. }
\label{figure:intro}
\end{figure}

In this paper, we consider modeling both the effect of unmeasured confounders and the temporal dependence of the replicates. 
Figure~\ref{figure:intro} shows a toy example on Gaussian graphical models with unmeasured confounders and correlated replicates, where we compare our proposed method with \cite{friedman2008sparse} that ignores both the unmeasured confounders and correlated replicates, and \cite{chandrasekaran2010latent} that models the unmeasured confounders but ignores the correlated replicates. The tuning parameters for all methods are selected such that  all methods yield six edges.
We see from Figure~\ref{figure:intro} that when there are correlated replicates and unmeasured confounders, our proposed method recovers the true graph whereas \cite{friedman2008sparse} and \cite{chandrasekaran2010latent} fail to recover the true graph.

Recently, \citet{tan2016replicates} proposed to estimate a semiparametric exponential family  graphical model with unmeasured confounders under the setting in which multiple replicates are collected for each subject.
The main crux of their proposed method is on the construction of a nuisance-free loss function that does not depend on the unmeasured confounders.
The proposed method relies on two crucial assumptions: (i) the unmeasured confounders are constant across replicates within each subject; (ii) given the unmeasured confounders, the observed replicates within each subject are mutually independent.  
However, in many scientific settings, these assumptions may be violated.
For instance, in the aforementioned fMRI study, unmeasured different states of mind will induce different latent effects across the brain scans and violate the constant unmeasured confounders assumption in \citet{tan2016replicates}.   
Moreover, brain scans are taken every 1.5 seconds and thus the replicates are correlated.

We relax the two aforementioned key assumptions in \citet{tan2016replicates}.
Instead of assuming the unmeasured confounders are the same for all replicates, we assume that the effect induced by the unmeasured confounders is piecewise constant across replicates within each independent subject. This is a reasonable assumption for fMRI data, since the latent effect can be always approximated by a constant in a small time interval (e.g., within 1.5 seconds). To model the correlation across replicates, we assume a one-lag  vector autoregressive model on the replicates.
Under the relaxed assumption, we propose a novel method for modeling exponential family graphical models with correlated replicates and unmeasured confounders. 
Our proposal incorporates a lasso penalty for estimating a sparse graph among the observed variables, a lasso penalty for modeling the correlation between two successive replicates, and a fused lasso penalty for modeling the piecewise constant latent effect induced by the unmeasured confounders.
The resulting convex optimization problem is then solved using a block coordinate descent approach.  

Theoretically, we establish the non-asymptotic error bound for the proposed estimator.  
Due to the use of both lasso and fused lasso penalty, the error bound consists of both the estimation error of the lasso term and the fused lasso term.  Thus, standard proof for lasso type problem in \citet{buhlmann2011statistics} will lead to a slower rate of convergence.   
 To obtain a sharp rate, one needs to carefully balance these two terms by selecting the respective tuning parameters in an optimal way. 
By selecting the appropriate set of tuning parameters, our theoretical results reveal an interesting phenomenon on the interplay between the number of independent samples $n$ and the number of replicates $T$.  
Finally, we show that the proposed estimator is adaptive to the absence of unmeasured confounders, i.e., our estimator matches the rate of convergence obtained by solving a lasso problem using the oracle knowledge that there are no unmeasured confounders.

An \texttt{R} package \texttt{latentgraph} will be made publicly available on CRAN.

%%%%%%%%%%%%%%%%%%%%%%%%%%%%%%%%%%%%
%%%%%%%%%%%%%%%%%%%%%%%%%%%%%%%%%%%%
% Proposed Method
%%%%%%%%%%%%%%%%%%%%%%%%%%%%%%%%%%%%
%%%%%%%%%%%%%%%%%%%%%%%%%%%%%%%%%%%%
\section{Latent Variable Graphical Models with Correlated Replicates}
\label{sec:proposed method}
\subsection{A Review on Exponential Family Graphical Models}
\label{subsec: EFGM}
We start with a brief overview of the exponential family graphical model.  Let $\bX = (X_1, \ldots, X_p)^{\T} \in \RR^p$ be a $p$-dimensional random vector, corresponding to $p$ nodes in a graph.  
Then, the pairwise exponential family graphical model has the following joint density function 
\begin{equation}
\label{equation1}
p\left(\bx\right) = \exp \left\{\sum^p_{j = 1} f_j(x_j) + \frac{1}{2} \sum^{p}_{j = 1} \sum_{k \ne j} \theta_{jk} x_j x_k - A\left(\boldsymbol{\Theta}, \boldsymbol{\zeta} \right)\right\},
\end{equation}
where $f_j(\cdot)$ is a node potential function, $A(\cdot)$ is the log-partition function such that the density in \eqref{equation1} integrates to one, $\boldsymbol{\Theta} = \{\theta_{jk}\}_{1\le j<k\le p}$ is a symmetric square matrix, and $\boldsymbol{\zeta}$ is a matrix of parameters for $f_j(\cdot)$.
The parameter $\theta_{jk}$ encodes the conditional dependence relationship between the $j$th and the $k$th variables, i.e., $\theta_{jk}=0$ if and only if the $j$th and the $k$th variables are conditionally independent.  
Thus, estimating the exponential family graphical models amounts to estimating $\theta_{jk}$.

In principle, given $n$ independent subjects, an estimator of $\theta_{jk}$ can be obtained by maximizing the joint density of \eqref{equation1} for $n$ independent subjects.
However, $A(\bTheta, \bzeta)$ is computationally intractable even for moderate $p$.
To avoid this issue, many authors have proposed to maximize the conditional distribution of each variable, and then combine the resulting estimates to form a single graphical model  \citep{meinshausen2006high, ravikumar2010high, allen2012log, yang2012graphical, chen2014selection}.

More specifically, for any node $j$, let $\bX_{-j} = (X_1, \ldots,X_{j-1}, X_{j+1}, \ldots,  X_p)^{\T} \in \RR^{p-1}$.  Then, $\bX$ follows the exponential family graphical model if for any node $j$, the conditional density of $X_j$ given $\bX_{-j}$ is  
\begin{equation}
\label{equation2}
p\left(x_j \mid \bx_{-j}\right) = \exp \left\{f_j(x_j) +  x_j \btheta^{\T}_{j,-j} \bx_{-j} - D_j\left(\btheta_{j,-j}, f_j\right)\right\},
\end{equation}
where $\btheta_{j,-j} = (\theta_{j1}, \ldots, \theta_{j(j-1)}, \theta_{j(j+1)}, \ldots, \theta_{jp})^{\T}$ and $D_j(\btheta_{j,-j}, f_j)$ is the log-partition function that depends on $\btheta_{j,-j}$ and $f_j$. 
The exponential family graphical model can then be constructed by estimating $\btheta_{j,-j}$ for $j\in \{1,\ldots,p\}$ through fitting $p$ generalized linear models.

%%%%%%%%%%%%%%%%%%%%%%%%%%%%%%%%%%%%%%%%
%%%%%%%%%%%%%%%%%%%%%%%%%%%%%%%%%%%%%%%%
%%%%%%%%%%%%%%%%%%%%%%%%%%%%%%%%%%%%%%%%
%%%%%%%%%%%%%%%%%%%%%%%%%%%%%%%%%%%%%%%%
\subsection{Exponential Family Graphical Models with Correlated Replicates and Unmeasured Confounders }
\label{subsec: EFGM extend}
The pairwise exponential family graphical model in~\eqref{equation1} assumes that all variables are observed and that there are no unmeasured confounders. 
Moreover, \eqref{equation1} does not accommodate correlated measurements or replicates.  
%As mentioned in the Introduction, ignoring correlation and not taking into account the unmeasured confounders can lead to biased estimate of the conditional independence graph.  
In this section, we propose an extension of the exponential family graphical model to accommodate both the correlated replicates and unmeasured confounders.  
Let $\bX_t \in \RR^p$ and $\bU_t \in \RR^q$ be vectors of the observed and unmeasured confounding random variables for the $t$th replicate, respectively.
For simplicity, we assume that there are a total of $T$ replicates.  
We start with the following assumption on the joint density of the replicates.  

%%%%%%%%%%%%%%%%%%%%%%%%%%%%%
%%%%%%%%%%%%%%%%%%%%%%%%%%%%%
% Assumption on the replicates
%%%%%%%%%%%%%%%%%%%%%%%%%%%%%
%%%%%%%%%%%%%%%%%%%%%%%%%%%%%
\begin{assumption}
\label{assumption2}
The joint conditional density of the $T$ replicates, given the unmeasured confounders, takes the form
\[
p\left( \bx_{1},\ldots,\bx_T \mid \bu_1,\ldots,\bu_T\right) = \prod_{t=1}^T p \left(\bx_{t} \mid \bx_{t-1}, \bu_{t}\right).
\] 
\end{assumption}
\noindent In other words, conditioned on the unmeasured confounders, the $T$ replicates are assumed to follow a one-lag  vector autoregressive model. 
That is, the $t$th replicate depends only on the $(t-1)$th replicate of the observed random variables.
Moreover, the observed variables are conditionally independent of the unmeasured confounders across different replicates.

\begin{definition}%[\sf Exponential Family Graphical Models with Correlated Replicates and Unmeasured Confounders]
\label{def: EFGM extend}
A $(p+q)$-dimensional random vector $(\bX_t^{\T},\bU_t^\T)^\T$ follows the exponential family graphical model with correlated replicates and unmeasured confounders if for each node $j$, the conditional distribution of $X_{tj}$ given $\bX_{t(-j)}$, $\bX_{t-1}$, and $\bU_t$ is
\begin{equation}
\label{eq:model1}
\begin{split}
&p\left(x_{tj}\mid\bx_{t(-j)}, \bx_{t-1},\bu_t\right)\\
&= \exp \left\{f_{tj}\left(x_{tj}\right) + \sum_{k \ne j}\theta_{jk}x_{tk}x_{tj} + \sum_{k = 1}^p\alpha_{jk}x_{(t-1)k}x_{tj}+\sum_{m =  1}^q\delta_{jm}u_{tm}x_{tj} - D_{tj}\left(\theta_{jk}, \alpha_{jk},\delta_{jm}, f_{tj}\right)\right\},
\end{split}
\end{equation}
where $f_{tj}(\cdot)$ is the node potential function and $D_{tj}(\cdot)$ is the log-partition function such that the conditional density integrates to one.
\end{definition}

In Definition~\ref{def: EFGM extend}, $\theta_{jk}$ encodes the conditional dependence relationship between the $k$th and $j$th nodes.  That is, $\theta_{jk}=0$ if and only if $X_{tj}$ and $X_{tk}$ are conditionally independent, given $\bX_{t(-j)}$, $\bX_{t-1}$, and $\bU_{t}$ for all replicates $t=1,\ldots,T$. 
The parameter $\alpha_{jk}$ models the correlation between $X_{(t-1)k}$ and $X_{tj}$.  
Finally, $\delta_{jm}$ encodes the conditional dependence relationship between the $m$th latent variable and the $j$th observed variable.  
The form of the node potential function $f_{tj}(\cdot)$ and the log-partition function $D_{tj}(\cdot)$ is specific to each exponential family distribution. Let $f_{tj}(x_{tj}) = B_{1tj}x_{tj} + B_{2tj}x_{tj}^2 + \sum_{k=3}^K B_{ktj}G_{ktj}(x_{tj})$ for some scalar $B_{ktj}$ and function $G_{ktj}(x_{tj})$. For notational simplicity, denote $\eta_{tj} = B_{1tj}+ \sum_{k \ne j}\theta_{jk}x_{tk} + \sum_{k = 1}^p\alpha_{jk}x_{(t-1)k} +\sum_{m =  1}^q\delta_{jm}u_{tm}$. 
In the following, we provide three special cases of the model in Definition~\ref{def: EFGM extend}.

%%%%%%%%%%%%%%%%%%%%%%%%%%%%%%%%%%%
% Example GGM
%%%%%%%%%%%%%%%%%%%%%%%%%%%%%%%%%%%
\begin{example}
\label{exa1}
The Gaussian graphical model with correlated replicates and unmeasured confounders. 
The conditional distribution of $X_{tj}$ given $\Xb_{t(-j)}$, $\Xb_{t-1}$ and $\bU_{t}$ with $B_{2tj} = -1/2$  is given by:
\begin{equation}
\label{equation3}
p\left(x_{tj}\mid \xb_{t(-j)},\xb_{t-1},\Delta_{tj}\right) = \exp\left\{-\frac{1}{2} x_{tj}^2 + \eta_{tj} x_{tj} - D_{tj}\left(\eta_{tj}\right)\right\}~~\left(x_{tj} \in \RR\right),
\end{equation}
where $f_{tj}(x_{tj}) = B_{1tj} x_{tj}-x_{tj}^2 / 2$ and $D_{tj}(\eta_{tj}) =  \eta_{tj}^2/2 + \log(2\pi)/2$.
\end{example}
%%%%%%%%%%%%%%%%%%%%%%%%%%%%%%%%%%%
% Example Ising Model
%%%%%%%%%%%%%%%%%%%%%%%%%%%%%%%%%%%
\begin{example}
\label{exa2}
The Ising model with correlated replicates and unmeasured confounders. 
The conditional distribution of $X_{tj}$ given $\Xb_{t(-j)}$, $\Xb_{t-1}$ and $\bU_t$ is:
\begin{equation}
\label{equation4}
p\left(x_{tj}\mid \xb_{t(-j)},\xb_{t-1},\Delta_{tj}\right) = \exp\left\{\eta_{tj} x_{tj} - D_{tj}\left(\eta_{tj}\right)\right\}~~\left(x_{tj} \in \left\{0,1\right\}\right),
\end{equation}
where $f_{tj}(x_{tj}) = 0$ and $D_{tj}(\eta_{tj}) = \log(1 + e^{\eta_{tj}})$.
\end{example}
%%%%%%%%%%%%%%%%%%%%%%%%%%%%%%%%%%%
% Example Poisson 
%%%%%%%%%%%%%%%%%%%%%%%%%%%%%%%%%%%
\begin{example}
\label{exa3}
The Poisson graphical model with correlated replicates and unmeasured confounders. 
The conditional distribution of $X_{tj}$ given $\Xb_{t(-j)}$, $\Xb_{t-1}$ and $\bU_{t}$ is:
\begin{equation}
\label{equation5}
p\left(x_{tj}\mid \xb_{t(-j)},\xb_{t-1},\Delta_{tj}\right) = \exp\left\{\eta_{tj} x_{tj} - \log\left(x_{tj}!\right)-D_{tj}\left(\eta_{tj}\right)\right\}~~\left(x_{tj} \in \left\{0,1,\dots\right\}\right),
\end{equation}
where $f_{tj}(x_{tj}) = B_{1tj} x_{tj} - \log(x_{tj}!)$ and $D_{tj}(\eta_{tj}) = \exp(\eta_{tj})$.
\end{example}

%%%%%%%%%%%%%%%%%%%%%%%%%%%%%%%%%%%%%%%%%
%%%%%%%%%%%%%%%%%%%%%%%%%%%%%%%%%%%%%%%%%
% Proposed Method
%%%%%%%%%%%%%%%%%%%%%%%%%%%%%%%%%%%%%%%%%
%%%%%%%%%%%%%%%%%%%%%%%%%%%%%%%%%%%%%%%%%
\section{Method}
\subsection{Problem Formulation and Parameter Estimation}
Suppose that there are $n$ independent subjects $i=1,\ldots,n$ and each subject has $t=1,\ldots,T$ replicates. 
For simplicity, we assume that all independent subjects have the same number of replicates; our proposed method can be easily modified to accommodate different number of replicates across the $n$ subjects. 
Let $\bX_{it}\in\RR^p$ and $\bU_{it}\in\RR^q $ be the random observed variables and unmeasured confounders corresponding to the $t$th replicate of the $i$th subject, respectively.  
The primary goal is to estimate the conditional dependence relationships among the observed variables given the latent variables.  A naive approach is to obtain a maximum likelihood estimator by maximizing the marginal  likelihood function of all the observed variables for $t=1,\ldots,T$ and $i=1,\ldots,n$.  However, the marginal  likelihood function involves the integral over the distributions of unmeasured confounders $\Ub_t$ and is computationally infeasible. 

%However, this is not possible since \eqref{eq:model1} depends on the number of unmeasured confounders $q$ and the unmeasured confounders $\Ub_t$. 

Inspired by the literature on measurement error models \citep{carroll2006measurement}, we use a functional approach to deal with the unmeasured confounders. To be specific, we treat the realization of the  unmeasured confounders $U_{itj}$ as nonrandom incidental nuisance parameters, which may differ from subject to subject. Such an approach is dated back to the so-called Neyman and Scott's problem in 1948; see \cite{lancaster2000incidental} for a survey. However, in this functional approach, the graphical model involves a large number of unknown nuisance parameters such that the estimation of $\theta_{jk}$ in (\ref{eq:model1}) is often inconsistent. To alleviate this problem, we further assume that for the same subject, the value of $U_{itj}$ is piecewise constant across $t=1,\ldots,T$. In theory, this assumption may improve the estimation accuracy by reducing intrinsic dimension of  the unknown incidental nuisance parameters. In practice, it is much less restrictive than assuming that the latent variables are constant as assumed in \citet{tan2016replicates} and is more appropriate for modeling fMRI data. 
%%%%%%%%%%%%%%%%%%%%%%%%%%%%%%%%%%%%
%%%%%%%%%%%%%%%%%%%%%%%%%%%%%%%%%%%%
% Assumption
%%%%%%%%%%%%%%%%%%%%%%%%%%%%%%%%%%%%
%%%%%%%%%%%%%%%%%%%%%%%%%%%%%%%%%%%%
\begin{assumption}
\label{assumption1}
The unmeasured confounders are piecewise constant across replicates.  That is, we assume for the $i$th sample, we have $l$ knots with unknown location denoted as $k_{i1}, k_{i2}, \ldots, k_{il}$ and let $k_{i0} = 1$, $k_{i(l+1)} = T$. Then the $j$th unmeasured confounder at the $t$th replicate for the $i$th subject satisfies
$$
U_{itj} = \sum_{a=1}^{l+1} g_{iaj} \mathbbm{1}{\left(k_{i(a-1)} \leq t \leq k_{ia}\right)},$$
where $g_{iaj}$ is an unknown constant and $\mathbbm{1}{(\cdot)}$ is an indicator function.
\end{assumption}

\begin{figure}[!htp]
\centering
   \subfigure[]{\includegraphics[scale=0.28]{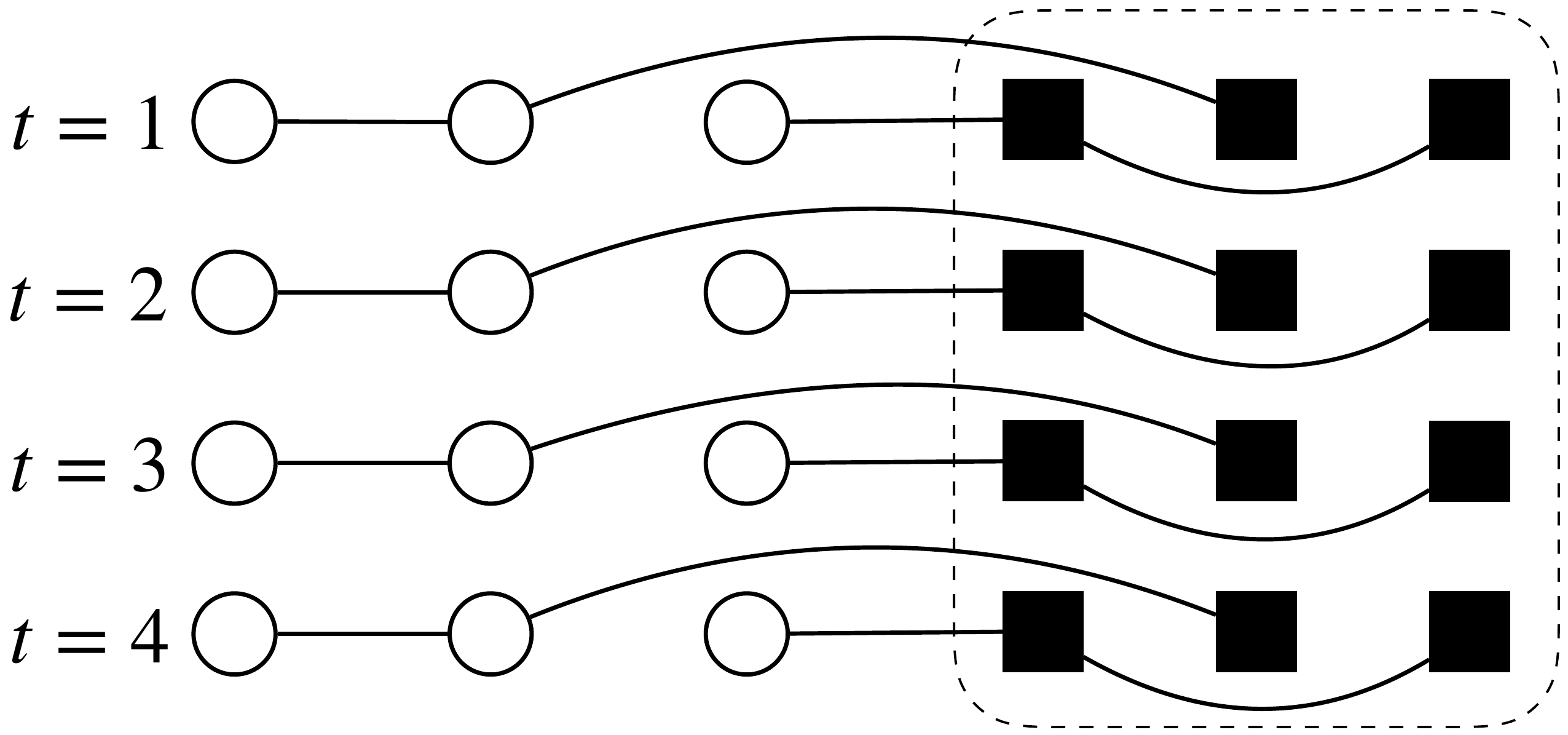}}\qquad\qquad
   \subfigure[]{\includegraphics[scale=0.28]{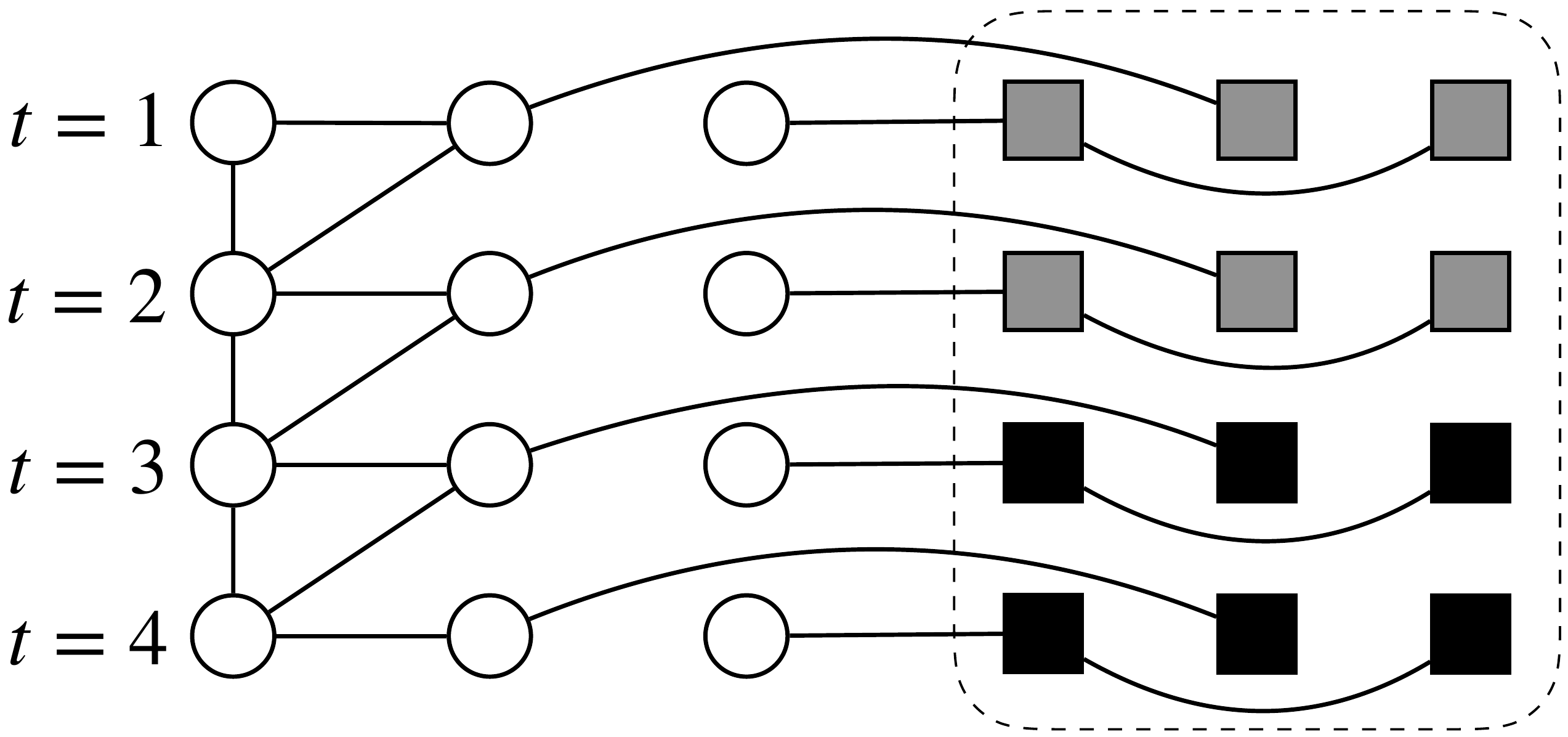}}
\caption{Panels (a) and (b) correspond to the assumptions on the replicates and unmeasured confounders of the method by \citet{tan2016replicates} and our proposed method, respectively. There are four replicates for each subject, i.e., $t = \{1, 2, 3, 4\}$. Hollow circles represent the observed variables, and squares represent the unmeasured confounders. In panel (b), the color of the unmeasured confounders changes from gray to black, indicating that the value of unmeasured confounders are allowed to change across replicates.}
\label{fig:f1}
\end{figure}

Figure~\ref{fig:f1} provides a schematic of the assumptions in \citet{tan2016replicates} and our proposal for the $i$th sample. 
Figure~\ref{fig:f1}(a) represents the assumptions in \citet{tan2016replicates}: the $t$th and $(t-1)$th replicates for the observed variables are independent and the unmeasured confounders are constant across replicates. 
Figure~\ref{fig:f1}(b) depicts the assumptions for our proposed method: the $t$th replicate of the observed variables are conditionally dependent on the $(t-1)$th replicate, and the  unmeasured confounders are piecewise constant that may change across replicates. 

%%%%%%%%%%%%%%%%%%%%%%%%%%%%%%%%%%%
% Estimation
%%%%%%%%%%%%%%%%%%%%%%%%%%%%%%%%%%%
%\subsection{Parameter Estimation}
%\label{subsec:estimation}

We now reformulate the conditional density in~\eqref{eq:model1} under Assumption~\ref{assumption1}.  
Let $\Delta_{tj} = \sum_{m=1}^q \delta_{jm}u_{tm}$.
Then, \eqref{eq:model1} in Definition~\ref{def: EFGM extend} can be rewritten as:
\begin{align}
\label{equation6}
&p\left(x_{tj}\mid\xb_{t(-j)}, \xb_{t-1},\Delta_{tj}\right)\nonumber\\
&= \exp \left \{f_{tj}\left(x_{tj}\right) + \sum_{k \ne j}\theta_{jk}x_{tk}x_{tj} + \sum_{k = 1}^p\alpha_{jk}x_{(t-1)k}x_{tj} + \Delta_{tj}x_{tj} - D_{tj}\left(\theta_{jk}, \alpha_{jk},\Delta_{tj}, f_{tj}\right)\right\}.
\end{align} 
We now construct a joint likelihood function for $n$ subjects, each of which has $T$ replicates using~\eqref{equation6}. 
For the $i$th subject, let 
$\balpha_{j} = (\alpha_{j1}, \alpha_{j2}, \ldots, \alpha_{jp})^{\T} \in \RR^{p}$, 
$\btheta_{j,-j} = (\theta_{j1}, \ldots, \theta_{j,j-1}, \theta_{j,j+1},\ldots, \theta_{jp})^{\T} \in \RR^{p-1}$, and
$\bDelta_{j} = (\Delta_{11j}, \Delta_{12j}, \ldots, \Delta_{1Tj}, \Delta_{21j}, \Delta_{22j}, \ldots, \Delta_{nTj})^{\T} \in \RR^{nT}$.
Thus, we estimate $\btheta_{j,-j}$, $\balpha_{j}$, and $\bDelta_{j}$ by solving
\begin{equation}
\label{equation7}
\underset{\boldsymbol{\theta}_{j,-j},\boldsymbol{\alpha}_j,\bDelta_{j}}{\mathrm{minimize}}\quad -\frac{1}{nT}l\left(\boldsymbol{\theta}_{j,-j},\boldsymbol{\alpha}_{j},\boldsymbol{\Delta}_{j}\right)+\lambda \big \|\boldsymbol{\theta}_{j,-j}\big\|_1+\beta \big \|\boldsymbol{\alpha}_{j} \big\|_1+\gamma \big \|\left(\Ib_n \otimes \Cb\right) \boldsymbol{\Delta}_{j}\big\|_1,
\end{equation}
where $l(\boldsymbol{\theta}_{j,-j},\boldsymbol{\alpha}_{j},\boldsymbol{\Delta}_{j}) = \sum_{i = 1}^n \sum_{t=1}^T \log p(x_{itj}|\bx_{it(-j)},\bx_{i(t-1)},\Delta_{itj})$.  
Here, $\lambda$, $\beta$, and $\gamma$ are the sparsity inducing tuning parameters, $\Ib_n$ is an $n$-dimensional identity matrix, and $\Cb \in \RR^{(T-1) \times T}$ is the discrete first derivative matrix defined as follows:
$$
\Cb = 
\left(
\begin{matrix}
 -1      & 1    &0  & \cdots & 0   &0  \\
 0       & -1   &1   & \cdots & 0  &0   \\
 \vdots                                             \\
 0      & 0     &0     & \cdots & -1 &1      \\
\end{matrix}
\right).
$$
Note that the penalty term $ \big \|\left(\Ib_n \otimes \Cb\right) \boldsymbol{\Delta}_{j}\big\|_1$ is essentially a fused lasso penalty on $\bDelta_{ij}$ for each subject, since we assume that the unmeasured confounders are piecewise constant in Assumption~\ref{assumption1}.

%%%%%%%%%%%%%%%%%%%%%%%%%%
%%%%%%%%%%%%%%%%%%%%%%%%%%%%%%%%%%%%
% Algorithm
%%%%%%%%%%%%%%%%%%%%%%%%%%%%%%%%%%%%
%%%%%%%%%%%%%%%%%%%%%%%%%%%%%%%%%%%%
\subsection{Algorithms for Solving~\eqref{equation7}}
\label{sec:algorithm}
In this section, we propose two algorithms for solving the convex optimization problem~\eqref{equation7} in the context of Gaussian graphical models, and exponential family graphical models.  
In the context of Gaussian graphical models, $l(\cdot)$ has a quadratic form and thus can be efficiently solved using a block coordinate descent algorithm.
In the context of exponential family graphical models, we instead employ the generalized gradient descent \citep{beck2009fast}, coupled with the block coordinate descent method.
The convergence of the block coordinate descent algorithm is studied in \citet{tseng2001convergence}.

\subsubsection{Block Coordinate Descent for Gaussian Graphical Models}
\label{subsec:gaussian graphical model}
We start with defining some notation.  
Let 
 $\xb_{j} = (x_{11j}, x_{12j}, \ldots, x_{1Tj}, x_{21j}, x_{22j}, \ldots, x_{nTj})^{\T} \in \RR^{nT}$, 
 $\Xb_{i(-j)} = (\xb_{i1(-j)}, \xb_{i2(-j)}, \ldots, \xb_{iT(-j)})^{\T} \in \RR^{T \times (p-1)}$, and 
 $\Xb_{ij} = (\xb_{i0}, \xb_{i1}, \ldots, \xb_{i(T-1)})^{\T}\in \RR^{T \times p}$. 
 In addition, let
 $\Xb_{-j}^{\otimes} = (\Xb_{1(-j)}^{\T}, \Xb_{2(-j)}^{\T}, \ldots, \Xb_{n(-j)}^{\T})^{\T} \in \RR^{(nT) \times (p-1)}$ and
$\Xb_{j}^{\otimes} = (\Xb_{1j}^{\T}, \Xb_{2j}^{\T}, \ldots, \Xb_{nj}^{\T})^{\T}\in \RR^{(nT) \times p}$.
Then, from Example~\ref{exa1}, the canonical parameter  $\boldsymbol{\eta}_{j} = \Bb_j + \Xb_{-j}^{\otimes}\boldsymbol{\theta}_{j,-j} + \Xb_j^{\otimes}\boldsymbol{\alpha}_j + \boldsymbol{\Delta}_j$, where 
$\Bb_{j} = \boldsymbol{1}_n \otimes (B_{11j}, B_{12j}, \ldots, B_{1Tj})^{\T}$ and $\boldsymbol{1}_n$ is an $n$-dimensional vector of ones.
For simplicity, we assume that the data is centered, such that $\Bb_j  = \mathbf{0}$. 
In the context of Gaussian graphical models, the optimization problem in (\ref{equation7}) reduces to
\begin{equation}
\label{equation14}
\underset{\btheta_{j,-j}, \balpha_j, \boldsymbol{\Delta}_{j}}{\mathrm{minimize}}\quad \frac{1}{2nT}\big\|\xb_{j} - \boldsymbol{\eta}_{j}\big\|_2^2 +\lambda\big\|\boldsymbol{\theta}_{j,-j}\big\|_1+\beta\big\|\boldsymbol{\alpha}_j\big\|_1+\gamma\big\|\left(\Ib_n \otimes \Cb\right)\boldsymbol{\Delta}_{j}\big\|_1.
\end{equation}

Optimization problem~\eqref{equation14} involves a fused lasso type penalty on $\bDelta_j$, and can be rewritten into a lasso problem by a change of variable.  To this end, let
$\Eb = \Ib_n \otimes \boldsymbol{1}_{T}^{\T} \in \RR^{n \times (nT)}$, 
$\widetilde{\Cb} = ((\Ib_n \otimes \Cb)^{\T}, \Eb^{\T})^{\T} \in \RR^{(nT) \times (nT)}$, 
and $\Hb_j = \widetilde{\Cb}\bDelta_j = [\{(\Ib_n \otimes \Cb)\bDelta_j\}^{\T}, (\Eb \bDelta_j)^{\T}]^{\T}\in \RR^{nT}$. 
Then, (\ref{equation14}) can be rewritten as 
\begin{equation}
\label{equation13}
\underset{\btheta_{j,-j}, \balpha_j, \Hb_j}{\mathrm{minimize}}\quad \frac{1}{2nT}\|\xb_{j} - \boldsymbol{\eta}_{j}\|_2^2 +\lambda\|\boldsymbol{\theta}_{j,-j}\|_1+\beta\|\boldsymbol{\alpha}_j\|_1+\gamma\|\Hb_{j1}\|_1,
\end{equation}
where $\boldsymbol{\eta}_{j} = \Xb_{-j}^{\otimes}\boldsymbol{\theta}_{j,-j} + \Xb_j^{\otimes}\boldsymbol{\alpha}_j + \widetilde{\Cb}^+\Hb_j$ and $\Hb_{j1} = (\Ib_n \otimes \Cb)\bDelta_j \in \RR^{n(T-1)}$. 
Problem~\eqref{equation13} is convex in $\btheta_{j,-j}$, $\balpha_j$, and $\Hb_j$, and thus can be solved using a block coordinate descent algorithm.  The details are presented in Algorithm~\ref{alg:gaussian}.  Specifically, our proposed algorithm solves three lasso problems iteratively, and can be solved using the $\texttt{glmnet}$ package in $\texttt{R}$.

%%%%%%%%%%%%%%%%%%%%%%%%%%%%%%%%%%%%
%%%%%%%%%%%%%%%%%%%%%%%%%%%%%%%%%%%%
% General Algorithm
%%%%%%%%%%%%%%%%%%%%%%%%%%%%%%%%%%%%
%%%%%%%%%%%%%%%%%%%%%%%%%%%%%%%%%%%%
\subsubsection{Generalized Gradient Descent for Exponential Family Graphical Models}
\label{sec:general algorithm}
Other than the Gaussian graphical models, the loss function $l(\cdot)$ in (\ref{equation7}) does not take the form of squared error loss, and thus Algorithm~\ref{alg:gaussian} cannot be applied directly.   
To this end, we employ the generalized gradient descent to provide a quadratic approximation for $l(\cdot)$ through the second-order Taylor expansion.   
That is, we instead consider solving the following optimization problem iteratively, starting with an initial value $\hat{\boldsymbol{\eta}}_j^{0}$:
\begin{align}
\label{equation16}
&(\btheta_{j,-j}^k, \balpha_{j,-j}^k, \Hb_{j}^k) = \nonumber \\
&\underset{\boldsymbol{\theta}_{j,-j},\boldsymbol{\alpha}_j,\Hb_{j}}{\mathrm{argmin}}\quad \frac{L}{2nT} \big \| \frac{1}{L} \xb_{j} -\boldsymbol{\eta}_{j} + \hat{\boldsymbol{\eta}}_{j}^{k-1} - \frac{1}{L}\boldsymbol{D}^{'}_{j}(\hat{\boldsymbol{\eta}}_{j}^{k-1}) \big\|_2^2+\lambda \big\|\boldsymbol{\theta}_{j,-j} \big\|_1+
\beta \big\|\boldsymbol{\alpha}_{j} \big\|_1+\gamma  \big \|\Hb_{j1} \big\|_1,
\end{align} 
where $\boldsymbol{\eta}_{j} = \Xb_{-j}^{\otimes}\boldsymbol{\theta}_{j,-j} + \Xb_j^{\otimes}\boldsymbol{\alpha}_j + \widetilde{\Cb}^+\Hb_j$  
 and $L$ is chosen such that $l''(\boldsymbol{\eta}_j) \preceq L \boldsymbol{I}$.
For instance, in the context of Ising model, it can be shown that $L=1$ will satisfy the above constraint.  
Note that at the $k$th iteration, $\hat{\boldsymbol{\eta}}_{j}^{k-1}$ and $\boldsymbol{D}^{'}_{j}(\hat{\boldsymbol{\eta}}_{j}^{k-1})/L$ are both constants. Thus, the loss function is quadratic in $\btheta_j$ and a block coordinate descent algorithm can be employed to solve~(\ref{equation16}).  The details are presented  in  Algorithm~\ref{alg:exponential}.

%%%%%%%%%%%%%%%%%%%%%%%%%%%%%%%%%%
%%%%%%%%%%%%%%%%%%%%%%%%%%%%%%%%%%
% Gaussian Algorithm
%%%%%%%%%%%%%%%%%%%%%%%%%%%%%%%%%%
%%%%%%%%%%%%%%%%%%%%%%%%%%%%%%%%%%
\begin{algorithm}[!t]  
  \caption{Block Coordinate Descent Algorithm for solving~\eqref{equation7} in the context of Gaussian Graphical Models.}  
  \label{alg:gaussian}  
 \begin{enumerate}
  \item Initialize the constant $\tau > 0$ and $\hat{\boldsymbol{\theta}}_{j,-j}, \hat{\boldsymbol{\alpha}}_{j}$, and  $\hat{\Hb}_j= \mathbf{0}$, respectively.
  \item Estimate $\boldsymbol{\theta}_{j,-j}$. 
   \begin{itemize}
   \item[a.]Compute $\rb_{j}^{\theta} =  \xb_{j}  - \Xb_j^{\otimes}\hat{\boldsymbol{\alpha}}_{j} - \widetilde{\Cb}^+\hat{\Hb}_j$.
  \item[b.] Set 
  $\hat{\btheta}_{j,-j} = \underset{\btheta_{j,-j}\in \RR^{p-1}}{\mathrm{argmin}}\quad (2nT)^{-1}\|\rb_{j}^{\theta} - \Xb_{-j}^{\otimes}\btheta_{j,-j}\|_2^2+\lambda\|\btheta_{j,-j}\|_1$.
  \end{itemize}
  \item Estimate $\balpha_j$. 
   \begin{itemize}
   \item[a.]Compute $\rb_{j}^{\alpha} = \xb_{j} - \Xb_{-j}^{\otimes}\hat{\boldsymbol{\theta}}_{j,-j} - \widetilde{\Cb}^+\hat{\Hb}_j$.
  \item[b.] Set
   $\hat{\balpha}_j = \underset{\boldsymbol{\alpha}_j\in \RR^p}{\mathrm{argmin}}\quad  (2nT)^{-1}\|\rb_{j}^{\alpha} - \Xb_j^{\otimes}\boldsymbol{\alpha}_j\|_2^2+\beta\|\boldsymbol{\alpha}_j\|_1$.
   \end{itemize}
   \item Estimate $\Hb_{j}$. 
    \begin{itemize}
   \item[a.]Compute $\rb_{j}^{H} = \xb_{j} - \Xb_{-j}^{\otimes}\hat{\boldsymbol{\theta}}_{j,-j} - \Xb_j^{\otimes}\hat{\boldsymbol{\alpha}}_{j}$.
  \item[b.] Set
   $\hat{\Hb}_j = \underset{\boldsymbol{\Hb}_{j}\in \RR^{nT}}{\mathrm{argmin}}\quad  (2nT)^{-1}\|\rb_j^{H} -\widetilde{\Cb}^+\Hb_j\|_2^2+\gamma\|\Hb_{j1}\|_1$.
    \end{itemize}
    \item Repeat Steps 2--4 until the stopping criterion
    $\min\{\|\hat{\btheta}_{j,-j}^k - \hat{\btheta}_{j,-j}^{k-1}\|_2^2, ~\|\hat{\balpha}_j^k - \hat{\balpha}_j^{k-1}\|_2^2,~\|\hat{\Hb}_j^k - \hat{\Hb}_j^{k-1}\|_2^2\} \le \tau$ is met.  Here,
     $\hat{\btheta}_{j,-j}^k$, $\hat{\balpha}_j^k$, and $\hat{\Hb}_j^k$ are the values of $\hat{\btheta}_{j,-j}$, $\hat{\balpha}_j$, and $\hat{\Hb}_j$ at the $k$th iteration. Output $\hat{\btheta}_{j,-j}$, $\hat{\balpha}_j$, and  $\hat{\bDelta}_j = \widetilde{\Cb}^+\hat{\Hb}_j$. 
 \end{enumerate}
 \end{algorithm}  
 
 %%%%%%%%%%%%%%%%%%%%%%%%%%%%%%%%%%
%%%%%%%%%%%%%%%%%%%%%%%%%%%%%%%%%%
% General Algorithm
%%%%%%%%%%%%%%%%%%%%%%%%%%%%%%%%%%
%%%%%%%%%%%%%%%%%%%%%%%%%%%%%%%%%%
\begin{algorithm}[!t]  
  \caption{Generalized Gradient Descent and Block Coordinate Descent Algorithm for solving~\eqref{equation7} in the context of Exponential Family Graphical Models.}  
  \label{alg:exponential}  
 \begin{enumerate}
  \item Initialize constant $\tau > 0$, $L>0$,  $\hat{\boldsymbol{\theta}}_{j,-j}, ~\hat{\boldsymbol{\alpha}}_{j},$ and $\hat{\Hb}_j= \mathbf{0}$, respectively.
  \item Estimate $\boldsymbol{\theta}_{j,-j}$. 
   \begin{itemize}
    \item[a.]Compute $\rb_{j}^{\theta} =  \xb_{j}/L + \Xb_{-j}^{\otimes}\hat{\boldsymbol{\theta}}_{j,-j}- \boldsymbol{D}^{'}_{j}(\Xb_{-j}^{\otimes}\hat{\boldsymbol{\theta}}_{j,-j} +\Xb_j^{\otimes}\hat{\boldsymbol{\alpha}}_j + \widetilde{\Cb}^+\hat{\Hb}_j)/L$.
   \item[b.]Set   $\hat{\btheta}_{j,-j} = \underset{\boldsymbol{\theta}_{j,-j} \in \RR^{p-1}}{\mathrm{argmin}}\quad (2nT)^{-1} \cdot L\|\rb_{j}^{\theta} - \Xb_{-j}^{\otimes}\boldsymbol{\theta}_{j,-j}\|_2^2+\lambda\|\boldsymbol{\theta}_{j,-j}\|_1$.
  \end{itemize}
  \item Estimate $\boldsymbol{\alpha}_j$. 
   \begin{itemize}
    \item[a.]Compute $\rb_{j}^{\alpha} = \xb_{j}/L +\Xb_j^{\otimes}\hat{\boldsymbol{\alpha}}_j - \boldsymbol{D}^{'}_{j}( \Xb_{-j}^{\otimes}\hat{\boldsymbol{\theta}}_{j,-j} +\Xb_j^{\otimes}\hat{\boldsymbol{\alpha}}_j + \widetilde{\Cb}^+\hat{\Hb}_j)/L$.
  \item[b.]Set 
   $\hat{\balpha}_j = \underset{\boldsymbol{\alpha}_j \in \RR^p}{\mathrm{argmin}}\quad  (2nT)^{-1} \cdot L\|\rb_{j}^{\alpha} - \Xb_j^{\otimes}\boldsymbol{\alpha}_j\|_2^2+\beta\|\boldsymbol{\alpha}_j\|_1$.
   \end{itemize}
   \item Estimate $\Hb_{j}$. 
    \begin{itemize}
     \item[a.]Compute $\rb_{j}^{H} = \xb_{j}/L + \widetilde{\Cb}^+\hat{\Hb}_j - \boldsymbol{D}^{'}_{j}(\Xb_{-j}^{\otimes}\hat{\boldsymbol{\theta}}_{j,-j} +\Xb_j^{\otimes}\hat{\boldsymbol{\alpha}}_j + \widetilde{\Cb}^+\hat{\Hb}_j)/L$.
  \item[b.]Set $\hat{\Hb}_j = \underset{\Hb_{j} \in \RR^{nT}}{\mathrm{argmin}}\quad  (2nT)^{-1}\cdot L\|\rb_j^{H} -\widetilde{\Cb}^+\Hb_j\|_2^2+\gamma\|\Hb_{j1}\|_1$.
    \end{itemize}
    \item Repeat steps 2--4 until the stopping criterion $\min\{\|\hat{\btheta}_{j,-j}^k - \hat{\btheta}_{j,-j}^{k-1}\|_2^2, ~\|\hat{\balpha}_j^k - \hat{\balpha}_j^{k-1}\|_2^2,~\|\hat{\Hb}_j^k - \hat{\Hb}_j^{k-1}\|_2^2\} \le \tau$ is met, 
    where $\hat{\btheta}_{j,-j}^k$, $\hat{\balpha}_j^k$, and $\hat{\Hb}_j^k$ are the values of $\hat{\btheta}_{j,-j}$, $\hat{\balpha}_j$, and $\hat{\Hb}_j$ at the $k$th iteration. Calculate $\hat{\bDelta}_j = \widetilde{\Cb}^+\hat{\Hb}_j$.
 \end{enumerate}
 \end{algorithm}  

%%%%%%%%%%

%%%%%%%%%%%%%%%%%%%%%%%%%%%%%%%%%%%%
%%%%%%%%%%%%%%%%%%%%%%%%%%%%%%%%%%%%
% Theoretical Analysis
%%%%%%%%%%%%%%%%%%%%%%%%%%%%%%%%%%%%
%%%%%%%%%%%%%%%%%%%%%%%%%%%%%%%%%%%%
\section{Theoretical Results}
\label{sec:theory}
In this section, we derive non-asymptotic upper bounds for the estimation error of $\hat{\btheta}_{j,-j}$, $\hat{\balpha}_{j}$, and $\hat{\bDelta}_{j}$.
In particular, we aim to provide upper bounds for $\big\| \hat{\btheta}_{j,-j} - \btheta_{j,-j}^*\big\|_1 + \big\| \hat{\balpha}_{j} - \balpha_{j}^* \big \|_1 + (nT)^{-1/2} \big \|\hat{\bDelta}_j - \bDelta_j^* \big \|_2$  under two scenarios in which the number of samples $n$ is less than and greater than the number of replicates $T$.  
Throughout this section, we analyze the theoretical properties of the proposed estimator in the context of Gaussian graphical models.  
Recall that, for the $i$th subject, $t$th replicate, and $j$th variable,  we assume the model 
\begin{align}
\label{GauModel}
X_{itj} = \Xb_{it(-j)} \btheta_{j,-j}^* + \Xb_{i(t-1)} \balpha_j^* + \Delta_{itj}^* + \epsilon_{itj},
\end{align}
where the random noise $\epsilon_{itj} \sim N\{0, (\sigma_{jj, t}^{\epsilon})^2\}$ is independent  of $ \Xb_{it(-j)}$ and $\Xb_{i(t-1)}$. 
Note that the random noise is independent but may not be identically distributed, i.e., the random noise in~\eqref{GauModel} can have different variance. 
For notational simplicity, throughout the manuscript, let  $(\sigma_{m}^{\epsilon})^2 = \max_{t,j} \{(\sigma_{jj, t}^{\epsilon})^2\}$. 
Let $\Delta_{m}=\max_{i,t,j}|\Delta_{itj}^* - \Delta_{i(t-1)j}^*|$ be the maximum difference between two consecutive elements of the sequence $\Delta_{i1j}^*, \Delta_{i2j}^*,\ldots,\Delta_{iTj}^*$, and let $\tau=\max_{i,j}\sum_{t=2}^TI(\Delta_{itj}^* \neq \Delta_{i(t-1)j}^*)$ be the maximum number of differences between the consecutive elements $\Delta_{itj}^*$ and $ \Delta_{i(t-1)j}^*$.
Let $\Delta_{\max} = \Delta_{m} + 1$.

We start with imposing an assumption on the mean and the covariance matrix of the replicates for each independent subject.

\begin{assumption}
\label{assumption3}
For the $i$th subject and $j$th variable, let $\bX_{ij}  = (X_{i1j}, \ldots, X_{iTj})^{\T} \sim N(\bmu_{ij}, \bSigma_{jj})$.
Assume that the mean of $X_{itj}$ is bounded by a constant, i.e.,  $|\mu_{itj}| \le \mu_m$. 
In addition, assume that the $\ell_2$-norm of $\bmu_{ij}$ satisfies  $\| \bmu_{ij}\big\|_2 \le \mu_m \min (c_4^{\frac{2}{3}} n^{\frac{1}{3}} T^{\frac{1}{6}}, \sqrt{T})$ with $c_4 = \{ 4\log (T) \Delta_{\max}^2 \tau^2 / \pi^2\}^{1 / 4}$. 
Finally, assume that there exists a constant $\kappa>0$ such that 
$\max_{1\le j \le p} \big \|\bSigma_{jj} \big \|_{\mathrm{op}} \le \kappa,$
where $\|\bSigma_{jj}\|_{\mathrm{op}}$ is the operator norm of $\bSigma_{jj}$.
\end{assumption}

Recall that the mean $\bmu_{ij}$ depends on the latent effect $\Delta_{itj}^*$ in (\ref{GauModel}). For technical convenience, similar to \cite{hall2016inference}, we assume that $\mu_{itj}$ is bounded in order to control $X_{itj}$. In addition, we further require that the $\ell_2$-norm of $\bmu_{ij}$ cannot grow too fast with $T$ and $n$, which is mainly used to control the magnitude of $\sum_{t=1}^TX_{itj}^T(\bar \Delta_{itj}-\Delta^*_{itj})$ with some intermediate estimator $\bar \Delta_{itj}$. We note that the bound on the $\ell_2$-norm of $\bmu_{ij}$ always holds, provided $|\mu_{itj}| \le \mu_m$ and $T=O(n)$.

Next, we state a compatibility-type condition similar to that of \citet{buhlmann2011statistics} for model~\eqref{GauModel}.  
To this end, we define some additional notation. 
Let $\bomega_j^* = \{(\btheta_{j, -j}^*)^{\T}, (\balpha_j^*)^{\T}\}^{\T}$ and let  $\cS_{j} = \{ k: \omega_{jk}^* \ne 0\}$ be the active set.
We denote $s_{j} = | \cS_{j} |$ as the cardinality of $\cS_{j}$. 
Let $\Yb_{itj} = (\Xb_{it(-j)}, \Xb_{i(t-1)})\in \mathbb{R}^{1\times (2p-1)}$ and $\Yb_{j} = (\Yb_{11j}^{\T}, \Yb_{12j}^{\T}, \ldots, \Yb_{1Tj}^{\T}, \Yb_{21j}^{\T}, \ldots, \Yb_{nTj}^{\T})^{\T}\in \mathbb{R}^{nT\times (2p-1)}$.
Moreover, let $\bomega_{j}^{\cS_{j}}$ and $\bomega_{j}^{\cS_{j}^c}$ be subvectors of $\bomega_{j}$ with indices $\cS_j$ and $\cS_j^c$, respectively.

\begin{assumption}
\label{cconditions}
Let $\hat{\bSigma}_j = \Yb_{j}^{\T}\Yb_{j}/(nT)$. 
For some constant $\phi_0>0$ and $\bomega_{j}^*$ satisfying $(\bomega_{j}^*)^{\cS_j^c} \le 7 \|(\bomega_{j}^*)^{\cS_j} \|_1$, we have
$$\big \| (\bomega_{j}^*)^{\cS_j} \big \|_1^2 \le \frac{\left(\bomega_{j}^*\right)^{\T} \hat{\bSigma}_j \bomega_{j}^*s_j}{\phi_0^2}.$$
\end{assumption}

In the following, we present our main results on the estimation error of our proposed estimator under two scenarios: (i) the case when the number of replicates exceeds the number of independent samples, i.e., $T > c_4^2 n$; (ii) the case when $T\le c_4^2 n$.
For notational simplicity, let $\sigma_{m}  = \max \{\sqrt{2 \kappa}, \sqrt{2}\sigma_m^{\epsilon},1\}$ where $\kappa$ is as defined in Assumption~\ref{assumption3}. Moreover, we will use the notation $\mathcal{C}_i$ for $i = 1, \ldots, 7$ to denote generic constants that do not depend on $n$, $p$, $T$, $\tau$, $\Delta_{\max}$, and $\sigma_m$; see the proof in Appendix~\ref{proof of Theorem} for specific values of $\mathcal{C}_i$.
\begin{theorem}
\label{th1}
Assume that $T > c_4^2 n$ holds. Set the tuning parameters as
$$
\gamma =  \mathcal{C}_1\sigma_m \sqrt{\log(T) / \max \{1, \lfloor c_4^{4/3} T^{1/3} n^{2/3}\rfloor\}} n^{-1} T^{-1/2},~~~
\lambda = \beta = 2 \log (T) \log (nTp) n^{-1/6} T^{-1/3},
$$
in (\ref{equation13}). 
When 
$$2\max \left(c_1 n^{-\frac{1}{6}} T^{-\frac{1}{3}}, c_2 n^{-\frac{1}{6}} T^{-\frac{1}{3}} + c_3 n^{-\frac{1}{2}} T^{-\frac{1}{2}}\right) \le 1,$$
 $n, T, p \ge 6$, and under Assumptions~\ref{assumption2}--\ref{cconditions}, we have
\begin{align*}
\big\| \hat{\btheta}_{j,-j} - \btheta_{j,-j}^*\big\|_1 + \big\| \hat{\balpha}_{j} - \balpha_{j}^* \big \|_1 + \frac{1}{\sqrt{nT}} \big \|\hat{\bDelta}_j - \bDelta_j^* \big \|_2 
\le 2\max \left(c_1 n^{-\frac{1}{6}} T^{-\frac{1}{3}}, c_2 n^{-\frac{1}{6}} T^{-\frac{1}{3}} + c_3 n^{-\frac{1}{2}} T^{-\frac{1}{2}}\right),
\end{align*}
with probability at least 
$1  - 2 \exp\{-\lfloor \mathcal{C}_2 (nT)^{2/3}\rfloor\} - 2/ \{(T-1) \sqrt{ \log(T-1)}\} - 4/(nTp) - 2T^{-1/(2 \mathcal{C}_2) \min \{\log(T)/\mathcal{C}_2 , 1\}}$, where
\begin{align*}
c_1 &=  \mathcal{C}_3 \log (T) \log (nTp) (s_j + \sqrt{s_j}\phi_0)/\phi_0^2,\\
c_2 &= \mathcal{C}_4 [\sigma_m^2 c_4^{4/3} (\mu_m + 1)^2 / \{\log(T) \log(nTp)\} + \sigma_m c_4^{2/3}],\\
c_3 &= \mathcal{C}_5 [\sigma_m \sqrt{\log(T)} + \sigma_m^2 \{\log(T) + c_4^{2/3}  \sqrt{\log(T)} (\mu_m + 1) \} / \{\log(T) \log(nTp)\}].
\end{align*}
\end{theorem}

\begin{theorem}
\label{th2}
Assume that  $T \le c_4^2 n$ holds.  Set the tuning parameters as
$$
\gamma = \mathcal{C}_1 \sigma_m \sqrt{\log(T) / (T-1)} n^{-1} T^{-1/2}, ~\mathrm{and}~ \lambda = \beta = 2 \log (T) \log (nTp) T^{-1/2},
$$
in (\ref{equation13}). When 
$$T^{\frac{1}{2}} \ge 2 \max \left(c_1^{'}, c_2^{'}\right),$$
 $n, T, p \ge 6$, and under Assumptions~\ref{assumption2}--\ref{cconditions}, we obtain
\begin{align*}
\big\| \hat{\btheta}_{j,-j} - \btheta_{j,-j}^*\big\|_1 + \big\| \hat{\balpha}_{j} - \balpha_{j}^* \big \|_1 + \frac{1}{\sqrt{nT}} \big \|\hat{\bDelta}_j - \bDelta_j^* \big \|_2 
\le 2 \max \left(c_1^{'}, c_2^{'}\right) T^{-\frac{1}{2}},
\end{align*}
with probability at least $1 - 2 \exp\{-(T-1)\} - 2/ \{(T-1) \sqrt{ \log(T-1)}\} - 4/(nTp) - 2T^{-1/(2 \mathcal{C}_2) \min \{\log(T)/\mathcal{C}_2 , 1\}}$, where
\begin{align*}
c_1^{'} &=  \mathcal{C}_3 \log (T) \log (nTp) (s_j + \sqrt{s_j}\phi_0)/\phi_0^2,\\
c_2^{'} &= \mathcal{C}_6 [\sigma_{m}^2 c_3^{'} (\mu_m + 3)^2 / \{\log (T) \log (nTp)\} +  \sigma_m (c_3^{'})^{1/2} (\mu_m + 4)],\\
c_3^{'} &= \Delta_{\max} \tau \log^{1/2} (T).
\end{align*}
\end{theorem}

\begin{remark}
Since the estimator $(\hat{\btheta}_{j,-j}, \hat{\balpha}_{j}, \hat{\bDelta}_j)$ is obtained by solving a lasso type  problem in (\ref{equation13}), one may follow the standard proof in \cite{buhlmann2011statistics} to establish the error bound of $(\hat{\btheta}_{j,-j}, \hat{\balpha}_{j}, \hat{\bDelta}_j)$. However, this will lead to slower rates of convergence than those obtained in the above Theorems~\ref{th1}--\ref{th2} due to the structure of the fused lasso penalty not being fully exploited. 
In a recent paper, \cite{wang2016trend} established the sharp rates for the fused lasso estimator based on the incoherence property of the discrete difference operator; see also \cite{tibshirani2014adaptive}. 
Our proof strategy is partially inspired by their technique. However, there are several important differences. First, we decouple the temporal dependence among random variables using martingales.  Second, due to the combination of lasso and fused lasso penalties in (\ref{equation13}), the error bound consists of both the estimation  error of lasso $(\hat{\btheta}_{j,-j} - \btheta_{j,-j}^*, \hat{\balpha}_{j} - \balpha_{j}^*)$ and the error of fused lasso $\hat{\bDelta}_j - \bDelta_j^*$. To obtain a sharp rate, one needs to carefully quantify and balance these two terms in the proof by choosing their tuning parameters $\gamma, \lambda$, and $\beta$ in an optimal way. Our theorems reveal that the optimal choices of $\gamma, \lambda$, and $\beta$ differ depending on whether $T$ exceeds $c_4^2 n$ and vice versa. 

\end{remark}

To further simplify the results in Theorem \ref{th1} and \ref{th2},   assume that $\phi_0$, $\sigma_m$, $\Delta_{\max}$, and $\tau$ are all constants. Consider the asymptotic regime $n,T\rightarrow\infty$. Then, Theorems \ref{th1} and \ref{th2} imply that with probability tending to 1,
\begin{equation}\label{eq_rate}
\big\| \hat{\btheta}_{j,-j} - \btheta_{j,-j}^*\big\|_1 + \big\| \hat{\balpha}_{j} - \balpha_{j}^* \big \|_1 + \frac{1}{\sqrt{nT}} \big \|\hat{\bDelta}_j - \bDelta_j^* \big \|_2 \lessapprox \bigg\{ 
  \begin{matrix}
    s_j n^{-1/6}T^{-1/3}, & ~~\textrm{if}~~ T \gtrapprox n, \\
    s_j T^{-1/2}, &  ~~\textrm{otherwise,}
  \end{matrix}
\end{equation}
where the notation $a_n \lessapprox b_n$ stands for $a_n = O(b_n)$  up to a logarithmic factor and $a_n \gtrapprox b_n$ is defined similarly.  Following the standard proof in \cite{buhlmann2011statistics}, we can show that, if the incidental nuisance parameter $ \bDelta_j $ is known, we can obtain the following error bound for the lasso estimator 
\begin{equation}\label{eq_rate2}
\big\| \bar{\btheta}_{j,-j} - \btheta_{j,-j}^*\big\|_1 + \big\| \bar{\balpha}_{j} - \balpha_{j}^* \big \|_1\lessapprox s_j(nT)^{-1/2},
\end{equation}
where $(\bar{\btheta}_{j,-j},\bar{\balpha}_{j})$ minimizes the loss function $(2nT)^{-1}\|\xb_{j} - \boldsymbol{\eta}_{j}\|_2^2 +\lambda\|\boldsymbol{\theta}_{j,-j}\|_1+\beta\|\boldsymbol{\alpha}_j\|_1$ with $\bDelta_j $ fixed and $\boldsymbol{\eta}_{j}$ defined in  (\ref{equation13}) and 
$nT$ can be viewed as the sample size. Due to the presence of a large amount of unknown incidental nuisance parameters $ \bDelta_j^* $, the rate in (\ref{eq_rate}) is nonstandard and slower than $s_j(nT)^{-1/2}$.

In the literature on incidental nuisance parameters, it is often of interest to study the estimator under the following two scenarios: (1) $N$ is fixed and $T\rightarrow\infty$; (2) $T$ is fixed and $N\rightarrow\infty$.  In the first case, we have $T \gtrapprox n$ and therefore the estimation error in (\ref{eq_rate}) is of order $s_jT^{-1/3}$. Moreover, if $s_j=O(1)$, the rate becomes $O_p(T^{-1/3})$, which agrees with the minimax optimal rate of the fused lasso estimator (ignoring the logarithmic factors) \citep{tibshirani2014adaptive}.  Thus, in the first case, the estimation error in (\ref{eq_rate}) is dominated by that from the fused lasso $\frac{1}{\sqrt{nT}} \big \|\hat{\bDelta}_j - \bDelta_j^* \big \|_2$ and given the results in \cite{tibshirani2014adaptive} the upper bound in (\ref{eq_rate}) is non-improvable. In the second case, we have $T \lessapprox n$ and the upper bound in (\ref{eq_rate}) becomes $O(s_j)$, which does not converge to 0. Therefore, the estimator is inconsistent. The current setting corresponds to the classical Neyman and Scott's problem, where the number of nuisance parameters $\bDelta_j$ increases too fast relative to the amount of data points $nT$. 

To conclude this section, we will show that our estimator is adaptive to the absence of unmeasured confounders. Recall that if we know a priori that there are no unmeasured confounders, i.e., $\bDelta_{j}^* = 0$, we can estimate $({\btheta}_{j,-j}, {\balpha}_{j})$ by the oracle lasso estimator with $\bDelta_{j}^* = 0$ leading to the error bound in (\ref{eq_rate2}). The following corollary shows that  if our approach is applied to the setting when there are no unmeasured confounders, the rate of convergence of our estimator is $O_p(s_j(nT)^{-1/2})$ (ignoring the logarithmic factors), which matches the oracle lasso estimator in (\ref{eq_rate2}). Therefore, our estimator provides the best possible rate even if there are no unmeasured confounders. 

\begin{corollary}
\label{pro1}
Assume that the model (\ref{GauModel}) does not contain any unmeasured confounders, i.e., $\bDelta_{j}^* = 0$. Set the tuning parameters as
$$
\gamma = \mathcal{C}_1 \sigma_m \sqrt{\log(T) /  \lfloor \log(T)\rfloor} n^{-1} T^{-1/2},~\mathrm{and}~ \lambda = \beta = 2 \log (T) \log (nTp) n^{-1/2} T^{-1/2},
$$
in (\ref{equation13}). When 
$$n^{\frac{1}{2}} T^{\frac{1}{2}} \ge 2 \max \left(c_1^{''}, c_2^{''}\right),$$
 $n, T, p \ge 6$, and under Assumptions~\ref{assumption2}, \ref{assumption1}, and \ref{cconditions}, we obtain
\begin{align*}
\big\| \hat{\btheta}_{j,-j} - \btheta_{j,-j}^*\big\|_1 + \big\| \hat{\balpha}_{j} - \balpha_{j}^* \big \|_1 + \frac{1}{\sqrt{nT}} \big \|\hat{\bDelta}_j - \bDelta_j^* \big \|_2 
\le 2 \max \left(c_1^{''}, c_2^{''}\right) (nT)^{-\frac{1}{2}},
\end{align*}
with probability at least $1  - 2 \exp\{- \lfloor \log(T)\rfloor\} - 2/ \{(T-1) \sqrt{ \log(T-1)}\} - 4/(nTp) - 2T^{-1/(2 \mathcal{C}_7) \min \{\log(T)/\mathcal{C}_7 , 1\}}$, where
\begin{align*}
c_1^{''} &=  \mathcal{C}_3 \log (T) \log (nTp) (s_j + \sqrt{s_j}\phi_0)/\phi_0^2,\\
c_2^{''} &= \mathcal{C}_8 [\sigma_{m}^2 / \log (nTp) +  \sigma_m \{\log (T)\}^{1/2}].
\end{align*}
\end{corollary}
\noindent The proof of Corollary~\ref{pro1} is similar to the proofs of Theorems~\ref{th1}--\ref{th2}, and is hence omitted. 
 
%%%%%%%%%%%%%%%%%%%%%%%%%%%%%%%%%%%%
%%%%%%%%%%%%%%%%%%%%%%%%%%%%%%%%%%%%
% Numerical Studies
%%%%%%%%%%%%%%%%%%%%%%%%%%%%%%%%%%%%
%%%%%%%%%%%%%%%%%%%%%%%%%%%%%%%%%%%%
\section{Numerical Studies}
\label{sec:numerical studies}
In this section, we conduct extensive numerical studies to evaluate the performance of our proposal on different types of conditional independence graph: (i) Gaussian graphical models, and (ii) binary Ising models.
For each model, we compare our proposed method to some existing methods on latent variable graphical models.
To evaluate the performance across different methods, we define the true and false positive rates as the proportion of correctly estimated edges and the proportion of incorrectly estimated edges in the underlying graph, respectively.  

%%%%%%%%%%%%%%%%%%%%%%%%%%%%%%%%%%%%
%%%%%%%%%%%%%%%%%%%%%%%%%%%%%%%%%%%%
% Gaussian Graph
%%%%%%%%%%%%%%%%%%%%%%%%%%%%%%%%%%%%
%%%%%%%%%%%%%%%%%%%%%%%%%%%%%%%%%%%%
\subsection{Gaussian Graphical Models}
\label{subsec:ggm}
For Gaussian graphical models, we compare our proposal with four different existing methods: the graphical lasso \citep{friedman2008sparse}; the neighborhood selection approach \citep{meinshausen2006high}; the low-rank plus sparse latent variable Gaussian graphical model \citep{chandrasekaran2010latent}; and latent variable graphical models with replicates \citep{tan2016replicates}. 
\cite{friedman2008sparse}, \cite{meinshausen2006high}, and \cite{chandrasekaran2010latent} do not explicitly model the replicates: we therefore apply these methods by treating the replicates as independent samples.
Moreover, our proposal, \cite{meinshausen2006high}, and \cite{tan2016replicates} yield asymmetric estimates of the edge set.  To obtain a symmetric edge set, we consider both the intersection and union rules described in \cite{meinshausen2006high}, and report the best results for the competing methods. We report our results using only the intersection rule.  

All of the aforementioned methods have a sparsity tuning parameter: we apply all methods using a fine grid of the sparsity tuning parameter values to obtain the curves shown in Figures~\ref{figure:f2}--\ref{fig:f4}.  
There is an additional tuning parameter for \cite{chandrasekaran2010latent}, which models the confounding bias introduced by the unmeasured confounders.  
We set this tuning parameter to equal a constant multiplied by the sparsity tuning parameter, and we consider different values of constants and report the best results for \cite{chandrasekaran2010latent}. 
Our proposal has two additional tuning parameters which model the correlated data and the effect introduced by the unmeasured confounders. We detail the choice of tuning parameters for different settings on replicates and unmeasured confounders in the corresponding sections.

To assess the effects of correlated data and latent variables on graph estimation, we consider three different data generating mechanisms: (i) correlated replicates without latent variables; (ii) independent replicates with latent variables; and (iii) correlated replicates with latent variables.
Out of the aforementioned approaches, our proposed method is the only method that models both correlated replicates and latent variables.  Both \citet{chandrasekaran2010latent} and \citet{tan2016replicates} model only the latent variables and do not take into account correlated replicates.

%%%%%%%%%%%%%%%%%%%%%%%%%%%%%%%%%%%%%%%%
%%%%%%%%%%%%%%%%%%%%%%%%%%%%%%%%%%%%%%%%
%%%%%%%%%%%%%%%%%%%%%%%%%%%%%%%%%%%%%%%%
%%%%%%%%%%%%%%%%%%%%%%%%%%%%%%%%%%%%%%%%
Recall that for Gaussian graphical models, the inverse covariance matrix encodes the conditional dependence relationships among the variables.  
Let $\bTheta = \bSigma^{-1}$. 
We generate the inverse covariance matrix $\boldsymbol{\Theta}$ by randomly setting $10\%$ of the off-diagonal elements in $\boldsymbol{\Theta}$ to equal 0.3, and setting the others to zero.
To ensure the positive definiteness of $\boldsymbol{\Theta}$, we set $\Theta_{jj} = |\Lambda_{\min}(\boldsymbol{\Theta})| + 0.1$ for $j = 1,2,\ldots,p$, where $\Lambda_{\min} (\bTheta)$ is the minimum eigenvalue of $\bTheta$.  We will use the aforementioned to generate $\bTheta$, unless otherwise is specified.
For all of the numerical studies, we set $n=50$, $T=20$, and $p=100$.  The results, averaged over 100 independent data sets, are summarized in Figures~\ref{figure:f2}--\ref{fig:f4}.

%%%%%%%%%%%%%%%%%%%%%%%%%%%%%%%%%%%%%%%%
%%%%%%%%%%%%%%%%%%%%%%%%%%%%%%%%%%%%%%%%
%%%%%%%%%%%%%%%%%%%%%%%%%%%%%%%%%%%%%%%%
%%%%%%%%%%%%%%%%%%%%%%%%%%%%%%%%%%%%%%%%
\subsubsection{Correlated Replicates without Unmeasured Counfounders}
\label{subsec:scenario2}
In this section, we evaluate the effect of correlated replicates on graph estimation.  We assume that the replicates within each subject are correlated under an $\mathrm{AR}(1)$ process, i.e., we assume that 
 \begin{equation}  
 \label{eq:scenario2}
             \Xb_{i1} \sim N_p(\mathbf{0},\mathbf{\Sigma}), \qquad
             \Xb_{it} \mid \Xb_{i(t-1)} \sim N_p( \Ab\Xb_{i(t-1)} ,\bSigma),\quad \mathrm{for~} t=2,\ldots,T,
\end{equation}  
where $\Ab$ is a transition matrix that quantifies the correlation between $\Xb_{it}$ and $\Xb_{i(t-1)}$. 
We consider two different types of transition matrix:
\begin{enumerate}[(i)]
	\item Diagonal transition matrix $\Ab$ with $A_{jj} = 0.9$ for $j=1,\ldots,p$.  In other words, each variable at the $t$th replicate is conditionally dependent only with itself for the $(t-1)$th replicate.
	\item Sparse transition matrix $\Ab$ with 5\% elements of $\Ab$ set to equal 0.3.  In other words, the $j$th variable at the $t$th replicate may be conditionally dependent with other variables at the $(t-1)$th replicate.
\end{enumerate}

\begin{figure}[!t]
\centering
   \subfigure[]{\includegraphics[scale=0.8]{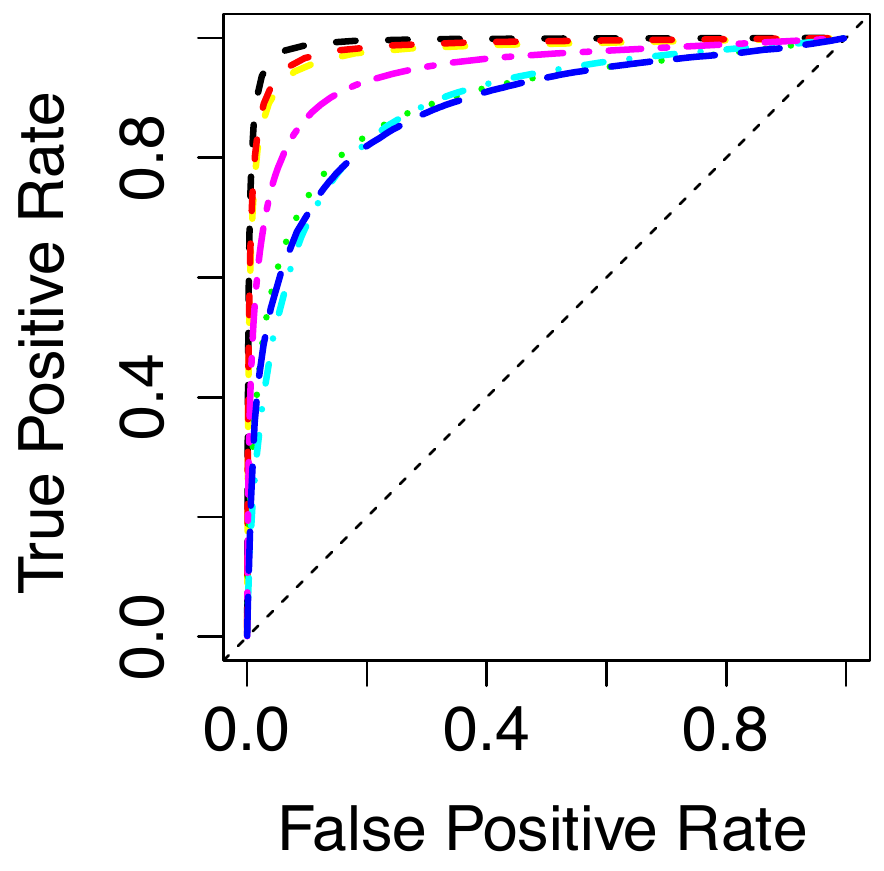}}\qquad\qquad
   \subfigure[]{\includegraphics[scale=0.8]{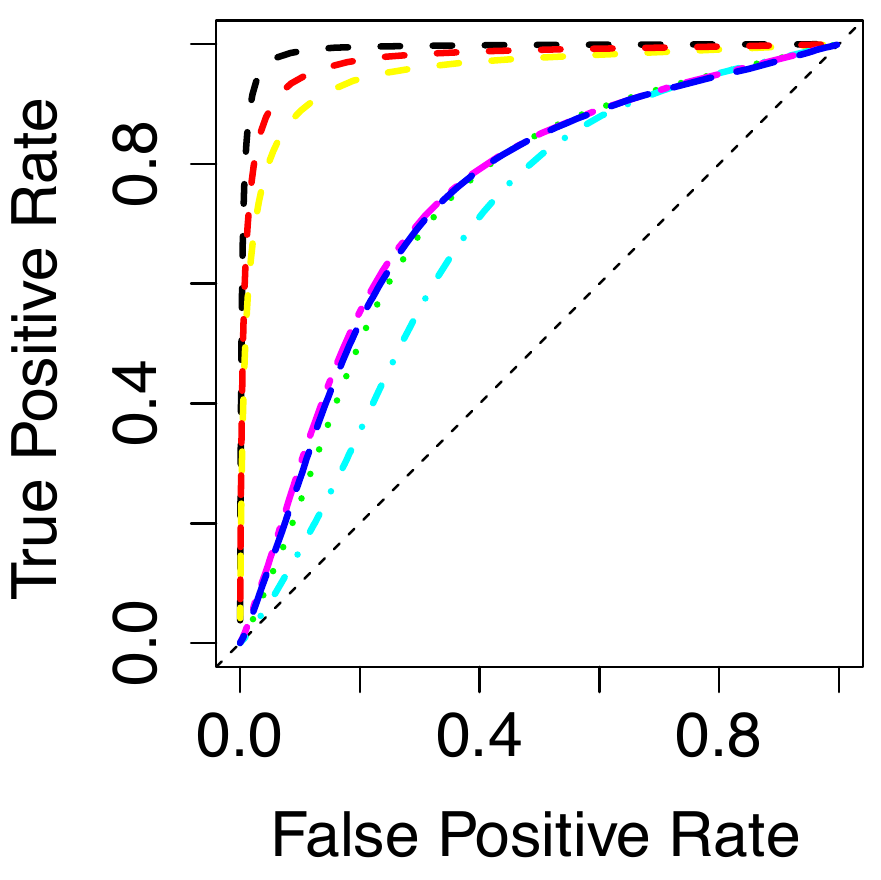}}
\caption{Results for correlated replicates without unmeasured confounders in Section~\ref{subsec:scenario2}. 
Panels (a) and (b) correspond to diagonal and sparse transition matrices, respectively. 
Since there are no unmeasured confounders, we set $\gamma$ to be arbitrary large, and consider three values of $\beta$: $\beta = 0.05$ (black short-dashed), $\beta=0.1$ (red short-dashed), and $\beta=0.15$ (yellow short-dashed). The other curves represent \cite{friedman2008sparse} (cyan dot-dashed); \cite{meinshausen2006high} (green dots); \cite{chandrasekaran2010latent} (blue long-dashed); and \cite{tan2016replicates} (pink short-long-dashed).}
\label{figure:f2}
\end{figure}

We generate the data according to \eqref{eq:scenario2}. 
For our proposal, we set $\gamma$ to be arbitrarily large since this simulation setting does not have unmeasured confounders.  We vary the tuning parameter $\beta$ to assess the performance of our proposal relative to existing methods across three values of $\beta$, i.e., $\beta\in \{0.05,0.1,0.15\}$.
The results are presented in Figure~\ref{figure:f2}. 
From Figure~\ref{figure:f2}, we see that our proposed method using different values of $\beta$ dominate all of the competing methods that assume independent replicates.
The results illustrate that not modeling the correlation among the replicates can have a significant impact on the estimated graph structure.   
This is especially apparent in Figure~\ref{figure:f2}(b) when the correlation between two replicates is modeled using a sparse transition matrix.

%%%%%%%%%%%%%%%%%%%%%%%%%%%%%%%%%%%%%%%%
%%%%%%%%%%%%%%%%%%%%%%%%%%%%%%%%%%%%%%%%
%%%%%%%%%%%%%%%%%%%%%%%%%%%%%%%%%%%%%%%%
%%%%%%%%%%%%%%%%%%%%%%%%%%%%%%%%%%%%%%%%

%%%%%%%%%%%%%%%%%%%%%%%%%%%%%%%%%%%%%%%%
%%%%%%%%%%%%%%%%%%%%%%%%%%%%%%%%%%%%%%%%
%%%%%%%%%%%%%%%%%%%%%%%%%%%%%%%%%%%%%%%%
%%%%%%%%%%%%%%%%%%%%%%%%%%%%%%%%%%%%%%%%
\subsubsection{Independent Replicates with Unmeasured Confounders}
\label{subsec:scenario3}
We now consider the case when there are unmeasured confounders with independent replicates.
Let $\Ub_{it}$ be unmeasured confounders for the $t$th replicate for subject $i$. 
We consider two settings:
\begin{enumerate}[(i)]
	\item The unmeasured confounders are constant across replicates within each subject, that is $\Ub_{i1} = \Ub_{i2} = \cdots = \Ub_{iT}$. This simulation setting is considered in \citet{tan2016replicates}. % \triangleq \Ub_i$.
	\item The unmeasured confounders are piecewise constant.  That is, we assume that 
	         \begin{numcases}{\Ub_{it}= }
                   \Ub_{it_1}, \quad \mathrm{when}~t \leq \lfloor T/2 \rfloor \notag,  \\
                   \Ub_{it_2}, \quad \mathrm{when}~t > \lfloor T/2 \rfloor  \notag, 
                  \end{numcases} 
where $\lfloor T/2 \rfloor$ is the largest integer that is less than or equal to $T/2$, and $ \Ub_{it_1} \neq  \Ub_{it_2}$.
\end{enumerate}

Similar to \citet{tan2016replicates}, we generate the data by first partitioning $\boldsymbol{\Sigma}$ and $\boldsymbol{\Theta}$ into 
\[
\boldsymbol{\Sigma} = 
\left(
\begin{matrix}
 \boldsymbol{\Sigma}_{XX}              &\boldsymbol{\Sigma}_{XU}  \\
 \boldsymbol{\Sigma}_{UX}               & \boldsymbol{\Sigma}_{UU}   \\
\end{matrix}
\right)\qquad  \mathrm{and} \qquad
\boldsymbol{\Theta} = 
\left(
\begin{matrix}
 \boldsymbol{\Theta}_{XX}              &\boldsymbol{\Theta}_{XU}  \\
 \boldsymbol{\Theta}_{UX}               & \boldsymbol{\Theta}_{UU}   \\
\end{matrix}
\right),
\]
where $\boldsymbol{\Theta}_{XX}$, $\boldsymbol{\Theta}_{XU}$, and $\boldsymbol{\Theta}_{UU}$ quantify the conditional independence relationships among the observed variables, between the observed variables and unmeasured confounders, and of the unmeasured confounders, respectively. 
We set $10\%$ of the off-diagonal entries in $\boldsymbol{\Theta}_{O,O}$ and $80\%$ of the off-diagonal entries in $\boldsymbol{\Theta}_{O,H}$ and $\boldsymbol{\Theta}_{H,H}$ to equal 0.3.
To ensure positive definiteness of $\bTheta$, we set $\Theta_{jj} = |\Lambda_{\min}(\boldsymbol{\Theta})| + 0.2$ for $j = 1,2,\ldots,p$. 

\begin{figure}[!t]
\centering
   \subfigure[]{\includegraphics[scale=0.8]{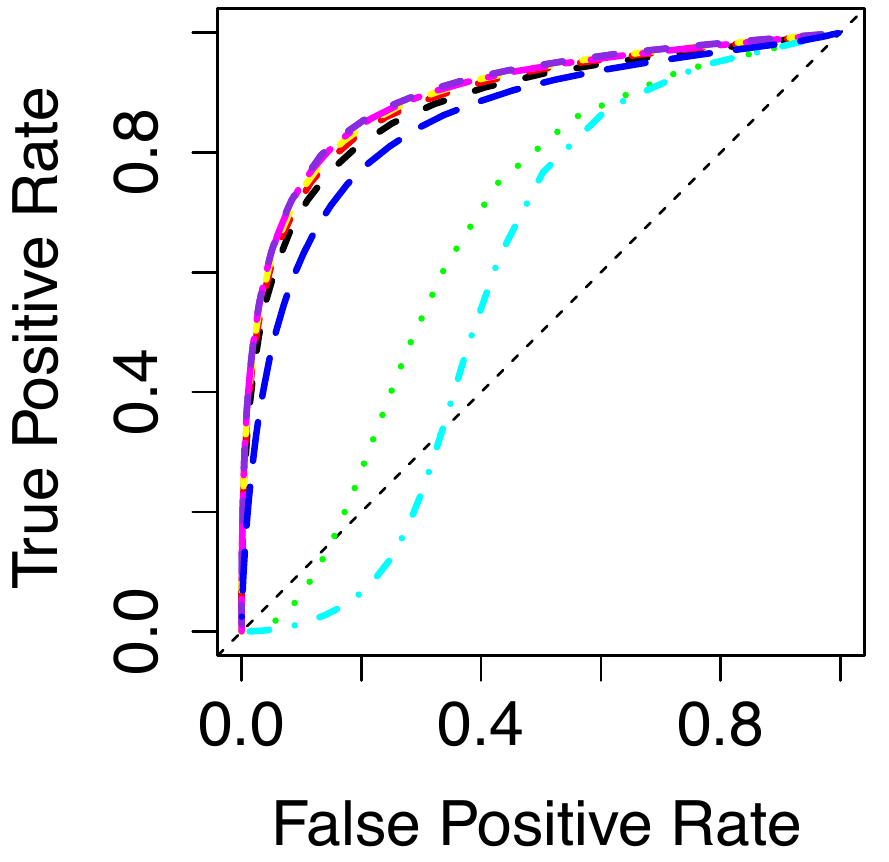}}\qquad\qquad
   \subfigure[]{\includegraphics[scale=0.8]{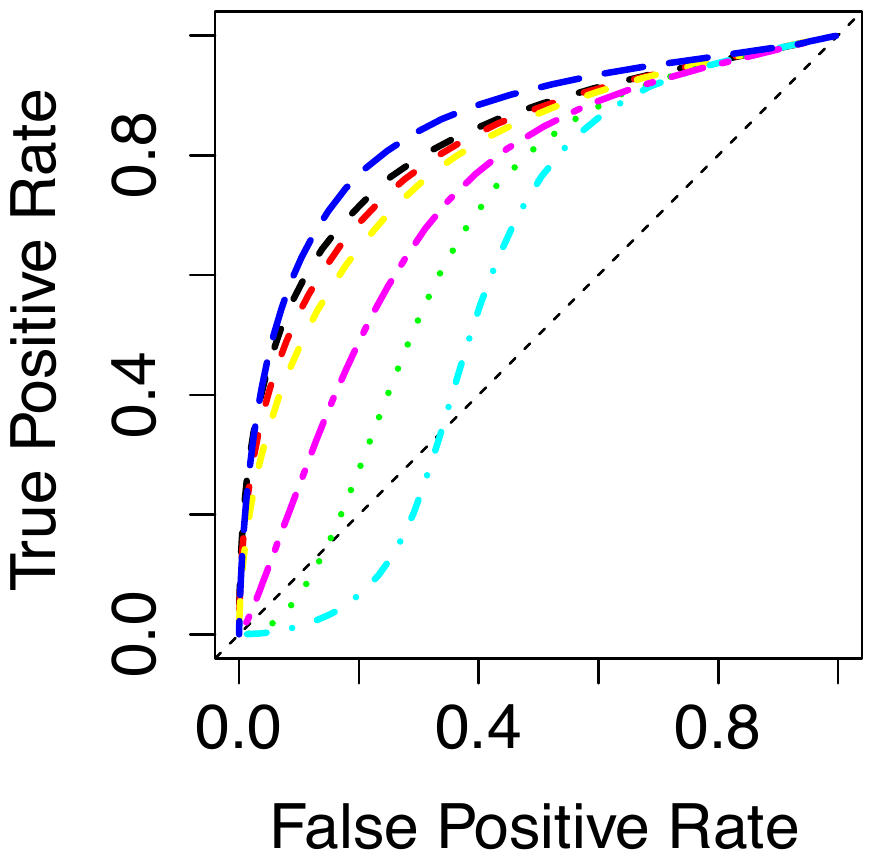}}
\caption{Results for independent replicates with unmeasured confounders in Section~\ref{subsec:scenario3}. 
Panels (a) and (b) correspond to the results for constant and piecewise constant unmeasured confounders, respectively.  
For our proposal, we set $\beta$ to be arbitrarily large since the replicates are independent. 
We consider three different values of $\gamma$: $\gamma = 1$ (black short-dashed), $\gamma = 1.5$ (red short-dashed), and $\gamma = 2$ (yellow short-dashed). For panel (a), we also consider the case when $\gamma$ is set to be arbitrarily large (purple short-dashed). 
Other details are as in Figure~\ref{figure:f2}.}
\label{fig:f3}
\end{figure}

For the scenario in which the unmeasured confounders are constant across replicates within each subject, we first generate $\Ub_i \sim N_q(\mathbf{0},\boldsymbol{\Sigma}_{UU})$.  
Then, we generate $T$ replicates for each subject from a conditional normal distribution, i.e., $\Xb_{it} \mid \Ub_i \sim N_p(\boldsymbol{\Sigma}_{XU} \boldsymbol{\Sigma}_{UU}^{-1}\Ub_i,\boldsymbol{\Sigma}_{XX}-\boldsymbol{\Sigma}_{XU} \boldsymbol{\Sigma}_{UU}^{-1}\boldsymbol{\Sigma}_{UX})$. 
For the second scenario in which the unmeasured confounders are piecewise constant within each subject, we generate $\Ub_i^1, \Ub_i^2 \sim N_q(\mathbf{0},\boldsymbol{\Sigma}_{UU})$.
Similarly to the first setting, when $t \leq \lfloor T/2 \rfloor$, we generate the $\lfloor T/2 \rfloor$ replicates for each subject from the conditional distribution depend on $\Ub_i^1$, then generate the rest replicates according to $\Ub_i^2$.
Recall that our proposal has two additional tuning parameters: we set $\beta$ to be arbitrary large since the replicates are independent, and consider three values of $\gamma \in \{1,1.5,2\}$. 
Besides, let $q = 5$, which means that we have 5 unmeasured confounders in total.
The results are summarized in Figure~\ref{fig:f3}. 

From Figure~\ref{fig:f3}, we see that methods that account for unmeasured confounders outperform methods that do not model the unmeasured confounders. Specifically, \citet{tan2016replicates} has the best performance in the case of independent replicates and constant unmeasured confounders in Figure~\ref{fig:f3}(a).
This is not surprising  since \citet{tan2016replicates} is explicitly designed to model such a setting.  
Our proposal reduces to that of \citet{tan2016replicates} as $\gamma,\beta \rightarrow \infty$. 
 Thus, our proposal has very similar performance to that of \citet{tan2016replicates}. 
However, when the unmeasured confounders are piecewise constant, our proposed method is much better than that of \citet{tan2016replicates} and is comparable to that of \citet{chandrasekaran2010latent}.

%%%%%%%%%%%%%%%%%%%%%%%%%%%%%%%%%%%%%%%%
%%%%%%%%%%%%%%%%%%%%%%%%%%%%%%%%%%%%%%%%
%%%%%%%%%%%%%%%%%%%%%%%%%%%%%%%%%%%%%%%%
%%%%%%%%%%%%%%%%%%%%%%%%%%%%%%%%%%%%%%%%

%%%%%%%%%%%%%%%%%%%%%%%%%%%%%%%%%%%%%%%%
%%%%%%%%%%%%%%%%%%%%%%%%%%%%%%%%%%%%%%%%
%%%%%%%%%%%%%%%%%%%%%%%%%%%%%%%%%%%%%%%%
%%%%%%%%%%%%%%%%%%%%%%%%%%%%%%%%%%%%%%%%
\subsubsection{Correlated Replicates with Unmeasured Confounders}
\label{subsec:scenario4}
In this section, we allow replicates within each subject to be correlated, and that there are unmeasured confounders.  
Throughout the numerical studies in this section, we assume that the correlated replicates are modeled according to the sparse transition matrix $\Ab$ as described in Section~\ref{subsec:scenario2}.
We consider constant and piecewise constant unmeasured confounders as described in Section~\ref{subsec:scenario3}.
Specifically, we assume the model
\begin{equation}
\label{eq:scenario4}
\begin{split}
\mathbf{X}_{i1}\mid \Ub_{i1} &\sim N_p(\boldsymbol{\Sigma}_{OH} \boldsymbol{\Sigma}_{HH}^{-1}\Ub_{i1}, \boldsymbol{\Sigma}_{XX}-\boldsymbol{\Sigma}_{XU} \boldsymbol{\Sigma}_{UU}^{-1}\boldsymbol{\Sigma}_{UX}),\\
\mathbf{X}_{it}\mid \mathbf{X}_{i(t-1)}, \Ub_{it} &\sim N_p(\Ab \Xb_{i(t-1)} + \boldsymbol{\Sigma}_{OH} \boldsymbol{\Sigma}_{HH}^{-1}\Ub_{it}, \boldsymbol{\Sigma}_{XX}-\boldsymbol{\Sigma}_{XU} \boldsymbol{\Sigma}_{UU}^{-1}\boldsymbol{\Sigma}_{UX}).
\end{split}
\end{equation}
We generate the data according to \eqref{eq:scenario4} using the same data generating mechanisms as described in Sections~\ref{subsec:scenario2}--\ref{subsec:scenario3}.

\begin{figure}[!htp]
\centering
   \subfigure[]{\includegraphics[scale=0.8]{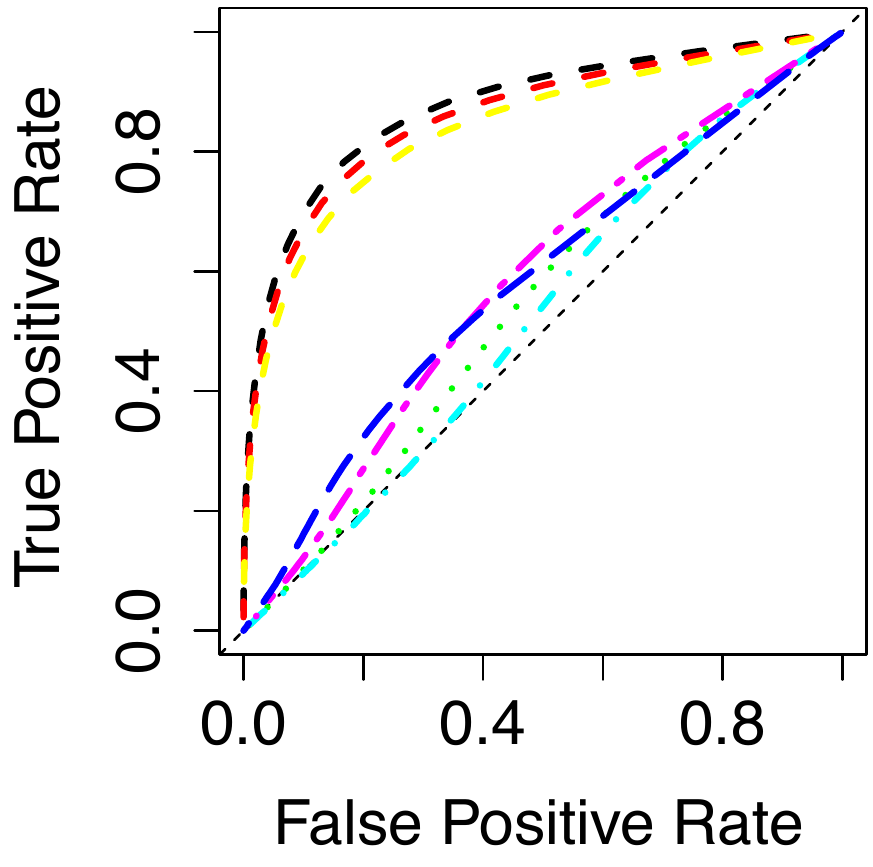}}\qquad\qquad
   \subfigure[]{\includegraphics[scale=0.8]{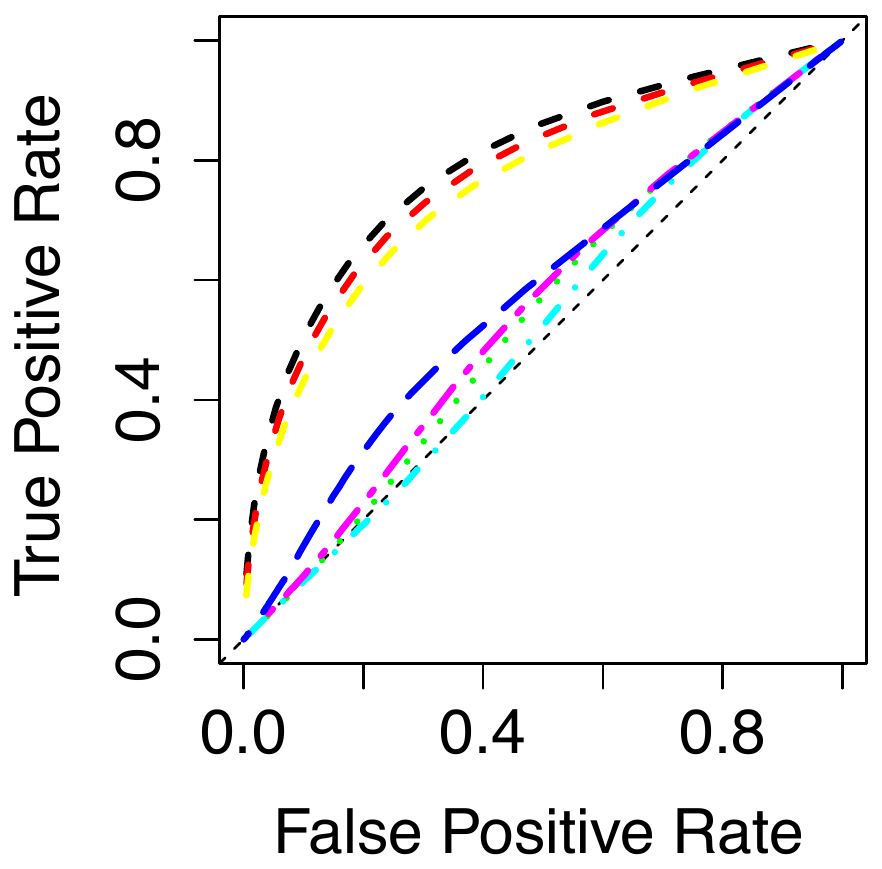}}
\caption{Results for constant and piecewise constant unmeasured confounders with sparse transition matrix $\Ab$ in Section~\ref{subsec:scenario4}. Panels (a) and (b) correspond to constant and piecewise constant unmeasured confounders, and thus we set $\gamma$ to be arbitrarily large and $\gamma=1$, respectively. 
The different curves for our proposal are with $\beta = 0.01$ (black short-dashed), $\beta = 0.02$ (red short-dashed), and $\beta = 0.03$ (yellow short-dashed).   
Other details are as in Figure~\ref{figure:f2}. }
\label{fig:f4}
\end{figure}

For the two additional tuning parameters in our proposal, we set $\gamma$ to be arbitrarily large for the case when the unmeasured confounders are constant, and consider $\beta\in \{0.01,0.02,0.03\}$.  
The results are shown in Figure~\ref{fig:f4}(a).
For the case when the unmeasured confounders are piecewise constant, we set $\gamma=1$ and consider $\beta\in \{0.01,0.02,0.03\}$.  
We have tried different values of $\gamma$ and have found that the results are not sensitive to different values of $\gamma$ in this simulation setting. 
The results are shown in Figure~\ref{fig:f4}(b).

We can see from both Figures~\ref{fig:f4}(a)--(b) that our proposal outperforms all existing methods when there are correlated replicates and unmeasured confounders. 
In fact, all existing methods have area under the curves of approximately 0.5.  
From Figure~\ref{fig:f4}(a), we see that even when the unmeasured confounders are constant, \citet{tan2016replicates} can no longer estimate the graph accurately since the conditional independent replicates assumption is violated.   
In short, we see that not modeling either the correlated replicates or unmeasured confounders can lead to biased estimation of the underlying graph.  

%%%%%%%%%%%%%%%%%%%%%%%%%%%%%%%%%%%%%%%%
%%%%%%%%%%%%%%%%%%%%%%%%%%%%%%%%%%%%%%%%
%%%%%%%%%%%%%%%%%%%%%%%%%%%%%%%%%%%%%%%%
%%%%%%%%%%%%%%%%%%%%%%%%%%%%%%%%%%%%%%%%

%%%%%%%%%%%%%%%%%%%%%%%%%%%%%%%%%%%%%%%%
%%%%%%%%%%%%%%%%%%%%%%%%%%%%%%%%%%%%%%%%
%% Ising Model
%%%%%%%%%%%%%%%%%%%%%%%%%%%%%%%%%%%%%%%%
%%%%%%%%%%%%%%%%%%%%%%%%%%%%%%%%%%%%%%%%
\subsection{Binary Ising Model}
\label{subsec:binaryising}
We now consider the binary Ising model with correlated replicates and  unmeasured confounders.  We compare our proposal to that of  \cite{ravikumar2011high}. 
We first generate $\bTheta$ described in Section~\ref{subsec:scenario3}, but set non-zero entries in $\bTheta$ from a Uniform distribution with support $[-0.5, -0.25] \cup [0.25, 0.5]$. 
Then, we generate the piecewise constant unmeasured confounders $\Ub_i$  as described in Section~\ref{subsec:scenario3}.
Given $\Ub_i$ and $\bTheta$, we apply apply Gibbs sampler to generate $\bX_{11}, \bX_{21}, \ldots, \bX_{n1}$, i.e., the first replicate for all subjects. 
Suppose that $x_{l11}, x_{l12}, \ldots, x_{l1p}$ are generated from the $l$th iteration of Gibbs sampler and we have obtained $\bX_{11}, \bX_{21}, \ldots, \bX_{(i-1)1}$, then
%% the following equation misses sth.
$$X_{(l+1) 1 j} \sim \mathrm{Bernoulli} \left\{ \frac{ \exp \left(\theta_{jj} +  \sum_{k\neq j} \theta_{jk} x_{l1k} + \sum_{m = p+1}^{p+q}\theta_{jm} u_{i1m}\right) }{1 + \exp \left(\theta_{jj} +  \sum_{k \neq j} \theta_{jk} x_{l1k} + \sum_{m = p+1}^{p+q}\theta_{jm} u_{i1m}\right) }\right\},
$$
where $j = 1, \ldots, p$.
Note that we take the first $10^4$ generated samples as burn-in samples, and collect one sample every $10^3$ iterations \citep{ravikumar2010high,tan2014learning}.

Then given the $i$th independent sample, we obtain $\bX_{i2}, \bX_{i3}, \ldots, \bX_{iT}$ using similar Gibbs sampler procedure but the distribution for $(l+1)$th iteration is now
$$X_{i (l+1) j} \sim \mathrm{Bernoulli} \left\{ \frac{ \exp \left(\theta_{jj} + \sum_{k \neq j} \theta_{jk} x_{l1k} + \sum_{m = p+1}^{p+q}\theta_{jm} u_{itm} + \sum_{k=1}^p \alpha_{jk} x_{i(t-1)k}\right) }{1 + \exp \left(\theta_{jj} + \sum_{k \neq j} \theta_{jk} x_{l1k} + \sum_{m = p+1}^{p+q}\theta_{jm} u_{itm} + \sum_{k=1}^p \alpha_{jk} x_{i(t-1)k}\right) }\right\},
$$
where $j = 1, \ldots, p$, $\xb_{il}$ are samples obtained from the $l$th iterations and $\balpha_j$ is the $j$th row of a diagonal transition matrix $\Ab$ described in Section~\ref{subsec:scenario2}.

\begin{figure}[!t]
\centering
   \includegraphics[scale=0.8]{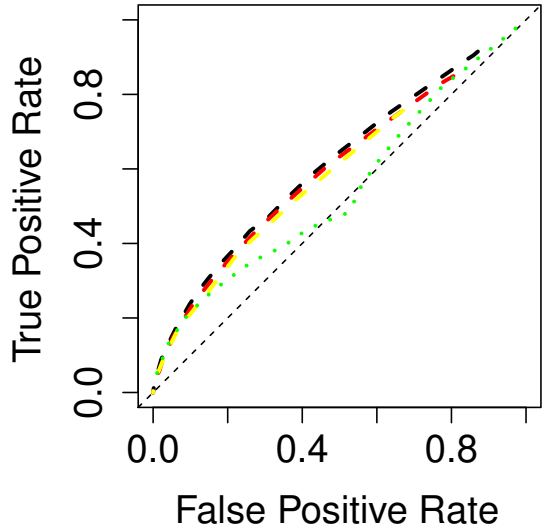}
\caption{Result for binary Ising model with correlated replicates and unmeasured confounders in Section~\ref{subsec:binaryising}. 
For our proposal, we set $\beta=0.01$ and three different values of $\gamma$: 
$\gamma = 0.5$ (yellow short-dashed), $\gamma = 1$ (red short-dashed), and $\gamma = 2$ (black short-dashed). 
The line with green dots represent \cite{ravikumar2011high}. }
\label{fig:f52}
\end{figure}

We set $n = 200$, $T = 10$, $p = 20$, and $q = 5$ and the results are shown in Figure \ref{fig:f52}.
For our proposal, we consider a fine-grid of $\lambda$, set $\beta = 0.01$, and vary $\gamma$ in three different values: 0.5, 1, and 2.
We see that our proposal outperforms \cite{ravikumar2011high}, which ignores the correlated replicates and unmeasured confounders.
%%%%%%%%%%%%%%%%%%%%%%%%%%%%%%%%%%%%%%%%
%%%%%%%%%%%%%%%%%%%%%%%%%%%%%%%%%%%%%%%%
%%%%%%%%%%%%%%%%%%%%%%%%%%%%%%%%%%%%%%%%
%%%%%%%%%%%%%%%%%%%%%%%%%%%%%%%%%%%%%%%%

%%%%%%%%%%%%%%%%%%%%%%%%%%%%%%%%%%%%%%%%
%%%%%%%%%%%%%%%%%%%%%%%%%%%%%%%%%%%%%%%%
%%%%%%%%%%%%%%%%%%%%%%%%%%%%%%%%%%%%%%%%
%%%%%%%%%%%%%%%%%%%%%%%%%%%%%%%%%%%%%%%%
\section{Data Application}
\label{sec:realdata}
In this section, we applied the proposed method to the ADHD-200 data \citep{biswal2010toward}. In this dataset, both resting state brain images and the phenotypic information of the subjects, such as age, gender, and intelligence quotient are available. 
After removing missing data from the original data set, we have 465 subjects, and each subject has between 76 and 276 images. 
We select 150 independent subjects from the groups of children and adolescent, respectively.  Moreover, for computational convenience, we select 10 consecutive images from each subject as replicates.  Similar to \citet{power2011functional}, we consider 264 brain regions of interest as nodes in Gaussian graphical models.

Although the data set consists of several phenotypic variables, there may also be some unmeasured phenotypic variables that can potentially serve as confounders.   
Ignoring the unmeasured confounders or the observed phenotypic variables and directly fitting a Gaussian graphical model using \citet{meinshausen2006high} may lead to a bias conditional independence graph.
In the following, we will compare the estimated graphs obtained from our proposed method with that of \citet{meinshausen2006high}.
Recall from Corollary~\ref{pro1} that our proposed estimator is adaptive to the absence of unmeasured confounders. 
In the event when there are no confounders, our method should yield a similar graph to that of  \citet{meinshausen2006high}.
On the other hand, if there are indeed confounders, the estimated graphs between our proposed method and \citet{meinshausen2006high} should be wildly different.

Our proposed method involves three tuning parameters, i.e., $\lambda$, $\beta$, and $\gamma$. 
As suggested by both Theorems~\ref{th1} and \ref{th2}, we set $\lambda=\beta$, reducing the tuning parameters from three to two. 
We consider a fine-grid of tuning parameters that yields the number of estimated edges in the range of $140-160$.
We then use the stability metric to select the two tuning parameters as suggested in \cite{lim2016estimation}.
Specifically, similar to a five-fold cross-validation, we split the independent subjects into five sub datasets, each of which consists of 80\% of the full data. 
We then estimate the parameters and calculate the estimation stability metric for each variable as follows:
\begin{equation*}
\label{RDA1}
\hat{\mb}_j = \frac{1}{5} \sum_{\ell= 1}^{5} \hat{\boldsymbol{\eta}}_j^\ell,\qquad
\mathrm{ES}_j = \frac{\frac{1}{5} \sum_{i = 1}^{5} \big\|  \hat{\boldsymbol{\eta}}_j^\ell - \hat{\mb}_j \big\|_2^2}{\big\| \hat{\mb}_j \big\|_2^2},
\end{equation*}
where $\hat{\boldsymbol{\eta}}_j^\ell = \Xb_{-j}^{\otimes}  \hat{\boldsymbol{\theta}}_{j,-j}^\ell + \Xb_j^{\otimes}  \hat{\boldsymbol{\alpha}}_j^\ell +  \hat{\boldsymbol{\Delta}}_j^\ell$ and $\hat{\boldsymbol{\theta}}_{j,-j}^\ell$, $\hat{\boldsymbol{\alpha}}_j^\ell$, and $\hat{\boldsymbol{\Delta}}_j^\ell$ are the estimates obtained using the $\ell$th subset of the data, for $\ell = 1, 2, \ldots, 5$.
Finally, we calculate the average estimation stability metric as
$\mathrm{ES} = \frac{1}{p} \sum_{j = 1}^{p} \mathrm{ES}_j$ and select the set of tuning parameters with the minimum ES.
The selected tuning parameters for the children subsets of data are $\lambda = \beta = 0.12$ and $\gamma = 0.5$, yielding a total number of $155$ edges.
On the other hand, the tuning parameters for the adolescent subsets of data are $\lambda = \beta = 0.1$ and $\gamma = 0.5$ which yield $159$ edges.

\begin{figure}[!t]
\centering
   \subfigure[]{\includegraphics[scale=0.19]{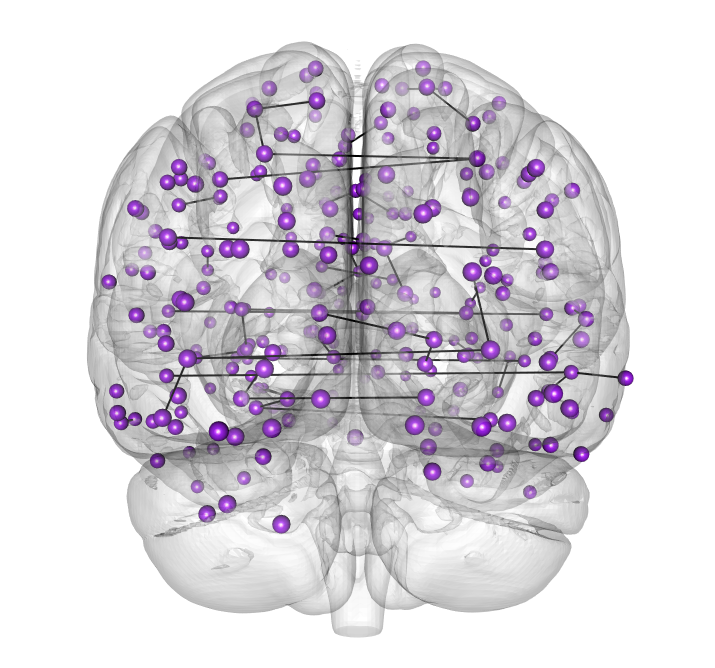}}\qquad
   \subfigure[]{\includegraphics[scale=0.19]{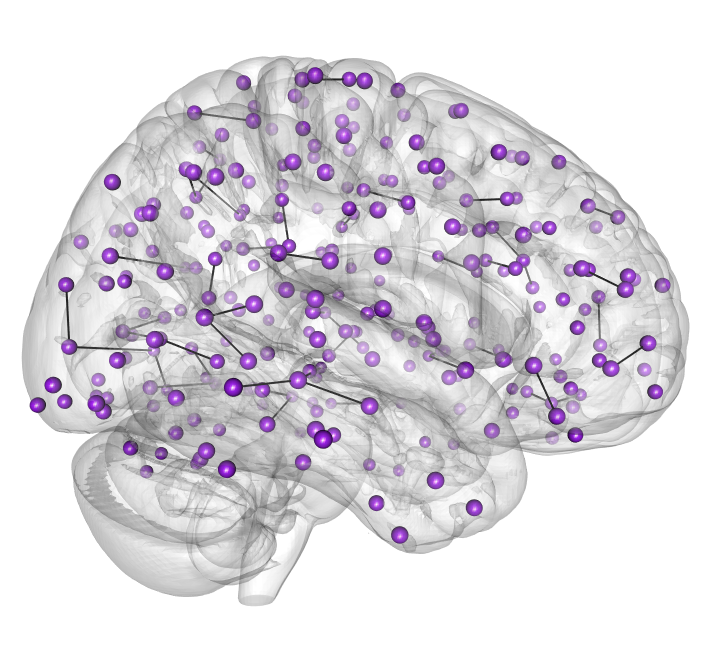}}\qquad
   \subfigure[]{\includegraphics[scale=0.19]{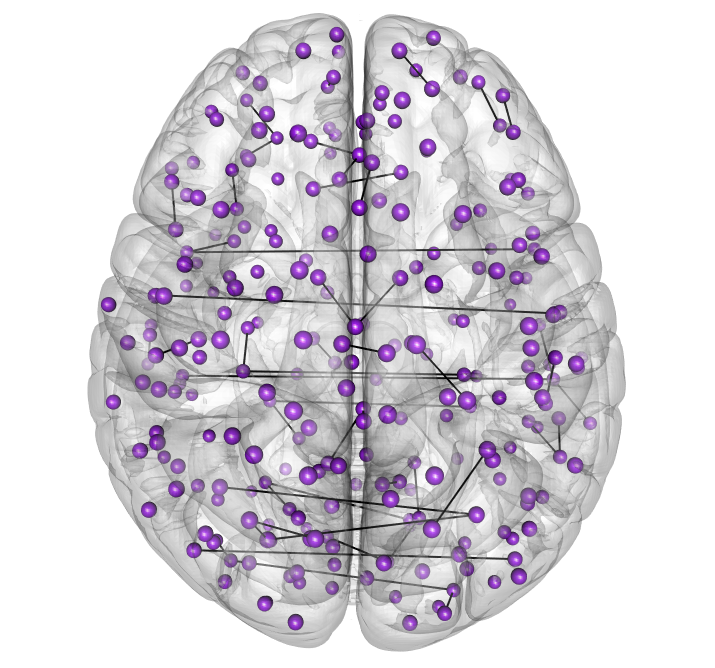}}\qquad
   \subfigure[]{\includegraphics[scale=0.19]{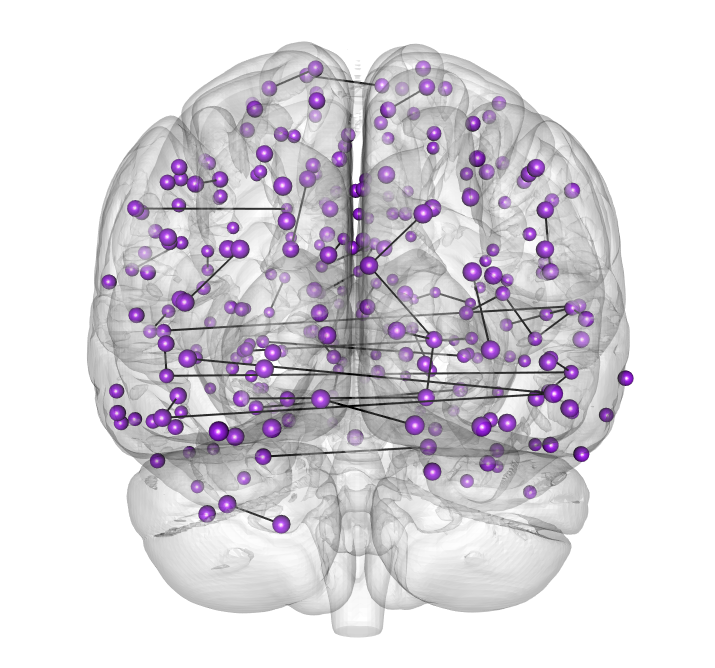}}\qquad
   \subfigure[]{\includegraphics[scale=0.19]{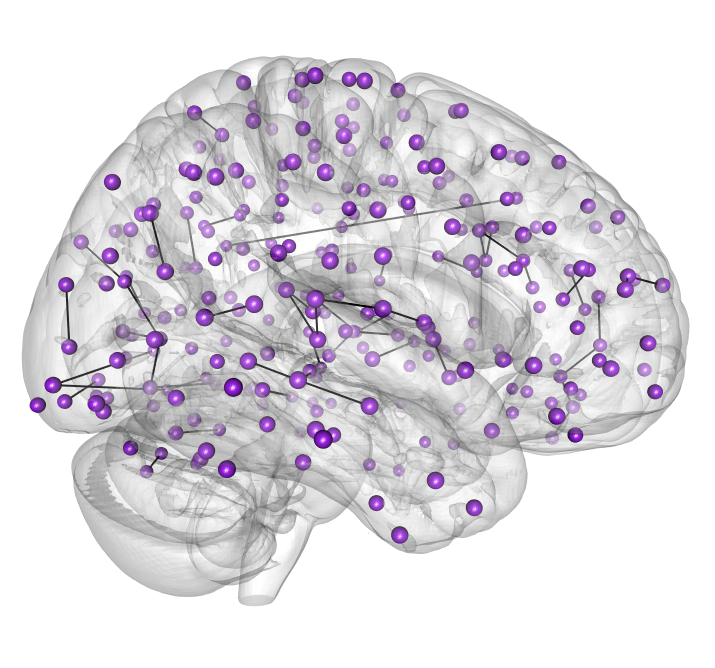}}\qquad
   \subfigure[]{\includegraphics[scale=0.19]{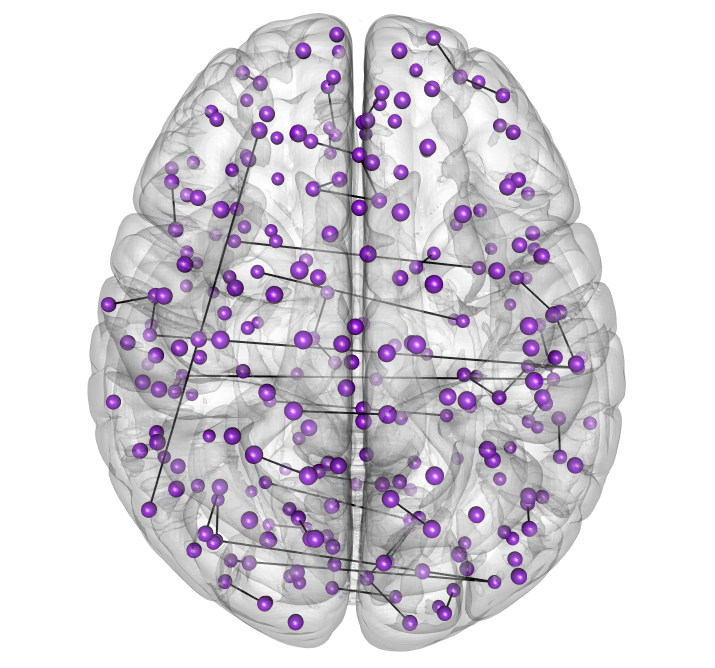}}
\caption{Coronal, sagittal, and transverse snapshots of the difference between our proposal and that of \cite{meinshausen2006high}. 
Panels (a--c) are coronal, sagittal, and transverse snapshots for children and panels (d--f) are coronal, sagittal, and transverse snapshots for adolescent.}
\label{fig:RDA2}
\end{figure}

We compare our proposed method to \cite{meinshausen2006high}, which ignores both correlated replicates and unmeasured confounders. 
We apply \cite{meinshausen2006high} by treating replicates as independent subjects.
We select the tuning parameter for \citet{meinshausen2006high} to yield 155 and 159 edges for children and adolescent, respectively.
We compare the estimated graphs obtained from our proposed method to that of \citet{meinshausen2006high} for both the children and adolescent datasets. The difference between the two estimated graphs are plotted in Figure~\ref{fig:RDA2}.
The estimated graphs between the two methods are drastically different for both children and adolescents. 
In particular, out of approximately 160 total number of edges, 50 edges are different between the two methods for both children and adolescents graphs.  
Our results suggest that the potential bias introduced by the correlation across replicates and unmeasured confounders can be large, and care must be taken when estimating a conditional dependence graph.

%%%%%%%%%%%%%%%%%%%%%%%%%%%%%%%%%%%%
%%%%%%%%%%%%%%%%%%%%%%%%%%%%%%%%%%%%
% Appendix
%%%%%%%%%%%%%%%%%%%%%%%%%%%%%%%%%%%%
%%%%%%%%%%%%%%%%%%%%%%%%%%%%%%%%%%%%
\appendix
\section{Technical Lemmas}
We first provide some technical lemmas to facilitate the proof of Theorem~\ref{th1}--\ref{th2}.
Lemmas~\ref{a1} and \ref{a3} control the tail behavior of interaction terms between the observed variable and the random noise $\epsilon_{itj}$. 
The proof of Lemma~\ref{a1} is provided in Section \ref{asub: proof of lemma 1}.
The proof of Lemma~\ref{a3} is similar to Lemma~\ref{a1} and is omitted.
Recall from Section~\ref{sec:theory} that $X_{itj} \sim N(\mu_{itj}, \sigma_{jj,t}^2)$ with $|\mu_{itj}| \le \mu_m$ and $(\sigma_{m}^X)^2  =  \max_{j,t} (\sigma_{jj,t}^2)$.
Besides, let $\epsilon_{itj} \sim N\{0, (\sigma_{jj,t}^{\epsilon})^2\}$ and $(\sigma_{m}^{\epsilon})^2 = \max_{j,t} \{(\sigma_{jj,t}^{\epsilon})^2\}$. 

\begin{lemma}
\label{a1} 
Assume $X_{itj} \sim N(\mu_{itj}, \sigma_{jj,t}^2)$ and $\epsilon_{itj} \sim N\{0, (\sigma_{jj,t}^{\epsilon})^2\}$.
We have
 $$ \underset{1 \le k \le p, k \ne j}{\max}~\frac{1}{nT}\left |\sum_{i = 1}^n \sum_{t = 1}^T \epsilon_{itj} X_{itk} \right |\le \lambda_0,$$ 
with probability at least $1 - \exp ( - \min [ \log^2 (T) / \{2 \sigma_{m}^{\epsilon}\sqrt{\mu_m^2 +(\sigma_{m}^{X})^2}\}^2, \log(T) / |2 \sigma_{m}^{\epsilon}\sqrt{\mu_m^2 +(\sigma_{m}^{X})^2}|] / 2) - \exp [\log\{2(p-1)\} - 3 \lambda_0^2 n T /  \{2 \lambda_0 \log (T) + 6 \log^2  (T)\}]$.
\end{lemma}

\begin{lemma}
\label{a3}
Assume $X_{i(t-1)k} \sim N(\mu_{i(t-1)k}, \sigma_{kk, t-1}^2)$ and  $\epsilon_{itj} \sim N\{0, (\sigma_{jj,t}^{\epsilon})^2\}$.
We have
$$ \underset{1 \le k \le p}{\max}~\frac{1}{nT} \left |\sum_{i = 1}^n \sum_{t = 1}^{T} \epsilon_{itj} X_{i(t-1)k} \right |\le \beta_0,$$
with probability at least $1 - \exp ( - \min [ \log^2 (T) / \{2 \sigma_{m}^{\epsilon}\sqrt{\mu_m^2 +(\sigma_{m}^{X})^2}\}^2, \log(T) / |2 \sigma_{m}^{\epsilon}\sqrt{\mu_m^2 +(\sigma_{m}^{X})^2}|] / 2) - \exp [\log(2p) - 3 \beta_0^2 n T /  \{2 \beta_0 \log (T) + 6 \log^2  (T)\}]$.
\end{lemma}

%\textcolor{red}{Recall that $\bDelta_{ij} = (\Delta_{i1j},\Delta_{i2j}, \ldots, \Delta_{iTj})^\T$ is the overall influence of unmeasured confounders on observed variables.}
Lemma~\ref{a2} establishes upper bounds for terms related to $\bDelta_{ij}$.
The proof is provided in Section~\ref{asub: proof of lemma 3}.
\begin{lemma}
\label{a2}
Let $\boldsymbol{\eta} \sim N \left(\boldsymbol{0}, \Qb \right)$ and $D = 8\sqrt{T\log(T) / (\pi^2 i_0)}$. 
For $i_0 \in \{1, \ldots, T-1\}$, we have
$$\frac{1}{nT} \boldsymbol{\eta}^{\T} \left (\bar{\bDelta}_{ij} - \bDelta_{ij} \right)
 \le \frac{\sqrt{2\big\|\Qb\big\|_{\mathrm{op}}} \left\{\sqrt{i_0} + \sqrt{\log\left(T\right)} \right\} }{nT} \big\| \bar{\bDelta}_{ij} - \bDelta_{ij} \big\|_2
  + \frac{\sqrt{\big \| \Qb \big\|_{\mathrm{op}}} D}{nT} \left( \big\| \Cb \bar{\bDelta}_{ij}\big\|_1 + \big\| \Cb \bDelta_{ij}\big\|_1 \right),$$
with probability at least $1 - \{2 (T-1) \sqrt{\log (T-1)}\}^{-1} - 2 \exp(-i_0) - 1 / \{T \sqrt{2\log{(T)}}\}$, where $\|\Qb\|_{\mathrm{op}}$ is the operator norm of $\Qb$.
\end{lemma}

%%%%%%%%%%%%%%%%%%%%%%%%%%%%%%%%%%%%
%%%%%%%%%%%%%%%%%%%%%%%%%%%%%%%%%%%%
% Proof of Theorem 1
%%%%%%%%%%%%%%%%%%%%%%%%%%%%%%%%%%%%
%%%%%%%%%%%%%%%%%%%%%%%%%%%%%%%%%%%%
\section{Proof of Theorems}
\label{proof of Theorem}
\subsection{Proof of Theorem~\ref{th1}}
\label{appendix:proof of theorem}
Let $\btheta_{j, -j}^*\in \RR^{p-1}$, $\balpha_{j}^* \in \RR^{p}$, and $ \bDelta_j^* \in \RR^{nT}$ be the true underlying  parameters, and let $\hat{\btheta}_{j, -j}$, $\hat{\balpha}_j$, and $\hat{\bDelta}_j$ be the solution obtained from solving~\eqref{equation7} under the Gaussian loss.
For notational convenience, we write $\bomega_j^* = \{(\btheta_{j, -j}^*)^{\T}, (\balpha_j^*)^{\T}\}^{\T}$ and $\hat{\bomega}_j = (\hat{\btheta}_{j, -j}^{\T}, \hat{\balpha}_j^{\T})^{\T}$. 
Let $\cS_{j} = \{ k: \omega_{jk}^* \ne 0\}$ be the active set and let $s_{j} = | \cS_{j} |$ be the cardinality of $\cS_{j}$. 
To establish an upper bound on the estimation error, we start with defining 
\[
N = \big \|\hat{\btheta}_{j, -j} - \btheta_{j, -j}^* \big \|_1 + \big \|\hat{\balpha}_{j} - \balpha_{j}^*\big \|_1 + \frac{1}{\sqrt{nT}} \big \|\hat{\bDelta}_j - \bDelta_j^*\big \|_2.
\]
The goal is to show that $N\le M$, where 
\begin{equation*}
\begin{split}
M &= 2 \max \left(c_1 n^{-\frac{1}{6}} T^{-\frac{1}{3}}, c_2 n^{-\frac{1}{6}} T^{-\frac{1}{3}} + c_3 n^{-\frac{1}{2}} T^{-\frac{1}{2}}\right);\\
c_1 &=  \frac{4 \log \left(T\right) \log \left(nTp\right) \left(4 s_j + \sqrt{s_j}\phi_0\right)}{\phi_0^2};\\
c_2 &= \frac{1792 \sigma_m^2 c_4^{\frac{4}{3}} \left(\mu_m + 1\right)^2 }{\log \left(T\right) \log \left(nTp\right)} + 448 \sigma_m c_4^{\frac{2}{3}};\\
c_3 &= 448 \sigma_m \sqrt{\log \left(T\right)} + \frac{1792 \sigma_m^2 \left\{\log \left(T\right) + 2 \left(\mu_m + 1\right) c_4^{\frac{2}{3}} \sqrt{\log \left(T\right)}\right\}}{\log \left(T\right) \log \left(nTp\right)};\\
c_4 &= \left[ \frac{4 \log \left(T\right) \Delta_{\max}^2 \tau^2}{\pi^2}\right]^{\frac{1}{4}}.
\end{split}
\end{equation*}
Note that the constant $\phi_0 > 0$ is the same compatability-type constant that appears in Assumption~\ref{cconditions}.
\iffalse
If $N\le M$, then we can conclude that 
$$\big \|\hat{\btheta}_{j,-j} - \btheta_{j,-j}^* \big \|_1 \le M;
\quad
\big \|\hat{\balpha}_{j, -j} - \balpha_{j, -j}^* \big \|_1 \le M;
\quad
\frac{1}{\sqrt{nT}}\big \|\hat{\bDelta}_j - \bDelta_j^* \big \|_2 \le  M.$$ 
\fi
%%%%%%%%%%%%%%%%%%%%%%%%%%%%%%%%%%%
% Proof Intuition
%%%%%%%%%%%%%%%%%%%%%%%%%%%%%%%%%%%
Let $ \zeta= M / (M+N)$ such that $0 <\zeta<1$. Set
\begin{equation*}
\begin{split}
\bar{\btheta}_{j, -j} &= \zeta\hat{\btheta}_{j, -j} + (1-\zeta)\btheta_{j, -j}^*; \\
\bar{\balpha}_{j} &= \zeta\hat{\balpha}_j + (1-\zeta)\balpha_j^*;\\
\bar{\bDelta}_{j} &= \zeta\hat{\bDelta}_j + (1-\zeta)\bDelta_j^*. 
\end{split}
\end{equation*}
Then, it can be shown that  $ \zeta N=\big \|\bar{ \btheta}_{j, -j} - \btheta_{j, -j}^* \big\|_1 + \big\|\bar{\balpha}_{j} - \balpha_{j}^*\big \|_1 +  \big\|\bar{\bDelta}_j - \bDelta_j^* \big\|_2 / \sqrt{nT}.$
In the following, we show that $\zeta N \le M/2$, which implies $N \le M$.

Let $Q \left(\bomega_j, \bDelta_{j} \right)$ be the loss function in \eqref{equation7} under the assumption that the random variables are Gaussian, that is,   
\begin{align}
\label{a3.00}
Q \left(\bomega_j, \bDelta_{j} \right) = \frac{1}{2nT} \big\| \Xb_{j} - \Yb_{j} \bomega_{j} - \bDelta_{j}\big \|_2^2+ \lambda \big\|\btheta_{j,-j}\big\|_1 + \beta \big\|\balpha_j \big\|_1 + \gamma \sum_{i = 1}^{n}\big\|\Cb \bDelta_{ij} \big \|_1, 
\end{align}
where
%where $\bDelta_{j} = (\bDelta_{1j}^{\T}, \bDelta_{2j}^{\T}, \ldots, \bDelta_{nj}^{\T})^{\T}$. 
$\bomega_j = (\btheta_{j, -j}^{\T}, \balpha_j^{\T})^{\T}$, 
$\Yb_{itj}= (\Xb_{it(-j)}, \Xb_{i(t-1)})$,
and $\Yb_{j}= (\Yb_{11j}^{\T}, \Yb_{12j}^{\T}, \ldots, \Yb_{1Tj}^{\T}, \Yb_{21j}^{\T}, \ldots, \Yb_{nTj}^{\T})^{\T}$. 
Since $Q(\cdot)$ is a convex loss, by convexity, we have 
\begin{align}
\label{a3.0}
Q \left (\bar{\bomega}_j, \bar{\bDelta}_j \right) 
&= Q \left \{ \zeta \hat{\bomega}_j +\left(1-\zeta\right) \bomega_j^* ,  \zeta\hat{\bDelta}_j + (1-\zeta)\bDelta_j^* \right\} \nonumber \\
&\le \zeta Q \left (\hat{\bomega}_j, \hat{\bDelta}_j \right) + \left (1-\zeta \right ) Q \left (\bomega_j^*, \bDelta_j^* \right ) \nonumber  \\
&\le Q\left (\bomega_j^*, \bDelta_j^* \right),
\end{align}
where the last inequality follows from the fact that $Q(\hat{\bomega}_j, \hat{\bDelta}_{j}) \le Q(\bomega_j^*, \bDelta_{j}^*)$.
Substituting (\ref{GauModel}) and (\ref{a3.00}) into (\ref{a3.0}), and upon rearranging the terms, we have 
\begin{align}
\label{a3.1}
&\frac{1}{2nT} \big\|\Yb_{j} \left (\bar{\bomega}_{j} - \bomega_{j}^* \right )\big\|_2^2 + \frac{1}{2nT} \big\|\bar{\bDelta}_j - \bDelta_j^* \big\|_2^2  \nonumber \\
&\le \underbrace{\frac{1}{nT} \bepsilon_{j}^{\T} \Yb_{j} \left (\bar{\bomega}_{j} - \bomega_{j}^* \right )  + \lambda \left (\|\btheta_{j,-j}^*\|_1 - \|\bar{\btheta}_{j,-j}\|_1\right )  
+ \beta \left (\|\balpha_{j}^*\|_1 - \|\bar{\balpha}_{j}\|_1\right )}_{\mathbb{I}_1} \nonumber\\
&+ \sum_{i = 1}^n \left \{ \underbrace{\frac{1}{nT} \bepsilon_{ij}^{\T} \left (\bar{\bDelta}_{ij} - \bDelta_{ij}^* \right ) 
 + 
 \frac{1}{nT} \left (\bar{\bomega}_{j} - \bomega_{j}^* \right )^{\T} \Yb_{ij}^{\T} \left (\bDelta_{ij}^* - \bar{\bDelta}_{ij} \right )}_{\mathbb{I}_2} 
 + 
 \gamma \left (\|\Cb \bDelta_{ij}^*\|_1 - \| \Cb \bar{\bDelta}_{ij}\|_1 \right ) 
 \right \}.
\end{align} 
We now establish upper bounds for $\mathbb{I}_1$ and $\mathbb{I}_2$, respectively.

\textbf{Upper Bound for $\mathbb{I}_1$:}
from the definition of $\Yb_j$, we have
\begin{align*}
\bepsilon_{j}^{\T} \Yb_{j} (\bar{\bomega}_{j} - \bomega_{j}^*) 
=  \underbrace{\bepsilon_{j}^{\T} \Xb_{-j}^{\otimes} (\bar{\btheta}_{j,-j} - \btheta_{j,-j}^*)}_{\mathbb{I}_{11}}
+  \underbrace{\bepsilon_{j}^{\T} \Xb_{j}^{\otimes} (\bar{\balpha}_{j} - \balpha_{j}^*)}_{\mathbb{I}_{12}}.
\end{align*} 
It suffices to obtain upper bounds for $\mathbb{I}_{11}$ and $\mathbb{I}_{12}$. 
By the Holder's~inequality, Lemma~\ref{a1}, and picking $\lambda = 2 \log (T) \log (nTp) n^{-1/6} T^{-1/3}$, 
when $n, T, p \ge 6$ we have  
\begin{equation}
\label{a3.2}
\frac{1}{nT} \bepsilon_{j}^{\T} \Xb_{-j}^{\otimes} \left (\bar{\btheta}_{j,-j} - \btheta_{j,-j}^* \right ) 
 \le  \frac{1}{nT} \big \| \bepsilon_{j}^{\T} \Xb_{-j}^{\otimes} \big \|_{\infty} \cdot \big \| \bar{\btheta}_{j,-j} - \btheta_{j,-j}^* \big \|_1 
 \le \frac{\lambda}{2} \big \| \bar{\btheta}_{j,-j} - \btheta_{j,-j}^* \big\|_1,
\end{equation} 
with probability at least $1 - \exp ( - \min [ \log^2 (T) / \{2 \sigma_{m}^{\epsilon}\sqrt{\mu_m^2 +(\sigma_{m}^{X})^2}\}^2, \log(T) / |2 \sigma_{m}^{\epsilon}\sqrt{\mu_m^2 +(\sigma_{m}^{X})^2}|] / 2) - 2 / (nTp)$. 
Similarly, by an application of Lemma~\ref{a3} and picking $\beta = \lambda$, when $n, T, p \ge 6$ we obtain
\begin{equation}
\label{a3.2-1}
\frac{1}{nT}\bepsilon_{j}^{\T} \Xb_{j}^{\otimes} (\bar{\balpha}_{j} - \balpha_{j}^*) \le  \frac{1}{nT} \big \| \bepsilon_{j}^{\T} \Xb_{j}^{\otimes}  \big \|_{\infty} \cdot \|  \bar{\balpha}_{j} - \balpha_{j}^*\|_1 \le \frac{\beta}{2} \| \bar{\balpha}_{j} - \balpha_{j}^*\|_1,
\end{equation}
with probability at least $1 - \exp ( - \min [ \log^2 (T) / \{2 \sigma_{m}^{\epsilon}\sqrt{\mu_m^2 +(\sigma_{m}^{X})^2}\}^2, \log(T) / |2 \sigma_{m}^{\epsilon}\sqrt{\mu_m^2 +(\sigma_{m}^{X})^2}|] / 2) -  2 / (nTp)$.
%Note that $\lambda \le \beta$.
%In addition, let $\omega_{j,k}^{S_{j}} = \omega_{jk} 1\{k \in S_j\}$, where $1\{k \in S_j\}$ is the indicator function of $S_j$. 
%Let $\bomega_{j}^{S_j} = (\omega_{j1}^{S_j}, \omega_{j2}^{S_j}, \ldots, \omega_{j(2p-1)}^{S_j})$.
Since $\lambda = \beta$, substituting~\eqref{a3.2} and~\eqref{a3.2-1} into $\mathbb{I}_1$ yields
\begin{align}
\label{a3.3-1}
\mathbb{I}_1 \le  
\frac{\beta}{2} \| \bar{\bomega}_{j} - \bomega_{j}^*\|_1 
+ \beta \left (\|\bomega_{j}^*\|_1 - \|\bar{\bomega}_{j}\|_1\right ),
\end{align} 
with probability at least $1 - 2\exp ( - \min [ \log^2 (T) / \{2 \sigma_{m}^{\epsilon}\sqrt{\mu_m^2 +(\sigma_{m}^{X})^2}\}^2, \log(T) / |2 \sigma_{m}^{\epsilon}\sqrt{\mu_m^2 +(\sigma_{m}^{X})^2}|] / 2) - 4/ (nTp)$.
Let $\bomega_{j}^{\cS_{j}}$ and $\bomega_{j}^{\cS_{j}^c}$ be subvectors of $\bomega_{j}$ with indices $\cS_j$ and $\cS_j^c$, respectively. 
Then, upon rearranging the terms, (\ref{a3.3-1}) can be rewritten as
\begin{align}
\label{a3.3}
\mathbb{I}_1\le
 \frac{3\beta}{2} \left \| \bar{\bomega}_{j}^{\cS_j} - (\bomega_{j}^*)^{\cS_j}\right \|_1 
 -  \frac{\beta}{2} \left \|\bar{\bomega}_{j}^{\cS_j^c} \right\|_1,
\end{align}  
with probability at least $1 - 2\exp ( - \min [ \log^2 (T) / \{2 \sigma_{m}^{\epsilon}\sqrt{\mu_m^2 +(\sigma_{m}^{X})^2}\}^2, \log(T) / |2 \sigma_{m}^{\epsilon}\sqrt{\mu_m^2 +(\sigma_{m}^{X})^2}|] / 2) - 4/ (nTp)$.

\textbf{Upper Bound for $\mathbb{I}_2$}:  we start with providing an upper bound for $\bepsilon_{ij}^{\T} (\bar{\bDelta}_{ij} - \bDelta_{ij}^*) / (nT)$. 
For $i_0 \in \{1, \ldots, T-1\}$, let $D = 8\sqrt{T\log(T) / (\pi^2 i_0)}$. 
Recall that $\epsilon_{itj} \sim N\{0, (\sigma^{\epsilon}_{tj})^2\}$ and $(\sigma^{\epsilon}_{m})^2 = \max_{t,j} \{(\sigma^{\epsilon}_{tj})^2\}$.
By Lemma~\ref{a2}, we have
\begin{align}
\label{a30}
\frac{1}{nT} \bepsilon_{ij}^{\T} \left (\bar{\bDelta}_{ij} - \bDelta_{ij}^* \right )  
\le \frac{\sqrt{2}\sigma^{\epsilon}_{m} \left\{\sqrt{i_0} + \sqrt{\log\left(T\right)} \right\} }{nT} \big\| \bar{\bDelta}_{ij} - \bDelta_{ij}^* \big\|_2
  + \frac{\sigma^{\epsilon}_{m}D}{nT} \left( \big\| \Cb \bar{\bDelta}_{ij}\big\|_1 + \big\| \Cb \bDelta_{ij}^*\big\|_1 \right),
\end{align}
with probability at least $1 - \{2 (T-1) \sqrt{\log (T-1)}\}^{-1} - 2 \exp(-i_0) - 1 / \{T \sqrt{2\log{(T)}}\}$. 

Next, we provide an upper bound for $(\bar{\bomega}_{j} - \bomega_{j}^* )^{\T} \Yb_{ij}^{\T} (\bDelta_{ij}^* - \bar{\bDelta}_{ij})/ (nT)$ in $\mathbb{I}_2$.
Let $\EE (\Yb_{ij}) = \Ub_{ij}$ and $B(n, T) = \mu_m \min (c_4^{2/3} n^{1/3} T^{1/6}, \sqrt{T})$. 
By Assumption~\ref{assumption3}, $\|\bmu_{ij}\|_2 \le B(n, T)$. 
Recall that $\| \bar{\bomega}_{j} - \bomega_{j}^*\|_1 + \| \bar{\bDelta}_{j} - \bDelta_{j}^*\|_2 / \sqrt{nT} = \zeta N \le M$ with $\zeta = M / (N + M)$.
This implies that $\big \| \bar{\bomega}_{j} - \bomega_{j}^* \big \|_1 \le M$.
Coupling the above with the Holder's inequality, we obtain
\begin{equation}
\begin{split}
\label{a31}
&\frac{1}{nT}  \left (\bar{\bomega}_{j} - \bomega_{j}^*  \right )^{\T}  \Yb_{ij}^{\T} \left (\bDelta_{ij}^* - \bar{\bDelta}_{ij} \right )\\
& \le \frac{1}{nT}  \big\| \bar{\bomega}_{j} - \bomega_{j}^*  \big\|_1 \cdot \big\|  \Yb_{ij}^{\T} \left (\bDelta_{ij}^* - \bar{\bDelta}_{ij} \right) \big\|_{\infty} \\
& \le \frac{M}{nT}\big \|  \Yb_{ij}^{\T} \left (\bDelta_{ij}^* - \bar{\bDelta}_{ij} \right) \big\|_{\infty} \\
& \le \frac{M}{nT}\big \|  \left(\Yb_{ij} - \Ub_{ij}\right)^{\T} \left (\bDelta_{ij}^* - \bar{\bDelta}_{ij} \right) \big\|_{\infty} +  \frac{M}{nT}\big \|  \Ub_{ij}^{\T} \left (\bDelta_{ij}^* - \bar{\bDelta}_{ij} \right) \big\|_{\infty} \\
& \le \frac{M}{nT}\big \|  \left(\Yb_{ij} - \Ub_{ij}\right)^{\T} \left (\bDelta_{ij}^* - \bar{\bDelta}_{ij} \right) \big\|_{\infty} +  \frac{M}{nT}\max_{k \neq j} \big\|\bmu_{ik}\big\|_2 \big \| \bDelta_{ij}^* - \bar{\bDelta}_{ij}  \big\|_2 \\
& \le  \frac{M\sqrt{2\kappa} \left\{\sqrt{i_0} + \sqrt{\log\left(T\right)} + B\left(n, T\right) \right\} }{nT} \big\| \bar{\bDelta}_{ij} - \bDelta_{ij}^* \big\|_2
  + \frac{M\sqrt{\kappa} D}{nT} \left( \big\| \Cb \bar{\bDelta}_{ij}\big\|_1 + \big\| \Cb \bDelta_{ij}^*\big\|_1 \right),
\end{split}
\end{equation}
with probability at least $1 - \{2 (T-1) \sqrt{\log (T-1)}\}^{-1} - 2 \exp(-i_0) - 1 / \{T \sqrt{2\log{(T)}}\}$.
Note that the last inequality follows by an application of Lemma~\ref{a2}, Assumption~\ref{assumption3}, and the fact that $\|  (\Yb_{ij} - \Ub_{ij})^{\T} (\bDelta_{ij}^* - \bar{\bDelta}_{ij} )\|_{\infty}
= \underset{k \ne \{j, j+2p\}}{\max}~ \widetilde{\bY}_{ik}^{\T} (\bDelta_{ij}^* - \bar{\bDelta}_{ij} )$, where $ \widetilde{\bY}_{ik}$ is the $k$th column of $\{(\Yb_{ij} - \Ub_{ij}), -(\Yb_{ij} - \Ub_{ij})\}$.
Let $\sigma_{m}  = \max \{\sqrt{2 \kappa}, \sqrt{2}\sigma_m^{\epsilon},1\}$. 
By (\ref{a30}) and (\ref{a31}), we have 
\begin{equation}
\label{a3.61}
\mathbb{I}_2 \le \frac{ \left( M + 1  \right)\left\{ \sqrt{i_0} + \sqrt{\log \left(T\right)} + B\left(n, T\right) \right\} \sigma_{m} }{nT} \big\| \bar{\bDelta}_{ij} - \bDelta_{ij}^* \big\|_2 + \frac{\left(M +1\right) \sigma_{m} D}{nT} \left( \big\|\Cb \bDelta_{ij}^*\big\|_1 +\big \|\Cb \bar{\bDelta}_{ij}\big\|_1 \right),
\end{equation}
with probability at least $1 - \{2 (T-1) \sqrt{\log (T-1)}\}^{-1} - 2 \exp(-i_0) - 1 / \{T \sqrt{2\log{(T)}}\}$.

Under the condition that $ 2\max (c_1 n^{-1/6} T^{-1/3}, c_2 n^{-1/6} T^{-1/3} + c_3 n^{-1/2} T^{-1/2}) \le 1$, we have $M \le 1$.
Moreover, recall that $\gamma$ is the tuning parameter for $\|(\Ib_n \otimes \Cb ) \bDelta_j^*\|_1$.
Let $\gamma = 2 \sigma_{m}  D/ (nT)$ and substituting (\ref{a3.3}) and (\ref{a3.61}) into (\ref{a3.1}), we have 

\begin{align}
\label{a3.6}
&\frac{1}{2nT} \big\|\Yb_{j} \left (\bar{\bomega}_{j} - \bomega_{j}^* \right )\big\|_2^2  + \frac{1}{2nT} \big\|\bar{\bDelta}_j - \bDelta_j^*\big\|_2^2 + \frac{\beta}{2} \big\|\bar{\bomega}_{j}^{\cS_j^c}\big\|\nonumber \\
&\le \frac{3\beta}{2} \big\| \bar{\bomega}_{j}^{\cS_j} - (\bomega_{j}^*)^{\cS_j}\big\|_1 +
\frac{2  \sigma_{m}  \left\{ \sqrt{i_0} + \sqrt{\log \left(T\right)}\right\}+ B\left(n, T\right)}{nT} \big\| \bar{\bDelta}_j - \bDelta_j^*\big\|_2 + \frac{4 \sigma_{m}  D}{nT} \big\| \left(\Ib_n \otimes \Cb \right) \bDelta_j^*\big\|_1,
\end{align}
with probability at least $1 - \{2 (T-1) \sqrt{\log (T-1)}\}^{-1} - 2 \exp(-i_0) - 1 / \{T \sqrt{2\log{(T)}}\}  - 4/ (nTp)  - \exp ( - \min [ \log^2 (T) / \{2 \sigma_{m}^{\epsilon}\sqrt{\mu_m^2 +(\sigma_{m}^{X})^2}\}^2, \log(T) / |2 \sigma_{m}^{\epsilon}\sqrt{\mu_m^2 +(\sigma_{m}^{X})^2}|] / 2)$.

We now consider (\ref{a3.6}) under the following two cases:
\begin{enumerate}[(i)]
	\item \label{step1}
                $\frac{2  \sigma_{m}  \left\{ \sqrt{i_0} + \sqrt{\log \left(T\right)}+ B\left(n, T\right)\right\}}{nT} \big\| \bar{\bDelta}_j - \bDelta_j^*\big\|_2 + \frac{4 \sigma_{m}  D}{nT} \big\| \left(\Ib_n \otimes \Cb \right) \bDelta_j^*\big\|_1 \le \frac{1}{4}\beta \big\|\bar{\bomega}_{j} - \bomega_{j}^*\big\|_1$;
	\item  \label{step2}
	$\frac{2  \sigma_{m}  \left\{ \sqrt{i_0} + \sqrt{\log \left(T\right)} + B\left(n, T\right) \right\}}{nT} \big\| \bar{\bDelta}_j - \bDelta_j^*\big\|_2 + \frac{4 \sigma_{m}  D}{nT} \big\| \left(\Ib_n \otimes \Cb \right) \bDelta_j^*\big\|_1 > \frac{1}{4}\beta \big\|\bar{\bomega}_{j} - \bomega_{j}^*\big\|_1$.
\end{enumerate}
Recall that $\zeta N = \| \bar{\bomega}_{j} - \bomega_{j}^*\|_1 + \|\bar{\bDelta}_j - \bDelta_j^*\|_2 / \sqrt{nT}$ and the goal is to obtain $\zeta N \le M /2$. 
To this end, we will derive upper bounds for $ \| \bar{\bomega}_{j} - \bomega_{j}^*\|_1$ and $ \|\bar{\bDelta}_j - \bDelta_j^*\|_2 / \sqrt{nT}$ separately.

\textbf{Case (i):} in this case, (\ref{a3.6}) can be simplified to
\begin{align}
\label{a3.7}
\frac{1}{2nT} \big\|\Yb_{j} \left (\bar{\bomega}_{j} - \bomega_{j}^* \right )\big\|_2^2 + \frac{1}{2nT} \big\|\bar{\bDelta}_j - \bDelta_j^*\big\|_2^2 + \frac{\beta}{4} \big\|\bar{\bomega}_{j}^{\cS_j^c}\big\|
\le \frac{7\beta}{4} \big\| \bar{\bomega}_{j}^{\cS_j} - (\bomega_{j}^*)^{\cS_j}\big\|_1.
\end{align} 
Since $\| \bar{\bomega}_{j} - \bomega_{j}^*\|_1 = \| \bar{\bomega}_{j}^{\cS_j} - (\bomega_{j}^*)^{\cS_j}\|_1 + \| \bar{\bomega}_{j}^{\cS_j^c}\|_1$, following an argument similar to Lemma 6.3 in \cite{buhlmann2011statistics}, we have
\begin{align}
\label{a3.8}
&\frac{2}{nT} \big\|\Yb_{j} \left (\bar{\bomega}_{j} - \bomega_{j}^* \right )\big\|_2^2 
+ 
\frac{2}{nT} \big\|\bar{\bDelta}_j - \bDelta_j^*\big\|_2^2 
+ 
\beta  \big\| \bar{\bomega}_{j} - \bomega_{j}^*\big\|_1 \nonumber \\
& =  \frac{2}{nT} \big\|\Yb_{j} \left (\bar{\bomega}_{j} - \bomega_{j}^* \right )\big\|_2^2 
+ 
\frac{2}{nT} \big\|\bar{\bDelta}_j - \bDelta_j^*\big\|_2^2 
+ \beta  \big\| \bar{\bomega}_{j}^{\cS_j} - (\bomega_{j}^*)^{\cS_j}\big\|_1 
+ \beta \big\| \bar{\bomega}_{j}^{\cS_j^c}\big\|_1 \nonumber \\
& \le 8 \beta \big\| \bar{\bomega}_{j}^{\cS_j} - (\bomega_{j}^*)^{\cS_j}\big\|_1 \nonumber \\
& \le \frac{8 \beta \sqrt{s_j}}{\phi_0 \sqrt{nT}} \big\|\Yb_{j} \left (\bar{\bomega}_{j} - \bomega_{j}^* \right )\big\|_2 \nonumber \\
& \le \frac{2}{nT} \big\|\Yb_{j} \left (\bar{\bomega}_{j} - \bomega_{j}^* \right )\big\|_2^2
+ 
\frac{8 \beta^2 s_j}{\phi_0^2},
\end{align}
where $\phi_0$ is a compatibility-type constant introduced in Assumption \ref{cconditions}. 
The first inequality follows from (\ref{a3.7}), the second inequality follows from Assumption \ref{cconditions}, and the last inequality follows from the fact that $uv\le u^2 + v^2/4$ for any $u,v\ge 0$.
Simplifying (\ref{a3.8}), we obtain 
$$\frac{2}{nT} \big\|\bar{\bDelta}_j - \bDelta_j^*\big\|_2^2 + \beta  \big\| \bar{\bomega}_{j} - \bomega_{j}^*\big\|_1
\le \frac{8 \beta^2 s_j}{\phi_0^2},$$
which directly implies  $\| \bar{\bomega}_{j} - \bomega_{j}^*\|_1 \le 8 \beta s_j / \phi_0^2$ and $ \|\bar{\bDelta}_j - \bDelta_j^*\|_2 / \sqrt{nT} \le 2 \beta \sqrt{s_j}  / \phi_0$.
Recall that $\beta = 2 \log(T) \log(nTp) n^{-1/6} T^{-1/3}$, and thus we have 
\begin{align}
\label{a3.10}
\zeta N = 
\big \| \bar{\bomega}_{j} - \bomega_{j}^*\big\|_1 + \frac{1}{\sqrt{nT}}\big\|\bar{\bDelta}_j - \bDelta_j^*\big\|_2 
\le c_1 n^{-\frac{1}{6}} T^{-\frac{1}{3}},
\end{align}
where $c_1 = 4 \log (T) \log (nTp) (4s_j + \sqrt{s_j}\phi_0)/ \phi_0^2$.
%Since $\zeta = M / (M + N)$, by (\ref{a3.10}), we obtain $N \le M$.

%%%%%%%%%%%%%%%%%%%%%%%%%%%%%%%%%%
%%%%%%%%%%%%%%%%%%%%%%%%%%%%%%%%%%
% Case Two
%%%%%%%%%%%%%%%%%%%%%%%%%%%%%%%%%%
%%%%%%%%%%%%%%%%%%%%%%%%%%%%%%%%%%
\textbf{Case (ii):} we first derive the upper bound of $\|\bar{\bDelta}_j - \bDelta_j^*\|_2 / \sqrt{nT}$.
From the condition of case (\ref{step2}), we have 
\begin{align}
\label{a3.101}
\frac{3\beta}{2} \big\|\bar{\bomega}_{j}^{\cS_j} - (\bomega_{j}^*)^{\cS_j}\big\|_1
<
\frac{12  \sigma_{m}  \left\{ \sqrt{i_0} + \sqrt{\log \left(T\right)} + B\left(n, T\right)\right\}}{nT} \big\| \bar{\bDelta}_j - \bDelta_j^*\big\|_2 + \frac{24 \sigma_{m}  D}{nT} \big\| \left(\Ib_n \otimes \Cb \right) \bDelta_j^*\big\|_1.
\end{align}
From (\ref{a3.6}), we obtain
\begin{align}
\label{a3.102}
\frac{1}{2nT} \big\|\bar{\bDelta}_j - \bDelta_j^*\big\|_2^2
\le & \frac{3\beta}{2} \big\| \bar{\bomega}_{j}^{\cS_j} - (\bomega_{j}^*)^{\cS_j}\big\|_1 +
\frac{2  \sigma_{m}  \left\{ \sqrt{i_0} + \sqrt{\log \left(T\right)}+ B\left(n, T\right)\right\}}{nT} \big\| \bar{\bDelta}_j - \bDelta_j^*\big\|_2 \nonumber\\
&+ \frac{4 \sigma_{m}  D}{nT} \big\| \left(\Ib_n \otimes \Cb \right) \bDelta_j^*\big\|_1.
\end{align}
Substituting (\ref{a3.101}) into (\ref{a3.102}), we have
\begin{align}
\label{a3.11}
\big \|\bar{\bDelta}_j - \bDelta_j^* \big \|_2^2  
 \le 28  \sigma_{m}  \left\{ \sqrt{i_0} + \sqrt{\log \left(T\right)}+ B\left(n, T\right)\right\} \big\| \bar{\bDelta}_j - \bDelta_j^*\big\|_2 + 56 \sigma_{m}  D \big\| \left(\Ib_n \otimes \Cb \right) \bDelta_j^*\big\|_1.
\end{align} 
Let $x = \|\bar{\bDelta}_j - \bDelta_j^*\|_2$, $b = 28 \sigma_{m} \{ \sqrt{i_0} + \sqrt{\log(T)}+ B(n, T)\}$ and $c = 56\sigma_{m} D \| (\Ib_n \otimes \Cb ) \bDelta_j^* \|_1$.
Then (\ref{a3.11}) can be rewritten as $x^2 - bx - c \le 0$. 
Since $x$ is bounded by the larger root of $x^2 - bx - c \le 0$, we have 
$$x \le \frac{b + \sqrt{b^2 + 4c}}{2} \le \frac{b + \sqrt{b^2} + \sqrt{4c}}{2} \le b + \sqrt{c}.$$
Thus, the upper bound for $\|\bar{\bDelta}_j - \bDelta_j^* \|_2 / \sqrt{nT}$ takes the form
\begin{align}
\label{a3.12}
\frac{1}{\sqrt{nT}} \big \|\bar{\bDelta}_j - \bDelta_j^*\big \|_2  
 \le \frac{ 28 \sigma_{m}  \left\{ \sqrt{i_0} + \sqrt{\log \left(T\right)}+ B\left(n, T\right)\right\}}{\sqrt{nT}} + \frac{ \sqrt{56\sigma_{m} D\big \| \left(\Ib_n \otimes \Cb \right) \bDelta_j^*\big \|_1}}{\sqrt{nT}}.
\end{align} 

Next, we derive the upper bound for $\|\bar{\bomega}_{j} - \bomega_{j}^*\|_1$ under case (\ref{step2}).
Recall $\sigma_{m} \ge 1$, we obtain
\begin{align}
\label{a3.13}
\big\|\bar{\bomega}_{j} - \bomega_{j}^* \big\|_1 
& < \frac{4}{\beta} \left [ \frac{2 \sigma_{m} \left\{ \sqrt{i_0} + \sqrt{\log \left(T\right)}+ B\left(n, T\right)\right\}}{nT}\big \| \bar{\bDelta}_j - \bDelta_j^*\big\|_2 + \frac{4 \sigma_{m} D}{nT} \big\| (\Ib_n \otimes \Cb) \bDelta_j^*\big\|_1\right ] \nonumber\\
& \le \frac{224 \sigma_{m}^2}{\beta nT} \left \{ \sqrt{i_0} + \sqrt{\log \left(T\right)} + B\left(n, T\right) + \sqrt{D\big\| (\Ib_n \otimes \Cb) \bDelta_j^*\big \|_1 }\right\}^2, 
\end{align} 
where the first inequality follows the assumption of case (\ref{step2}) and the last inequality follows from (\ref{a3.12}).
Let $\Delta_{\max}=\max_{i,t,j}~|\Delta_{itj}^* - \Delta_{i(t-1)j}^*| + 1$ and assume that $\bDelta_{ij}^*$ are piecewise constants with at most $\tau$ different  constants across the $T$ replicates for each subject.
Thus, $\| (\Ib_n \otimes \Cb) \bDelta_j^*\|_1 \le \Delta_{\max} \tau n$.
Combining (\ref{a3.12}) and (\ref{a3.13}), we have
\begin{align}
\label{a3.14}
\zeta N  
&=\big\|\bar{\bomega}_{j} - \bomega_{j}^*\big\|_1  + \frac{1}{\sqrt{nT}} \big\|\bar{\bDelta}_j - \bDelta_j^*\big\|_2 \nonumber \\
& \le \frac{224 \sigma_{m}^2}{\beta nT} \left \{ \sqrt{i_0} + \sqrt{\log \left(T\right)}  + B\left(n, T\right) + \sqrt{D \Delta_{\max} \tau n} \right \}^2  \nonumber \\
&+ \frac{112 \sigma_{m}}{\sqrt{nT}} \left \{ \sqrt{i_0} + \sqrt{\log \left(T\right)}  + B\left(n, T\right) + \sqrt{D \Delta_{\max} \tau n} \right \},
\end{align} 
where $D = 8\sqrt{T\log(T) / (\pi^2 i_0)}$. 
Since (\ref{a3.14}) holds for any value of $i_0$, next, we identify $i_0$ such that the upper bound $\zeta N$ at (\ref{a3.14}) is tight. 
For notational convenience, let $h = i_0^{1/4}$ and $z(h) = h^2 + 2 c_4 T^{1/4} n^{1/2} h^{-1} + \sqrt{\log (T)}+ B(n, T)$, where $c_4 = [ 4\log (T) \Delta_{\max}^2 \tau^2 / \pi^2]^{1/4}$. 
Then, (\ref{a3.14}) can be rewritten as
\begin{align*}
%\label{a3.15}
\zeta N \le f(z) = \frac{224 \sigma_{m}^2}{\beta nT} z^2 + \frac{112\sigma_{m}}{\sqrt{nT}}z.
\end{align*} 
The fact that $f^{'}(z) = 448 \sigma_{m}^2 z / (\beta nT) + 112\sigma_{m} / \sqrt{nT}> 0$ implies that $f(z)$ is an increasing function of $z$, and thus it suffices to find the of $h$ such that the value of $z(h)$ is minimized.
Since $z^{''} (h) = 2 + 4c_4 T^{1/4} n^{1/2} h^{-3} > 0$, $z$ is a strictly convex function of $h$. 
It can be shown that the minimum of $z(h)$ is achieved when $ h = (c_4 T^{1/4} n^{1/2})^{1/3}$.
Since $i_0 \in \{1, 2, \ldots, T-1\}$, we need to carefully select the value of $i_0$ on its range. 
When $T > c_4^2 n$, we have $i_0 = \max \{1, \lfloor (c_4 T^{1/4} n^{1/2})^{4/3} \rfloor \} \le T-1$.  
Thus, it can be shown that 
\begin{equation}
\label{a3.151}
 \sqrt{i_0} + \sqrt{\log \left(T\right)}  + B\left(n, T\right) + \sqrt{D \Delta_{\max} \tau n}  \le 
 4 c_4^{\frac{2}{3}}n^{\frac{1}{3}}T^{\frac{1}{6}} + 4\sqrt{\log(T)} + 4 B\left(n, T\right).
\end{equation}
Recall that $B(n,T) = \mu_m \min(c_4^{2/3} n^{1/3} T^{1/6}, \sqrt{T})$. 
Then the upper bound for $\zeta N$ at (\ref{a3.14}) is
\begin{equation}
\label{a3.152}
\min f\left(z\right) \le
\frac{224 \sigma_m^2}{\beta nT} \left\{  4\left(\mu_m + 1\right) c_4^{\frac{2}{3}}n^{\frac{1}{3}}T^{\frac{1}{6}} + 4\sqrt{\log(T)}\right\}^2  + \frac{112 \sigma_m}{\sqrt{nT}} \left\{4\left(\mu_m + 1\right) c_4^{\frac{2}{3}}n^{\frac{1}{3}}T^{\frac{1}{6}} + 4\sqrt{\log(T)}\right\}.
\end{equation}

%we impose the following scaling condition \textcolor{red}{This choice of $i_0$ gives bad scaling condition. We can pick something like $T^{1/3}n^{1/3}$ to get $T^2>n $ type condition at a cost of our rate of convergence}.
%\textcolor{red}{
%$$\frac{T}{n} > \frac{\Delta_{\max} \tau \sqrt{\log \left\{ 2 \left(T-1\right) \right\}}}{\pi}.$$}

%Besides, $z(i_0^*) \le z\{2 (c_3 T^{1/4} n^{1/2} /2)^{4/3} \} \le 4 c_3^{2/3} T^{1/6} n^{1/3} + \sqrt{\log (T)}$. 
Recall that $\beta = 2 \log (T) \log (nTp) n^{-1/6} T^{-1/3}$,
then (\ref{a3.152}) can be written as
\begin{align}
\label{a3.16}
\min f\left(z\right) 
\le c_2 n^{-\frac{1}{6}} T^{-\frac{1}{3}} + c_3 n^{-\frac{1}{2}} T^{-\frac{1}{2}},
%& \le \frac{224 \sigma_{m}^2}{\beta nT} \left \{ 4 c_3^{\frac{2}{3}} T^{\frac{1}{6}} n^{\frac{1}{3}} + \sqrt{\log \left(T\right)}\right\}^2 + \frac{112\sigma_{m}}{\sqrt{nT}}\left \{ 4 c_3^{\frac{2}{3}} T^{\frac{1}{6}} n^{\frac{1}{3}} + \sqrt{\log \left(T\right)}\right \} \nonumber \\
%& \le 14336 \sigma_m^2 \left \{c_3^{\frac{4}{3}}c_4 n^{-\frac{1}{2}} T^{-\frac{1}{6}}
%+ c_3^{\frac{2}{3}}\left (c_4 + 1\right) n^{-\frac{1}{2}} T^{-\frac{1}{3}}
%+ \left (c_4 + 1 \right) n^{-\frac{1}{2}} T^{-\frac{1}{2}}
%\right \} \nonumber\\
%& \le c_2 \left(\frac{n}{T}\right)^{\frac{1}{6}},
\end{align} 
where $c_2 = 1792 \sigma_m^2 (\mu_m + 1)^2 c_4^{4/3} / \{\log(T) \log(nTp)\} + 448 \sigma_m (\mu_m + 1) c_4^{2/3}$ and $c_3 = 448 \sigma_m \sqrt{\log(T)} + 1792 \sigma_m^2 \{\log(T) + 2 (\mu_m + 1) c_4^{2/3} \sqrt{\log(T)}\} / \{\log(T) \log(nTp)\}$.
%$c_2 = 112 \sigma_{m}^2 [ \{4c_3^{2/3} + \sqrt{\log (T)} \}^2/ \{\log (T) \log (nTp)\} + \{4c_3^{2/3} + \sqrt{\log (T)}\}]$.
%$c_4 = 224 \sigma_{m}^2 [ \{2 + 2 \sqrt{\Delta_{\max} \tau \log^{1/2} (T)}\}^2/ \{\log (T) \log (nTp)\} + \{2 + 2 \sqrt{\Delta_{\max} \tau \log^{1/2} (T)}\}]$. 
%$\frac{1}{\beta} = c_4 n^{\frac{1}{2}} T^{\frac{1}{2}}$.

Recall the definition of $M$. Combining the upper bound for $\zeta N$ in (\ref{a3.10}) and (\ref{a3.16}) yields
\begin{align}
\label{a3.161}
\zeta N  
\le \max \left(c_1 n^{-\frac{1}{6}} T^{-\frac{1}{3}}, c_2 n^{-\frac{1}{6}} T^{-\frac{1}{3}} + c_3 n^{-\frac{1}{2}} T^{-\frac{1}{2}}\right) 
\le \frac{M}{2}.
\end{align}
Since $\zeta = M / (M + N)$, by (\ref{a3.161}), we obtain $N \le M$.
Recall that $N = \| \hat{\btheta}_{j,-j} - \btheta_{j,-j}^*\|_1 + \| \hat{\balpha}_{j} - \balpha_{j}^*  \|_1 +  \|\hat{\bDelta}_j - \bDelta_j^* \|_2 / \sqrt{nT} $, then 
\begin{align*}
%\label{a3.17}
\big\| \hat{\btheta}_{j,-j} - \btheta_{j,-j}^*\big\|_1 + \big\| \hat{\balpha}_{j} - \balpha_{j}^* \big \|_1 + \frac{1}{\sqrt{nT}} \big \|\hat{\bDelta}_j - \bDelta_j^* \big \|_2 
\le M
\le 2\max \left(c_1 n^{-\frac{1}{6}} T^{-\frac{1}{3}}, c_2 n^{-\frac{1}{6}} T^{-\frac{1}{3}} + c_3 n^{-\frac{1}{2}} T^{-\frac{1}{2}}\right),
\end{align*} 
with probability at least $1 - \{2 (T-1) \sqrt{\log (T-1)}\}^{-1} - 2 \exp(-\lfloor [\log \{2(T-1)\} \Delta_{\max}^2 \tau^2 T n^2/ \pi^2]^{1/3}\rfloor) - 1 / \{T \sqrt{2\log{(T)}}\}  - 4/ (nTp)  - 2\exp ( - \min [ \log^2 (T) / \{2 \sigma_{m}^{\epsilon}\sqrt{\mu_m^2 +(\sigma_{m}^{X})^2}\}^2, \log(T) / |2 \sigma_{m}^{\epsilon}\sqrt{\mu_m^2 +(\sigma_{m}^{X})^2}|] / 2)$, as desired.

%%%%%%%%%%%%%%%%%%%%%%%%%%%%%%%%%%%%
%%%%%%%%%%%%%%%%%%%%%%%%%%%%%%%%%%%%
% Proof of Theorem 2
%%%%%%%%%%%%%%%%%%%%%%%%%%%%%%%%%%%%
%%%%%%%%%%%%%%%%%%%%%%%%%%%%%%%%%%%%
\subsection{Proof of Theorem~\ref{th2}}
\label{appendix:proof of theorem2}
The proof of Theorem~\ref{th2} is similar to the proof of Theorem~\ref{th1}. 
In particular, let
\[
N = \big \|\hat{\btheta}_{j, -j} - \btheta_{j, -j}^* \big \|_1 + \big \|\hat{\balpha}_{j} - \balpha_{j}^*\big \|_1 + \frac{1}{\sqrt{nT}} \big \|\hat{\bDelta}_j - \bDelta_j^*\big \|_2.
\]
The goal is to show that $N\le M^{'}$, where
\begin{equation*}
\begin{split}
M^{'} &= 2 \max \left(c_1^{'}, c_2^{'}\right) T^{-\frac{1}{2}};\\
c_1^{'} &=  \frac{2 \log \left(T\right) \log \left(nTp\right) \left(8 s_j + 2\sqrt{s_j}\phi_0\right)}{\phi_0^2};\\
c_2^{'} &= \frac{1792 \sigma_{m}^2 c_3^{'}  \left(\mu_m + 3\right)^2 }{\log \left(T\right) \log \left(nTp\right)} + 224 \sigma_{m} \left(c_3^{'}\right)^{\frac{1}{2}} \left(\mu_m + 4\right);\\
c_3^{'} &=  \Delta_{\max} \tau \log^{\frac{1}{2}} \left(T\right) .
\end{split}
\end{equation*}
Similar to the proof of Theorem~\ref{th1}, we require $M^{'} \le 1$.  Thus, we assume the condition 
$T^{\frac{1}{2}} \ge 2 \max \left(c_1^{'}, c_2^{'}\right).$

The proof is similar to that of Theorem~\ref{th1} with the main difference being the choice of  $\beta$, $\lambda$, and $i_0$. 
First, we choose $\beta = \lambda = 2 \log(T) \log(nTp) T^{-1/2}$ to obtain the optimal upper bound of $\zeta N$, then (\ref{a3.10}) will reduce to
\begin{equation}
\label{a3.201}
\zeta N \le c_1^{'} T ^{-\frac{1}{2}}, 
\end{equation}
where $c_1^{'} = 2 \log(T) \log(nTp) (8 s_j + 2 \sqrt{s_j} \phi_0) / \phi_0^2$. 
Besides, under the condition $T \le 2\log^{1/2}(T) \Delta_{\max} \tau n / \pi$, we choose $i_0 = T-1$, then (\ref{a3.151}) can be rewritten as
\begin{equation}
\label{a3.251}
 \sqrt{i_0} + \sqrt{\log \left(T\right)}  + B\left(n, T\right) + \sqrt{D \Delta_{\max} \tau n}  \le 
 \left(\mu_m + 2 \right) \sqrt{T} + 4 \sqrt{\Delta_{\max} \tau n \log^{\frac{1}{2}} \left(T\right)}.
\end{equation}
Thus,  the upper bound for $\zeta N$ in (\ref{a3.14}) will be 
\begin{equation}
\label{a3.252}
\zeta N \le c_2^{'} T^{-\frac{1}{2}},
\end{equation}
where $c_2^{'} = 1792 \sigma_m^2 c_3^{'} (\mu_m + 3)^2  / \log(T) \log(nTp) + 224 \sigma_m(c_3^{'})^{1/2}  (\mu_m + 4) $ and $c_3^{'} = \Delta_{\max} \tau \log^{1/2} (T)$. 

Combining (\ref{a3.201}) and (\ref{a3.252}), the upper bound for $\zeta N$ is 
\begin{equation}
\label{a3.253}
\zeta N \le \max(c_1^{'}, c_2^{'}) T^{-\frac{1}{2}} \le \frac{M^{'}}{2}.
\end{equation}
Recall the definition of $\zeta$ and $N$, we can obtain that 
\begin{equation}
\label{a3.254}
 \big \|\hat{\btheta}_{j, -j} - \btheta_{j, -j}^* \big \|_1 + \big \|\hat{\balpha}_{j} - \balpha_{j}^*\big \|_1 + \frac{1}{\sqrt{nT}} \big \|\hat{\bDelta}_j - \bDelta_j^*\big \|_2 \le 2 \max\left(c_1^{'}, c_2^{'}\right) T^{-\frac{1}{2}},
\end{equation}
with probability at least $1 - \{2 (T-1) \sqrt{\log (T-1)}\}^{-1} - 2 \exp\{-(T-1)\} - 1 / \{T \sqrt{2\log{(T)}}\}  - 4/ (nTp)  - 2\exp ( - \min [ \log^2 (T) / \{2 \sigma_{m}^{\epsilon}\sqrt{\mu_m^2 +(\sigma_{m}^{X})^2}\}^2, \log(T) / |2 \sigma_{m}^{\epsilon}\sqrt{\mu_m^2 +(\sigma_{m}^{X})^2}|] / 2)$. 
%%%%%%%%%%%%%%%%%%%%%%%%%%%%%%%%%%%%
%%%%%%%%%%%%%%%%%%%%%%%%%%%%%%%%%%%%
% Proof of Technical Lemmas
%%%%%%%%%%%%%%%%%%%%%%%%%%%%%%%%%%%%
%%%%%%%%%%%%%%%%%%%%%%%%%%%%%%%%%%%%
\section{Proof of Technical Lemmas}
\label{asub: proof of lemmas}

%%%%%%%%%%%%%%%%%%%%%%%%%%%%%%%%%%%%
%%%%%%%%%%%%%%%%%%%%%%%%%%%%%%%%%%%%
% Proof of Lemma 1
%%%%%%%%%%%%%%%%%%%%%%%%%%%%%%%%%%%%
%%%%%%%%%%%%%%%%%%%%%%%%%%%%%%%%%%%%
\subsection{Proof of Lemma~\ref{a1}}
\label{asub: proof of lemma 1}
This proof is similar to the proof of Lemma 6 in \cite{hall2016inference}. 
Recall that for $k \neq j$, $X_{itk}\sim N(\mu_{itk}, \sigma_{kk, t}^2)$ with $(\sigma_{m}^X)^2 = \max_{k, t} (\sigma_{kk, t}^2)$, and $\epsilon_{itj} \sim N\{ 0, (\sigma_{jj, t}^{\epsilon})^2 \}$ with $(\sigma_{m}^{\epsilon})^2 = \max_{t,j} \{ (\sigma_{jj,t}^{\epsilon})^2 \}$.
Let $\bepsilon_j = (\epsilon_{11j}, \epsilon_{12j}, \ldots, \epsilon_{1Tj}, \epsilon_{21j}, \ldots, \epsilon_{nTj})^{\T}$ and let $\bX_k = (X_{11k}, X_{12k}, \ldots, X_{1Tk}, X_{21k}, \ldots, X_{nTk})^{\T}$.
For simplicity, we rewrite $\bepsilon_j$ and $\bX_k$ as $\bepsilon_j = (\epsilon^{'}_{1j}, \epsilon^{'}_{2j}, \ldots, \epsilon^{'}_{(nT)j})^{\T}$ and $\bX_k = (X^{'}_{1k}, X^{'}_{2k}, \ldots, X^{'}_{(nT)k})^{\T}$, where $\epsilon^{'}_{lj} = \epsilon_{itj}$ and $X^{'}_{lk} = X_{itk}$ with $l = (i - 1)T + t$. 
Then, we have 
\begin{align}
\label{a3.330}
\underset{1 \le k \le p, k \ne j}{\max}~ \frac{1}{nT} \left |\sum_{i = 1}^{n} \sum_{t = 1}^{T} \epsilon_{itj} X_{itk} \right| 
=
\underset{1 \le k \le p, k \ne j}{\max} \frac{1}{nT} \left|\sum_{l = 1}^{nT} \epsilon_{lj}^{'} X_{lk}^{'} \right | .
\end{align}
In this proof, our goal is to bound (\ref{a3.330}) by Lemma~\ref{ao6}. 
First, we define notation $Z_m$, $G_m^k$ and $R_m$ needed by Lemma~\ref{ao6}. 
Denote the sequence $Z_m$ as 
$$Z_m = \frac{1}{nT} \sum_{l = 1}^{m} \epsilon_{lj}^{'} X_{lk}^{'}.$$
%Then $Z_m - Z_{m-1}= \epsilon_{mj}^{'} X_{mk}^{'} / (nT)$.
Then we have 
\begin{align*}
\EE \left (Z_m - Z_{m-1} \big| \epsilon_{1j}^{'}, \ldots, \epsilon_{(m-1)j}^{'},  X_{1k}^{'}, \ldots, X_{(m-1)k}^{'} \right)
&= \frac{1}{nT} \EE \left (\epsilon_{mj}^{'} X_{mk}^{'} \big| \epsilon_{1j}^{'}, \ldots, \epsilon_{(m-1)j}^{'},  X_{1k}^{'}, \ldots, X_{(m-1)k}^{'} \right) \\
&=\frac{1}{nT}  \EE \left (\epsilon_{mj}^{'} \right) \EE \left (X_{mk}^{'} \big| \epsilon_{1j}^{'}, \ldots, \epsilon_{(m-1)j}^{'},  X_{1k}^{'}, \ldots, X_{(m-1)k}^{'} \right)\\
&= 0,
\end{align*}
where the second equality holds since $\epsilon_{mj}^{'}$ is independent with $\bX_{k}$ for $j \neq k$ and $\epsilon_{mj}^{'}$ is independent with $\epsilon_{lj}^{'}$, for $l = 1, \ldots, m-1$. 
Thus, we conclude that $Z_m$ is a martingale. 
Recall that $X_{mk}^{'}  \sim N (\mu_{mk}^{'}, \sigma_{kk,m}^2)$.
Let $|r| \le 1/|2 \sigma_m^{\epsilon} \sqrt{\mu_m^2 + (\sigma_m^X)^2}|$, then by smoothing we have 
\begin{align}
\label{a3.3304}
\EE \left(e^{r \epsilon_{mj}^{'} X_{mk}^{'}}\right)
&= \EE \left\{ \EE \left(e^{r \epsilon_{mj}^{'} X_{mk}^{'}} \big | \epsilon_{mj}^{'}\right)\right\} \nonumber\\ 
&= \EE \left\{e^{r \mu_{mk}^{'} \epsilon_{mj}^{'} + \frac{1}{2}\left(r \sigma_{kk,m} \epsilon_{mj}^{'}\right)^2}\right\} \nonumber\\ 
&\le \sqrt{\EE \left(e^{2 r \mu_{mk}^{'} \epsilon_{mj}^{'} }\right)}
\sqrt{\EE \left\{e^{\left(r \sigma_{kk, m} \epsilon_{mj}^{'}\right)^2}\right\}} \nonumber\\ 
&= \frac{\exp\left\{2 \left(r \mu_{mk}^{'} \sigma_{jj, m}^{\epsilon}\right)^2\right\}}{\sqrt{1 - 2 \left(r \sigma_{jj, m}^{\epsilon}\sigma_{kk, m}\right)^2}}\nonumber\\ 
&\le e^{\frac{1}{2} \left\{2 r  \sigma_{m}^{\epsilon}\sqrt{\mu_m^2 +\left(\sigma_{m}^{X}\right)^2}\right\}^2},
\end{align} 
where the second equality holds with $\epsilon_{mj}^{'}$ is independent with $X_{mk}^{'}$ and the third equality holds with $\epsilon_{mj}^{'} \sim N \{0, (\sigma_{jj,m}^{\epsilon})^2\}$ and $(\epsilon_{mj}^{'}  / \sigma_{jj, m}^{\epsilon})^2 \sim \chi_1^2$.
Therefore, we obtain that $\epsilon_{mj}^{'} X_{mk}^{'}$ follows sub-exponential with parameter $|2 \sigma_{m}^{\epsilon}\sqrt{\mu_m^2 +(\sigma_{m}^{X})^2}|$, denoted as $\epsilon_{mj}^{'} X_{mk}^{'} \sim \mathrm{subE} (|2 \sigma_{m}^{\epsilon}\sqrt{\mu_m^2 +(\sigma_{m}^{X})^2}|)$. 
By Lemma~\ref{ao1}, we have
\begin{align}
\label{a3.3305}
\PP \left \{ \big |\epsilon_{mj}^{'} X_{mk}^{'} \big | \ge \log (T) \right \} 
\le \exp \left( - \frac{1}{2} \min \left[ \frac{\log^2\left(T\right)}{\left\{2   \sigma_{m}^{\epsilon}\sqrt{\mu_m^2 +\left(\sigma_{m}^{X}\right)^2}\right\}^2}, \frac{\log\left(T\right)}{\bigg |2 \sigma_{m}^{\epsilon}\sqrt{\mu_m^2 +\left(\sigma_{m}^{X}\right)^2}\bigg|}\right]\right).
\end{align}

Let $B = \log (T) / (nT)$ and define sequence $G_m^k$ as 
\begin{align}
\label{a3.331}
G_m^k = \sum_{l = 1}^m \EE \left \{ \left( \frac{1}{nT} \epsilon_{lj}^{'} X_{lk}^{'} \right)^k \big| \epsilon_{1j}^{'}, \ldots, \epsilon_{(l-1)j}^{'},  X_{1k}^{'}, \ldots, X_{(l-1)k}^{'} \right \} 
\le \big | G_m^k \big|
\le m B^{k},
\end{align}
with probability at least $1 - \exp ( - \min [ \log^2 (T) / \{2 \sigma_{m}^{\epsilon}\sqrt{\mu_m^2 +(\sigma_{m}^{X})^2}\}^2, \log(T) / |2 \sigma_{m}^{\epsilon}\sqrt{\mu_m^2 +(\sigma_{m}^{X})^2}|] / 2)$. The inequality in (\ref{a3.331}) follows (\ref{a3.3305}).
Then for $\rho >  0$, let 
$$R_m = \sum_{k = 2}^{\infty} \frac{\rho^k G_m^k}{k!} ; 
\quad
R_m^{'} = \sum_{k = 2}^{\infty} \frac{\left( -1 \right)^k \rho^k G_m^k}{k!}  ;
\quad
R_m^{''} = m \left(e^{\rho B} - 1 - \rho B \right). 
$$
Besides, we have
\begin{align}
\label{a3.3315}
R_m^{'} ~\mathrm{and}~R_m 
\le m \sum_{k = 2}^{\infty} \frac{\left(\rho B\right)^k}{k!} 
= R_m^{''},
\end{align}
where the second inequality follows (\ref{a3.331}).
The upper bound of $|Z_m|$ is
\begin{align}
\label{a3.332}
\PP (|Z_m| \ge z) & = \PP(Z_m \ge z) + \PP( - Z_m \ge z) \nonumber\\
& \le \EE \left( e^{\rho Z_m} \right)e^{- \rho z} + \EE \left( e^{ - \rho Z_m} \right)e^{- \rho z} \nonumber\\
& = \EE \left( e^{\rho Z_m - R_m + R_m} \right)e^{- \rho z} + \EE \left( e^{ - \rho Z_m - R_m^{'} + R_m^{'}} \right)e^{- \rho z} \nonumber\\
& \le \EE \left( e^{\rho Z_m - R_m} \right)e^{R_m^{''} - \rho z} + \EE \left( e^{ - \rho Z_m - R_m^{'}} \right)e^{R_m^{''} - \rho z} \nonumber\\
& \le 2 e^{R_m^{''} - \rho z},
\end{align}
where the first inequality follows Markov's inequality, the second inequality follows (\ref{a3.3315}) and the last inequality follows Lemma~\ref{ao6}. 
The next step of this proof is to find the value of $\rho$ to minimize the right hand side of (\ref{a3.332}). 
Recall the $R_m^{''} = m(e^{\rho B} - 1 - \rho B)$. Denote the right hand side of (\ref{a3.332}) as 
$$f\left (\rho \right ) = 2 e^{R_m^{''} - \rho z} = 2 \exp \left \{ m \left(e^{\rho B} - 1 - \rho B\right) - \rho z \right\}.$$
Since $f(\rho)$ is strictly convex, $f(\rho)$ obtains its minimizer at the root of $f^{'}(\rho) = 0$, which is $\rho^{*} =  \log\{ z / (mB) + 1\} / B$. 
Then (\ref{a3.332}) will be
\begin{align*}
\PP \left ( \big |Z_m \big | \ge z \right) 
\le f \left (\rho^{*} \right)
= 2 \exp \left \{ - m g \left(\frac{z}{mB}\right) \right \},
\end{align*}
where $g(x) = (1+x) \log (1+x) - x$. 
Since $g(x) \ge 3 x^2 / \{2 (x+3)\}$ for $x \ge 0$,  we have
\begin{align}
\label{a3.3325}
\PP \left( \big |Z_m \big | \ge z\right) 
\le 2 \exp \left( - \frac{3 z^2}{ 2 z B + 6 m B^2}\right)
= 2 \exp \left \{ - \frac{3 z^2 n^2 T^2}{ 2 z n T \log \left(T\right) + 6 m \log^2 \left (T\right)}\right \},
\end{align}
where the equality follows the fact that $B = \log (T) / (nT)$. 
Let $m = nT$ and $z = \lambda_0$. 
By (\ref{a3.3325}) we have
\begin{align}
\label{a3.3326}
\PP \left[  \underset{1 \le k \le p, k \ne j}{\max} \big |\frac{1}{nT}\sum_{l = 1}^{nT}  \epsilon_{lj}^{'} X_{lk}^{'} \big | 
\ge \lambda_0 \right] 
\le 2 \left(p-1\right)\exp \left \{ - \frac{3 \lambda_0^2 n T}{ 2 \lambda_0 \log \left(T\right) + 6 \log^2 \left (T\right)}\right \}.
\end{align}
Since (\ref{a3.3326}) happens when (\ref{a3.3305}) holds, we have 
$$\underset{1 \le k \le p, k \ne j}{\max} \big |\frac{1}{nT}\sum_{l = 1}^{nT}  \epsilon_{lj}^{'} X_{lk}^{'} \big | < \lambda_0,$$
with probability at least $1 - \exp ( - \min [ \log^2 (T) / \{2 \sigma_{m}^{\epsilon}\sqrt{\mu_m^2 +(\sigma_{m}^{X})^2}\}^2, \log(T) / |2 \sigma_{m}^{\epsilon}\sqrt{\mu_m^2 +(\sigma_{m}^{X})^2}|] / 2) - \exp [\log\{2(p-1)\} - 3 \lambda_0^2 n T /  \{2 \lambda_0 \log (T) + 6 \log^2  (T)\}]$.
%%%%%%%%%%%%%%%%%%%%%%%%%%%%%%%%%%%%
%%%%%%%%%%%%%%%%%%%%%%%%%%%%%%%%%%%%
% Proof of Lemma 3
%%%%%%%%%%%%%%%%%%%%%%%%%%%%%%%%%%%%
%%%%%%%%%%%%%%%%%%%%%%%%%%%%%%%%%%%%
\subsection{Proof of Lemma \ref{a2}}
\label{asub: proof of lemma 3}
%This proof is similar to the proof of Lemma 5 and Theorem 6 in \cite{wang2016trend}.
Let $\boldsymbol{\eta} \sim N(\boldsymbol{0}, \bQ)$. 
The goal is to obtain an upper bound for $\boldsymbol{\eta}^{\T}(\bar{\bDelta}_{ij} - \bDelta_{ij})/(nT)$. 
Recall that $\Cb$ is the discrete first derivative matrix, 
$$
\Cb = 
\left(
\begin{matrix}
 -1      & 1    &0  & \cdots & 0   &0  \\
 0       & -1   &1   & \cdots & 0  &0   \\
 \vdots                                             \\
 0      & 0     &0     & \cdots & -1 &1      \\
\end{matrix}
\right).
$$
Let $S_c(1) = \{\bW \in \mathrm{row}(\Cb): \|\Cb \bW\|_1 \le 1\}$, where $\mathrm{row}(\Cb)$ is the row space of $\Cb$. 
The steps in this proof are: 
\begin{enumerate}[(i)]
\item get the upper bound of $\boldsymbol{\eta}^{\T} \bW$ for $\forall \bW \in S_c(1)$.
\item substitute $\bW$ with a specific value related to $\bar{\bDelta}_{ij}$ and $\bDelta_{ij}$. 
\end{enumerate}

Step (i): first, we would like to introduce some new notation. 
The singular value decomposition of $\Cb$ is 
$$\Cb = \Ub \bXi \Vb^{\T},$$ 
where both $\Ub \in \RR^{(T-1) \times (T-1)}$ and $\Vb \in \RR^{T \times (T-1)}$ are orthogonal matrixes and $\bXi \in \RR^{(T-1) \times (T-1)}$ is a diagonal matrix with diagonal $\xi_i$, $i \in \{1, 2, \ldots, T-1\}$.
%The diagonal of $\bSigma$ are the singular values of $\Cb$, which are denoted as $\xi_i = 2 | \sin\{\pi (i-1)/(2T) \}  |$, $i \in \{1, 2, \ldots, T-1\}$. 
Then the pseudoinverse of $\Cb$ is
$$\Cb^{+} = \Vb \bXi^{-1} \Ub^{\T}.$$ 
%Let $\Pb_{E} = \Cb^+ \Cb$ be the projection onto $\Eb$, and for $\forall \bW \in S_c(1)$, $\exists~\mathrm{vector}~\bL$ that $\bW = \Pb_{E} \bL$.  
For $i_0 \in \{1, \ldots, T-1\}$, let $[i_0] = \{ 1, \ldots, i_0 \}$ and $\Pb_{[i_0]} = \Vb_{[i_0]}\Vb_{[i_0]}^{\T}$, where $\Vb_{[i_0]}$ is a matrix containing the first $i_0$ columns of $\Vb$.
Then $\boldsymbol{\eta}^{\T} \bW$ can be written as
\begin{align}
\label{a3.337}
\boldsymbol{\eta}^{\T} \bW 
%= \boldsymbol{\eta}^{\T} \Pb_{E}\bL 
= \underbrace{\boldsymbol{\eta}^{\T} \Pb_{[i_0]} \bW}_{\mathbb{I}_1}
+ \underbrace{\boldsymbol{\eta}^{\T} \left(\Ib - \Pb_{[i_0]}\right) \bW}_{\mathbb{I}_2}.
\end{align}
To bound the term $\boldsymbol{\eta}^{\T} \bW$, we would consider $\mathbb{I}_1$ and $\mathbb{I}_2$ separately. 
Upper Bound for $\mathbb{I}_1$ in (\ref{a3.337}): by Holder's inequality, we have 
\begin{align}
\label{a3.31}
\mathbb{I}_1 
\le  \big \| \Pb_{[i_0]}  \boldsymbol{\eta} \big \|_2 \cdot \big\| \bW  \big\|_2
\overset{d}{=} \sqrt{\sum_{i = 1}^{i_0} \eta_i^2}  \big\| \bW  \big\|_2.
\end{align} 
Now, we would like to further bound the term $\sum_{i = 1}^{i_0} \eta_i^2$ in (\ref{a3.31}). 
Recall $\boldsymbol{\eta} \sim N(\boldsymbol{0}, \Qb)$. 
Let $\boldsymbol{\eta}_{[i_0]} = (\eta_1, \eta_2, \ldots, \eta_{i_0})^{\T}$ and $\Qb_{[i_0]} = \mathrm{Cov} (\boldsymbol{\eta}_{[i_0]})$. 
Then $\sum_{i = 1}^{i_0} \eta_i^2$ can be written as
 $$\sum_{i = 1}^{i_0} \eta_i^2 = \boldsymbol{\eta}_{[i_0]}^{\T}\boldsymbol{\eta}_{[i_0]} = \Zb^{\T} \Qb_{[i_0]} \Zb,$$
where $\Zb \sim N (\boldsymbol{0}, \Ib)$. 
By Lemma~\ref{ao5}, we have 
%Following the fact that $\sum_{i = 1}^{i_0} \eta_i^2 = \boldsymbol{\eta}_{[i_0]}^{\T}\boldsymbol{\eta}_{[i_0]} = \Zb^{\T} \Qb_{[i_0]} \Zb$ with $\Zb \sim N (\boldsymbol{0}, \Ib)$, 
\begin{align}
\label{a3.32}
\PP \left \{ \sum_{i = 1}^{i_0} \eta_i^2  - \sum_{i = 1}^{i_0} \EE\left(\eta_i^2\right) > i_0 \nu \right \}
& \le \PP \left ( \big | \Zb^{\T} \Qb_{[i_0]} \Zb  - \EE\Zb^{\T} \Qb_{[i_0]} \Zb \big | > i_0 \nu \right ) \nonumber\\
& \le 2 \exp \left \{ -  \min \left ( \frac{i_0^2 \nu^2}{\big \|\Qb_{[i_0]}\big\|_\mathrm{F}^2}, \frac{i_0 \nu}{\big\|\Qb_{[i_0]}\big\|_{\mathrm{op}}}\right) \right \} ,
\end{align}
where $\|\Qb\|_{F}$ is the Frobenius norm and $\| \Qb \|_{\mathrm{op}}$ is the operator norm.
Set $\nu =  \|\Qb_{[i_0]}\|_{\mathrm{op}}$. 
Since $\sum_{i = 1}^{i_0} \EE(\eta_i^2) = \mathrm{tr}(\Qb_{[i_0]}) \le i_0 \|\Qb_{[i_0]}\|_{\mathrm{op}}$ and $\|\Qb_{[i_0]}\|_\mathrm{F}^2 \le i_0 \|\Qb_{[i_0]}\|^2_{\mathrm{op}}$, (\ref{a3.32}) will be
\begin{align}
\label{a3.33}
\PP \left ( \sum_{i = 1}^{i_0} \eta_i^2 \ge 2 i_0  \big\|\Qb_{[i_0]}\big\|_{\mathrm{op}} \right )
\le 2 \exp \left ( - i_0 \right).
\end{align}
Substituting (\ref{a3.33}) into (\ref{a3.31}), we have 
\begin{align}
\label{a3.20}
\mathbb{I}_1   
\le 
%\sqrt{2i_0\big \|\Qb_{[i_0]}\big \| } \big \| \Pb_{E}\Lb \big \|_2
\sqrt{2i_0\big \|\Qb_{[i_0]}\big \|_{\mathrm{op}}} \big \| \bW \big \|_2,
\end{align}
with probability at least $1 - 2 \exp ( - i_0 )$.
 
Upper Bound for $\mathbb{I}_2$ in (\ref{a3.337}): 
Recall that $\mathrm{row}(\Cb)$ is the row space of $\Cb$. 
Let $\Pb_{\mathrm{row}(\Cb)} = \Cb^+ \Cb$ be the projection onto $\mathrm{row}(\Cb)$, and for $\forall \bW \in S_c(1)$, $\exists~\mathrm{vector}~\bL$ that $\bW = \Pb_{\mathrm{row}(\Cb)} \bL$. 
Then we have
%let $\eb_j$ be the $j$th canonical basis vector and $\gb_j = (\Ib - \Pb_{[i_0]}) \Cb^+ \eb_j$, then 
\begin{align}
\label{a3.34}
\mathbb{I}_2
= \boldsymbol{\eta}^{\T} \left(\Ib - \Pb_{[i_0]}\right) \Pb_{\mathrm{row}(\Cb)} \bL
\le \big \|\boldsymbol{\eta}^{\T} \left(\Ib - \Pb_{[i_0]}\right) \Cb^+\big\|_{\infty} \cdot \big\| \Cb \bL \big \|_1 
\le \big \|\boldsymbol{\eta}^{\T} \left(\Ib - \Pb_{[i_0]}\right) \Cb^+\big \|_{\infty},
%\le \underbrace{\|\boldsymbol{\eta}\|_{\infty}}_{\mathbb{I}_{21}} \cdot \underbrace{\underset{1 \le j \le T-1}{\max} \|\gb_j\|_2}_{\mathbb{I}_{22}},
\end{align}
where the first inequality holds with Holder's inequality and the second inequality holds with the fact that $\|\Cb \bL\|_1 = \| \Cb  \Cb^+ \Cb  \bW\|_1 = \| \Cb \bW\|_1 \le 1$.
To further bound $\mathbb{I}_2$, let $\be_j$ be the $j$th canonical basis vector and $\bg_j = (\Ib - \Pb_{[i_0]}) \Cb^+ \be_j$.
let $\bu_j = (u_{j1}, u_{j2}, \ldots, u_{j(T-1)})^{\T}$, $j = 1, \ldots, T-1$, as the $j$th column of $\Ub$, then we obtain
\begin{align}
\label{a3.335}
\big \|\bg_j \big \|_2^2
=\big \| \left [\boldsymbol{0}, \Vb_{\left [T-1 \right ]/ \left [i_0 \right]} \right] \bXi^{-1} \Ub^{\T} \be_j \big \|_2^2  
= \sum_{i = i_0+1}^{T-1} \frac{u_{ji}^2 }{\xi_i^2},
%\le \frac{2}{T} \sum_{i = i_0+1}^{T-1} \frac{1}{\xi_i^2} 
%\le \frac{4T}{\pi^2 i_0},
\end{align}
where $[\boldsymbol{0}, \Vb_{[T-1]/  [i_0]}]$ can be obtained by substituting first $i_0$ columns of $\Vb$ with $\boldsymbol{0}$.
By relating $\Cb$ with finite difference operator,  \cite{wang2016trend} shows that $u_{ij} = \sqrt{2/T} \sin(\pi i j /T)$ and $\xi_i = 2\sin\{\pi (i-1)/(2T)\}$.
Then the upper bound for $\|\bg_j \|_2^2$ is 
\begin{align}
\label{a3.336}
 \sum_{i = i_0+1}^{T-1} \frac{u_{ji}^2 }{\xi_i^2}
&\le \frac{2}{T} \sum_{i = i_0+1}^{T-1} \frac{1}{\xi_i^2} \nonumber \\
%&\le \frac{4T}{\pi^2 i_0}\\
%&\le \frac{2}{T} \sum_{i = i_0+1}^{T-1} \frac{1}{\xi_i^2} \\
& = \frac{2}{T} \sum_{i = i_0+1}^{T-1} \frac{1}{4 \sin^2 \left( \pi \left(i - 1\right)/\left(2T\right)\right)} \nonumber\\
&\le 2\int_{\left(i_0 - 1\right)/T}^{\left(T-2\right)/T} \frac{1}{4 \sin^2 \left( \pi x/2\right)} dx \nonumber\\
& = \frac{\cot \left\{\pi \left(i_0 - 1\right)/\left(2T\right)\right\}}{\pi} \nonumber \\
& \le \frac{4T}{\pi^2 i_0},
\end{align}
where the first equality holds by $\sin(\pi i j /T) \le 1$ and the last inequality holds by $\cot\left(x\right) \le 1 / x$ and $ i_0 / (i_0 -1) \le 2$.
Recall that $\boldsymbol{\eta} \sim N(\boldsymbol{0}, \Qb)$, then $\bg_j^{\T}  \boldsymbol{\eta} \sim N(\boldsymbol{0}, \bg_j^{\T} \Qb \bg_j)$. 
Since $\|\boldsymbol{\eta}^{\T} (\Ib - \Pb_{[i_0]}) \Cb^+ \|_{\infty} 
= \underset{1 \le j \le T-1}{\max} |\boldsymbol{\eta}^{\T} \bg_j |
= \underset{1 \le j \le T-1}{\max} | \bg_j^{\T} \boldsymbol{\eta}|$,
\begin{align}
\label{a3.21}
P \left \{ \max_{1 \le j \le T-1} \big| \bg_j^{\T} \boldsymbol{\eta} \big| > 4 \sqrt{ \frac{\big \| \Qb \big\|_{\mathrm{op}} T \log{(T-1)}}{\pi^2 i_0}}  \right \} 
&\le \sum_{j = 1}^{T-1} P \left \{\big| \bg_j^{\T} \boldsymbol{\eta} \big| > 2 \sqrt{ \big \| \Qb \big\|_{\mathrm{op}} \big\| \bg_j \big\|_2^2 \log{(T-1)}}  \right \} \nonumber \\
&\le \sum_{j = 1}^{T-1} P \left \{\big| \bg_j^{\T} \boldsymbol{\eta} \big| > 2 \sqrt{ \bg_j^{\T} \Qb \bg_j \log{(T-1)}}  \right \} \nonumber \\
&\le \frac{1}{2 (T-1) \sqrt{\log(T-1)}},
\end{align}
where the first inequality follows that $\bg_j^{\T} \Qb \bg_j = \|\Qb^{\frac{1}{2}} \bg_j\|_2^2 \le \|\Qb\|_{\mathrm{op}} \|\bg_j\|_2^2$ and the last inequality follows Lemma~\ref{ao4}. 

Let $D = 8\sqrt{T\log(T) / (\pi^2 i_0)}$, then the upper bound for $\mathbb{I}_2$ is 
\begin{align}
\label{a3.28}
\mathbb{I}_2 \le \sqrt{\big \| \Qb \big\|_{\mathrm{op}}} D,
\end{align}
with probability at least $1 - 1 / \{2 (T-1) \sqrt{\log (T-1)}\}$. 
Substitute (\ref{a3.20}) and (\ref{a3.28}) into (\ref{a3.337}
), we have
\begin{align}
\label{a3.22}
\boldsymbol{\eta}^{\T} \bW 
\le \sqrt{2i_0 \big \|\Qb_{[i_0]}\big\|_{\mathrm{op}}}\big\|\bW\big\|_2 + \sqrt{\big \| \Qb \big\|_{\mathrm{op}}} D
\le \sqrt{2i_0 \big \|\Qb\big\|_{\mathrm{op}}}\big\|\bW\big\|_2 + \sqrt{\big \| \Qb \big\|_{\mathrm{op}}} D,
\end{align}
with probability at least $1 - \{2 (T-1) \sqrt{\log (T-1)}\}^{-1} - 2 \exp(-i_0)$. 
The second inequality follows Lemma~\ref{ao3}.
%In the following proof, we would use (\ref{a3.22}) to get the tail behavior of $\frac{1}{nT} \boldsymbol{\eta}^{\T} (\bar{\bDelta}_{ij} - \bDelta_{ij})$. 

Step (ii): since
\begin{align}
\label{a3.29}
\frac{1}{nT} \boldsymbol{\eta}^{\T} \left (\bar{\bDelta}_{ij} - \bDelta_{ij} \right)
 = \underbrace{\frac{1}{nT} \boldsymbol{\eta}^{\T}\Pb_{\mathrm{row}(\Cb)} \left (\bar{\bDelta}_{ij} - \bDelta_{ij} \right )}_{\mathbb{I}_3}
 + \underbrace{\frac{1}{nT} \boldsymbol{\eta}^{\T}\left(\Ib - \Pb_{\mathrm{row}(\Cb)}\right) \left (\bar{\bDelta}_{ij} - \bDelta_{ij} \right)}_{\mathbb{I}_4}.
\end{align}
We bound $\mathbb{I}_3$ and $\mathbb{I}_4$ in (\ref{a3.29}) separately. 
Upper Bound for $\mathbb{I}_3$ in (\ref{a3.29}):
%For the inequality $\boldsymbol{\eta}^{\T} \Xb \ge \sqrt{i_0} \|\Xb\|_2 + \delta D$, 
substituting $\bW = \Pb_{\mathrm{row}(\Cb)} (\bar{\bDelta}_{ij} - \bDelta_{ij}) / \| \Cb (\bar{\bDelta}_{ij} - \bDelta_{ij})\|_1$ into (\ref{a3.22}) and applying Holder's inequality several times, we have
\begin{align}
\label{a3.30}
\mathbb{I}_3
&\le \frac{\sqrt{\big \| \Qb \big\|_{\mathrm{op}}} D}{nT} \big \| \Cb \bar{\bDelta}_{ij} \big \|_1 + \frac{\sqrt{\big \| \Qb \big\|_{\mathrm{op}}} D}{nT} \big \| \Cb \bDelta_{ij} \big \|_1 + \frac{\sqrt{2i_0 \big \|\Qb \big \|_{\mathrm{op}}}}{nT} \big \|\Cb^+ \Cb \big\|_2 \cdot \big \| \bar{\bDelta}_{ij} - \bDelta_{ij} \big \|_2 \nonumber \\
& \le \frac{\sqrt{\big \| \Qb \big\|_{\mathrm{op}}} D}{nT}\big \| \Cb \bar{\bDelta}_{ij}\big\|_1 + \frac{\sqrt{\big \| \Qb \big\|_{\mathrm{op}}} D}{nT} \big\| \Cb \bDelta_{ij}\big\|_1 + \frac{\sqrt{2i_0\big\|\Qb\big\|_{\mathrm{op}}}}{nT} \big\| \bar{\bDelta}_{ij} - \bDelta_{ij} \big\|_2,
\end{align}
with probability at least $1 - \{2 (T-1) \sqrt{\log (T-1)}\}^{-1} - 2 \exp(-i_0)$.
The second inequality follows the fact that $\Cb^+ \Cb$ is idempotent. 

Upper Bound for $\mathbb{I}_4$ in (\ref{a3.29}):
\begin{align}
\label{a3.36}
\mathbb{I}_4
\le \frac{1}{nT} \big \| \boldsymbol{\eta}^{\T}\left(\Ib - \Pb_{\mathrm{row}(\Cb)} \right) \big \|_2 \cdot \big \| \bar{\bDelta}_{ij} - \bDelta_{ij} \big\|_2 
\le \frac{\sqrt{2\big\| \Qb\big\|_{\mathrm{op}}\log\left(T \right)}}{nT} \big \| \bar{\bDelta}_{ij} - \bDelta_{ij} \big\|_2,
\end{align}
with probability at least $1 - 1 / \{T \sqrt{2\log{(T)}}\}$. 
The second inequality is obtained by,
\begin{align*}
\PP \left\{ \| \boldsymbol{\eta}^{\T}(\Ib - \Pb_{\mathrm{row}(\Cb)})\|_2 > \sqrt{2\big\| \Qb\big\|_{\mathrm{op}}\log\left(T \right)} \right \}  
&= \PP \left\{ \frac{\big | \boldsymbol{1}^{\T} \boldsymbol{\eta} \big |}{\sqrt{T}} > \sqrt{2\big\| \Qb\big\|_{\mathrm{op}}\log\left(T \right)} \right\} \\
& = \PP \left\{ \big | \boldsymbol{1}^{\T} \boldsymbol{\eta} \big | > \sqrt{2\big \|\boldsymbol{1}\|_2^2\big\| \Qb\big\|_{\mathrm{op}}\log\left(T \right)} \right\} \\
& \le \PP \left\{ \big | \boldsymbol{1}^{\T} \boldsymbol{\eta} \big | > \sqrt{2 \boldsymbol{1}^{\T} \Qb \boldsymbol{1} \log\left(T \right)} \right\}\\
& \le \frac{1}{T \sqrt{2 \log{\left(T\right)}}}.
\end{align*}
Substituting (\ref{a3.30}) and (\ref{a3.36}) into (\ref{a3.29}), we have the desired conclusion
\begin{align*}
\frac{1}{nT} \boldsymbol{\eta}^{\T} \left (\bar{\bDelta}_{ij} - \bDelta_{ij} \right)
 \le \frac{\sqrt{2\big\|\Qb\big\|_{\mathrm{op}}} \left\{\sqrt{i_0} + \sqrt{\log\left(T\right)} \right\} }{nT} \big\| \bar{\bDelta}_{ij} - \bDelta_{ij} \big\|_2
  + \frac{\sqrt{\big \| \Qb \big\|_{\mathrm{op}}} D}{nT} \left( \big\| \Cb \bar{\bDelta}_{ij}\big\|_1 + \big\| \Cb \bDelta_{ij}\big\|_1 \right),
\end{align*}
with probability at least $1 - \{2 (T-1) \sqrt{\log (T-1)}\}^{-1} - 2 \exp(-i_0) - 1 / \{T \sqrt{2\log{(T)}}\}$.

%%%%%%%%%%%%%%%%%%%%%%%%%%%%%%%%%%%%
%%%%%%%%%%%%%%%%%%%%%%%%%%%%%%%%%%%%
% Additional Lemmas
%%%%%%%%%%%%%%%%%%%%%%%%%%%%%%%%%%%%
%%%%%%%%%%%%%%%%%%%%%%%%%%%%%%%%%%%%
\subsection{Some Technical Lemmas}
\label{asub: additional lemmas}
In this section, we provide several lemmas that are useful for proving 
Theorems~\ref{th1}--\ref{th2} and Lemmas~\ref{a1}--\ref{a2}.

\begin{lemma}[\sf Lemma 3.3 in \cite{houdre2003exponential}]
\label{ao6}
Let $(Z_m, m \in N)$ be a martingale. For all $k \ge 2$, let
$G_m^k = \sum_{l = 1}^m \EE  \{  (Z_l - Z_{l-1} )^k | \mathcal{F}_{l-1} \}$, where $\mathcal{F}_{l-1}$ is the filter containing all the information up to $l-1$. 
For $\forall \rho > 0$, let
$R_m = \sum_{k = 2}^{\infty} \rho^k G_m^k / k!$ and $R_m^{'} = \sum_{k = 2}^{\infty} (-1)^k \rho^k G_m^k / k!$.
%Then for integers $n \ge 1$ and for all $\eta$ such that for all $i \le n$, $\EE \{ \exp(|\eta(Z_i - Z_{i-1})|) \} \le \infty$ and let
%which is a super-martingale. 
If $Z_0 = 0$, then 
$$\EE \left\{ \exp \left( \rho Z_m  - R_m\right) \right\} \le 1;
\quad
\EE \left\{ \exp \left( -\rho Z_m  - R_m^{'}\right) \right\} \le 1.$$
\end{lemma}

\begin{lemma}[\sf Bernstein's inequality in \cite{rigollet2015high}]
\label{ao1}
Let $X \sim \mathrm{subE}(\nu)$ and $\EE(X) = 0$, then for any $t > 0$,
$$\PP \left( \big |X \big | > t\right) \le \exp\left \{ - \frac{1}{2} \min \left(\frac{t^2}{\nu^2}, \frac{t}{\nu}\right)\right \}.$$
\end{lemma}

\begin{lemma}[\sf Hanson-Wright inequality in \cite{rudelson2013hanson}]
\label{ao5}
Let $\Zb = (Z_1, \ldots, Z_n) \in \RR^n$ be a random vector with independent components $Z_i$ such that  $\EE (Z_i) = 0$ and $\|Z_i\|_{\psi_2} \le K$, where $\|\cdot\|_{\psi_2}$ is the  sub-gaussian norm. 
Let $\Qb$ be an $n \times n$ matrix. 
Then, for every $t \ge 0$, we have
$$\PP \left \{ \big |\Zb^{\T} \Qb \Zb - \EE (\Zb^{\T} \Qb \Zb) \big | > t \right \} \le 2 \exp \left \{ -c \min \left ( \frac{t^2}{K^4 \big \|\Qb\big \|_\mathrm{F}^2},  \frac{t}{K^2 \big \|\Qb \big\|_{\mathrm{op}}} \right)\right \},$$
where $\|\Qb\|_\mathrm{F}$ is the Frobenius norm and $\| \Qb \|_{\mathrm{op}}$ is the operator norm.
\end{lemma}

\begin{lemma}[\sf Proposition 1.1 in \cite{rigollet2015high}]
\label{ao4}
Let $X \sim N(\mu, \sigma^2)$, then for any $t > 0$,
$$\PP \left( \big |X - \mu \big | > t\right) \le \frac{\sigma}{t} \exp\left ( - \frac{t^2}{2\sigma^2}\right ).$$
\end{lemma}

\begin{lemma}[\sf Theorem 8.1.7 in \cite{golub2012matrix}] 
\label{ao3}
If $\Qb \in \RR^{T \times T}$ is symmetric and $\Qb_{t} = \Qb(1:t, 1:t)$, then
$$\lambda_{t+1}(\Qb_{t+1}) \le \lambda_{t}(\Qb_{t}) \le \lambda_{t}(\Qb_{t+1}) \le \cdots \le \lambda_{2}(\Qb_{t+1}) \le \lambda_{1}(\Qb_{t}) \le \lambda_{1}(\Qb_{t+1}),$$
for $t = 1,\ldots,T - 1$.

\end{lemma}

\bibliographystyle{ims}
\bibliography{reference} 
\end{document}